\documentclass[a4paper, 12pt]{article}
\usepackage[ansinew]{inputenc}
\usepackage{amssymb}
\usepackage{amsfonts}
\usepackage{amsmath}
\usepackage{hyperref}
\usepackage{dsfont} 
\usepackage{graphicx} 
\numberwithin{equation}{section}
\numberwithin{figure}{section}
\numberwithin{table}{section}
\newcommand{\captionfonts}{\small}
\makeatletter 
\long\def\@makecaption#1#2{%
\vskip\abovecaptionskip
\sbox\@tempboxa{{\captionfonts #1: #2}}%
\ifdim \wd\@tempboxa >\hsize
{\captionfonts #1: #2\par}
\else
\hbox to\hsize{\hfil\box\@tempboxa\hfil}%
\fi
\vskip\belowcaptionskip}
\makeatother 
\newcommand{\pa}{\paragraph}
\newcommand{\up}{\uparrow}
\newcommand{\down}{\downarrow}
\newcommand{\qed}{\begin{flushright}$\hfill \ensuremath{\Box}$\end{flushright}}
\newtheorem{theorem}{Theorem}[section]

\newtheorem{prop}{Proposition}[section]

\newtheorem{defi}{Definition}[section]

\begin{document}

\begin{titlepage}
\vspace*{-3cm}\hspace{7.5cm}\includegraphics{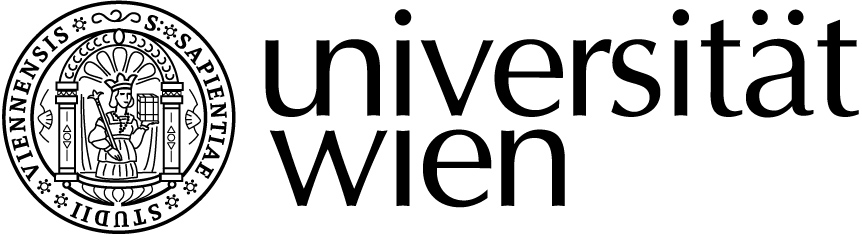}
\vspace{3cm}
\begin{center}
\begin{huge}
		\textbf{\sffamily{DIPLOMARBEIT}}\end{huge}\\
\vspace{20mm}
\sffamily{Titel der Diplomarbeit}\\
\vspace{7mm}
\begin{Large}\textbf{Relativistic Effects in Quantum Entanglement }\end{Large}	\\
 \vspace{30mm}
\sffamily{angestrebter akademischer Grad}\\
\vspace{5mm}
		\begin{Large}
		Magister der Naturwissenschaften (Mag.\:rer.\,nat.)
		\end{Large}\\
\vspace{3cm}		

\begin{tabular}{ll}
Verfasser: 	&		Nicolai Friis \\[3mm]
Matrikel-Nummer:	&	0402875 \\[3mm]
Studienrichtung (lt. Studienblatt): & A 411 Diplomstudium Physik UniStG \\[3mm]
Betreuer: & Ao. Univ. Prof. Dr. Reinhold A. Bertlmann\\[2cm]
Wien, am 3.Februar 2010 &
\end{tabular}
\end{center}
\setcounter{page}{1}
\pagenumbering{Roman}
\end{titlepage}
\newpage\thispagestyle{empty}\vspace*{20cm}\newpage
\section*{Acknowledgements}\setcounter{page}{3}

I want do dedicate this work to the people, without whom I could not have achieved this. Over the months there were numerous persons who had the right words at the right time for me and kept me going. I want to thank all of these people for their support. Clearly all of them have many good qualities more, than I list here, but I want to mention some of those, that were particularly important to me, when writing this thesis. I want to thank Verena Hofstätter, for always being right, and still listening to my musings about physics, my parents, Elisabeth and Hans Petter, for their patience and benevolence in stressful times, my close friends, Julia Kühne, Michael Albrecht, and Michaela Graf, for believing in me, and for their inimitable humor, Andreas Gabriel, for his t-shirt printing skills, Patrick Ludl, for his cooky-supply, Karin Picek and Ruth Bogoevski, for their coffee-supply, Patricia Schmidt, Georg Kopsky, and Helmuth Hüffel, for Trieste, Marcus Huber, for his mathematica skills, Beatrix Hiesmayr, Jakob Yngvason, and Frank Verstraete, for their counsel, and constructive criticism, Ivette Fuentes-Schuller, for her enthusiasm, and encouragement, and last, but not least, I want to thank Reinhold Bertlmann, for his trust, experience and unflinching verve.

\newpage\thispagestyle{empty}\vspace*{20cm}\newpage
\tableofcontents
\setcounter{tocdepth}{2}
\newpage\thispagestyle{empty}\vspace*{20cm}\newpage
\newpage
\addtocontents{toc}{\vspace{-2ex}}
\section*{Introduction}\label{introduction}\addcontentsline{toc}{section}{Introduction}

What is it, that makes quantum mechanics so special, so fascinating as a physical theory? What lies at the heart of this construct, that is applicable in such a huge variety of situations? From the early days of quantum physics on, scientists such as Erwin Schr\"odinger \cite{schroedinger35}, Albert Einstein \cite{EPR35}, and many others, were fascinated by the phenomenon of \emph{entanglement}. It certainly is the key element to the mysteries contained in quantum mechanics and after over seventy years of struggle to understand and interpret this peculiar feature of quantum physics, it still is a concept hard to grasp. The term ``entanglement" itself hints at an intimate relationship between physical systems, an inseparability of objects, and properties thereof, that in a classical world of everyday life seems inconceivable.\\

We are used to a world, where objects can at will be separated and be treated individually, creating images of clearly distinguished particles in our mind. However, this mind-set can not be kept, if we wish to interpret the world in terms of quantum physics. Even more so, when quantum mechanics is combined with special relativity, dictating us that the naive view of individual particles must be abandoned. Although the unison of these great theories has been studied for a long time in the form of quantum field theories, the analysis of quantum entanglement in the framework of relativistically moving observers has only for a short time been a subject of interest.\\

Aim of this work. The main interest of this work is to investigate how a change of inertial frame, as implemented by a Lorentz boost, affects the amount and distribution of entanglement in bipartite quantum systems and the resulting consequences for entanglement applications, such as tests of reality in Bell-type experiments. Furthermore we aim at unifying the different views on this subject into a cohesive picture, satisfying the vantage points of quantum information theory, as well as quantum field theory. The main obstacles in this agenda are the contrasting notations and the different quantities of interest of both fields, which need to be clarified and carefully compared to ensure understanding of the topic. Therefore this work is structured in the following way.\\

In Chapter \ref{chap:QM description} we will introduce the formalism of (non-relativistic) quantum mechanics, establishing the fundamentals of the theory necessary for later investigation, as well as fix notations and their interpretations. We will then proceed with presenting the concept of entanglement, possible ways of detection, and quantification, and its applications to Bell inequalities, covering the aspects of quantum information theory in Chapter \ref{chap:entanglement and nonlocality}, before continuing to special relativity, and the resulting consequences for a relativistic quantum theory in Chapter \ref{chap:relativistic quantum}, which will establish the notational connections to allow relativistic treatment of quantum entanglement in the representation of ordinary quantum information theory.\\

At this point we will have gathered all tools, and insights to study various situations of entangled spin $\tfrac{1}{2}$ particles, as observed from different inertial frames, which are presented in Chapter \ref{chap:entanglement and wigner rotations}. The main results about entanglement between different partitions of the Hilbert space, and the possible changes thereof, also found in Ref.~\cite{friisetal09}, are discussed for a variety of states and different observers.\\

\newpage
\addtocontents{toc}{\vspace{-1ex}}
\setcounter{page}{1}
\pagenumbering{arabic}
\section[Quantum Mechanical Description of Physical Systems]{Quantum Mechanical Description of\newline Physical Systems}\label{chap:QM description}

	\subsection{Principles of Quantum Mechanics}\label{sec:principles of QM}
	
In quantum mechanics the characterizing features of physical systems are most often referred to as the \emph{states} of these systems. Classically, the state of a system at a given time would be described by a point (or for statistical ensembles, a distribution) in phase space, or appropriate trajectories therein if time evolution was considered, specifying position and momentum of the system. At any time, the system would appear to be in a well defined state, while the dynamics of the system would be governed by the Hamilton \mbox{function} and the corresponding equations of motion.\\

We will not investigate the classical description any further in this work, but do now want to show how a quantum description, although it departs from such well imaginable constructs, allows for very interesting designs, namely entanglement, on a very basic level of the theory. To fully understand, why entanglement is a consequence of the quantum mechanical model, and how it can be implemented in terms of physical quantities, we need to establish the basic corpus of quantum theory, which in more detail can be found in textbooks, such as \cite{cohentannoudjiQM}, \cite{griffithsQM}, or \cite{sakuraiQM}.\\

The term ``state" will appear in different contexts throughout this whole work, often implemented via different mathematical objects, such as vectors, operators or spinors. Therefore we establish an interpretation of this term, applicable to any of the concepts mentioned above, even for the classical \mbox{description}.\\
\begin{center}\parbox{0.80\textwidth}{The \emph{state} of physical system is the collection of all accessible information about that system, encoded in an appropriate mathematical object.}\end{center}\ \\

		\subsubsection{Hilbert Space}\label{subsec:hilbert space}
	
The space underlying the physical states in quantum mechanics is called \emph{Hilbert space}, a vector space, which we will define as follows.

\begin{defi}\label{def:hilbert space}
	\begin{tabbing} \hspace*{3.5cm}\=\kill
			\> A \emph{Hilbert Space} $\mathcal{H}$ is a complete function space with\\
			\> scalar product in $\mathds{C}$.
	\end{tabbing}
\end{defi}

The property of \emph{completeness} ensures, that all \emph{Cauchy-sequences}, sequences whose elements lie ever closer to each other, converge to elements of the vector space. Together with the linearity of the vector space we can conclude that all linear combinations of state vectors are again state vectors, i.e. the physical states obey the \emph{superposition principle}.\\

The elements of the Hilbert space, such as defined in Def.~\ref{def:hilbert space}, are called \emph{wave functions}, usually denoted by the letters $\psi$ or $\phi$, which we require to be normalizable for reasons of interpretation, see Eq.~\eqref{eq:probability interpretation}. Defining the inner product $\left\langle\,.\,\right|\left..\,\right\rangle$ in $\mathcal{H}$ as

\begin{defi}\label{def:scalar product}
\hspace*{1cm}$\left\langle \psi\right.\left|\,\phi\right\rangle\,:=
			\,\int\limits^{\infty}_{-\infty}\!d^{3}\!x\,\psi^{*}(\vec{x})\,\phi(\vec{x})$ , with $\psi,\phi\in\mathcal{H}$
\end{defi}
and where the asterisk indicates complex conjugation, we immediately see that the \emph{normalization} requirement,

\begin{equation}
||\psi||^{2}\,=\,\int\limits_{-\infty}^{\infty}\!d^{3}\!x\ \psi^{*}(\vec{x})\,\psi(\vec{x})\,=
\,\int\limits_{-\infty}^{\infty}\!d^{3}\!x\ |\psi(\vec{x})|^{2}\,<\,\infty\ \ ,
\label{eq:normalization condition}
\end{equation}
restricts us to square integrable functions, $\psi\in\mathcal{L}_{2}$. Although it would suffice to require the norm of the vectors to be finite, it is convenient to choose it equal to one for all Hilbert state vectors. Later on, in Section \ref{subsubsec:spin in quantum mechanics}, we will expand this current restriction to include spin in our discussion.\\

At this point we need to make a clear distinction between wave functions and what we will call state vectors. Let us denote a \emph{state vector}, also called ``ket", of our Hilbert space $\mathcal{H}$, corresponding to a wave function $\psi(x)$, by $\left|\,\psi\,\right\rangle$, and the co-vector, referred to as ``bra", an element of the dual vector space $\mathcal{H}^{*}$, i.e. the space of linear functionals over $\mathcal{H}$, by $\left\langle\,\psi\,\right|$. This notation, called \emph{Dirac notation}, allows a simple relabeling of the basis vectors of any given basis in $\mathcal{H}$, usually a complete orthonormal basis, by the simple exchange

\begin{equation}
\left|\,\psi_{\mathrm{n}}\,\right\rangle\ \ \rightarrow\ \ \left|\,n\,\right\rangle \ \ .
\label{eq:dirac notation labels}
\end{equation}

Furthermore, in a slight abuse of notation, we can now regard the inner product (Def.~\ref{def:scalar product}) as a product of``bra" and ``ket" with the following properties.

\begin{equation}
\left\langle \phi\right.\left|\psi\right\rangle^{*}\,=\,\left\langle \psi\right.\left|\phi\right\rangle
\label{eq:c.c. of scalar product}
\end{equation}\vspace*{1mm}
\begin{equation}
\left\langle \phi\right.\left|c_{1}\psi_{1}\,+\,c_{2}\psi_{2}\right\rangle\,=
\,c_{1}\left\langle \phi\right.\left|\psi_{1}\right\rangle\,+
\,c_{2}\left\langle \phi\right.\left|\psi_{2}\right\rangle
\ \ \ \mbox{linear in ``ket"}
\label{eq:ket lineartiy}
\end{equation}\vspace*{1mm}
\begin{equation}
\left\langle c_{1}\phi_{1}\,+\,c_{2}\phi_{2}\right.\left|\psi\right\rangle\,=
\,c_{1}^{*}\left\langle \phi_{1}\right.\left|\psi\right\rangle\,+
\,c_{2}^{*}\left\langle \phi_{2}\right.\left|\psi\right\rangle
\ \ \ \mbox{antilinear in ``bra"}
\label{eq:bra antilineartiy}
\end{equation}\vspace*{1mm}
\begin{equation}
\left\langle \psi\right.\left|\psi\right\rangle\,>\,0\ \ \ \ \forall\,\psi\,\neq\,0,\ \ \ \
\left\langle \psi\right.\left|\psi\right\rangle\,=\,0 \ \ \ \Leftrightarrow \ \ \ \psi\,\equiv\,0
\ \ \ \mbox{positive definite}\\
\label{eq:positive definiteness}
\end{equation}\vspace*{1mm}

Due to the reflexivity of the Hilbert space the dual vectors $\left\langle\,\psi\,\right|$, are related to their corresponding vectors $\left|\,\psi\,\right\rangle$ by Hermitian conjugation, i.e. transposition and complex conjugation, such that the bi-dual (co-)vector is again the same (co-)vector.

\begin{equation}
\left|\,\psi\,\right\rangle^{\dagger}\,=\,\left\langle\,\psi\,\right|\ \ ,\ (\left|\,\psi\,\right\rangle^{\dagger}\,)^{\dagger}\,=\,\left|\,\psi\,\right\rangle
\label{eq:hermitian conjugation ket to bra}
\end{equation}

This correspondence between a ket, which represents a physical state, and its dual vector, allows us to regard the inner product as a \emph{transition amplitude} between the states of the Hilbert space. This suggests the following interpretation.\\

\begin{center}\parbox{0.80\textwidth}{The \emph{probability} for a transition between two physical states $\psi$ and $\phi$ is given by the modulus squared of their scalar product.}\end{center}
\begin{equation}
\textit{P}(\psi\rightarrow\phi)\,=
\,\left|\,\left\langle\,\phi\right.\left|\,\psi\right\rangle\,\right|^{2}
\label{eq:probability interpretation}
\end{equation}
\vspace*{3mm}

We can act upon the states of $\mathcal{H}$ with a linear operator A, where

\begin{defi}\label{def:linear operator}
 	\begin{tabbing}
		\hspace*{3.5cm}\=\hspace*{1cm}\=\kill
			\>A is called a \textbf{\textit{linear operator}}, if for $A\,\psi_{1}(x)\,=\phi_{1}(x)$ \\
			\>and $A\,\psi_{2}(x)\,=\phi_{2}(x)$, where  $\psi_{1},\,\psi_{2},\,\phi_{1},\,\phi_{2}\,\in\,\mathcal{L}_{2}$,
				follows \\[2mm]
			\> that $A(c_{1}\psi_{1}\,+\,c_{2}\psi_{2})\;=\;c_{1}\phi_{1}\,+\,c_{2}\phi_{2} \ \ \ \ c_{1},c_{2} \in \mathbb{C}$\,,\\
	\end{tabbing}
\end{defi}
such that A maps a state $\psi$ to a state $\phi$.
\begin{equation}
A\left|\,\psi\,\right\rangle\,=\,\left|\,A\psi\,\right\rangle\,=\,\left|\,\phi\,\right\rangle
\label{eq:operator in hilbert space}
\end{equation}

The notation of Eq.~\eqref{eq:operator in hilbert space} suggests to define the \emph{adjoint operator} in the following way:

\begin{defi}\label{def:adjoint operator}
 	\begin{tabbing}
		\hspace*{3.5cm}\=\hspace*{2cm}\=\kill
			\>$A^{\dagger}$ is called the \textbf{\textit{adjoint operator}} to $A$,
				if $\ \forall \ \psi,\,\phi\,\in\,L_{2}$\\[3mm]
			\> \> \begin{large}$\left\langle A^{\dagger}\psi\right.\left|\phi\right\rangle\,=
						 \,\left\langle \psi\right.\left|A\,\phi\right\rangle$\end{large}\;.
	\end{tabbing}
\end{defi}
A special class of operators, which is of particular interest, is the class of \emph{Hermitian operators}\footnote{We will not distinguish between Hermitian and self-adjoint operators, for which the domains of the operator and its adjoint are identical, since the difference is not essential to this work.}, which includes all physical \emph{observables}.

\begin{defi}\label{def:hermitian operator}
 	\begin{tabbing}
		\hspace*{3.5cm}\=\hspace*{3cm}\=\kill
			\>An Operator $A$ is called \textbf{\textit{Hermitian}}, if \\[3mm]
			\> \> \begin{large}$A^{\dagger}\,=\,A$\end{large}\\[2mm]
			\> and the domains satisfy  $\textit{D}\,(A^{\dagger})\supset\textit{D}\,(A)$\ \ .\\
	\end{tabbing}
\end{defi}

We can further exploit the Dirac notation to construct the outer product of vector and co-vector,
\begin{equation}
P\,=\,\left|\,\psi\,\right\rangle\left\langle\,\psi\,\right|\ \ ,
\label{eq:projection operators}
\end{equation}
resulting in a \emph{projection operator}, defined as

\begin{defi}\label{def:projection operator}
 	\begin{tabbing}
		\hspace*{3.5cm}\=\hspace*{3cm}\=\kill
			\>An Operator $P$ is called \textbf{\textit{projection operator}}, if \\[3mm]
			\> \> \begin{large}$P^{2}\,=\,P$\end{large}\ ,
	\end{tabbing}
\end{defi}
which is automatically satisfied for normalized state vectors. The vectors of the Hilbert space can be written as a linear combination of basis vectors of any complete orthonormal system of $\mathcal{H}$.

\begin{equation}
\left|\,\psi\,\right\rangle\,=\,\sum\limits_{\mathrm{n}}\,c_{\mathrm{n}}\,\left|\,\psi_{\mathrm{n}}\,\right\rangle\,=\,\sum\limits_{\mathrm{n}}\,\left\langle\,\psi_{\mathrm{n}}\,\right|\left.\psi\,\right\rangle\,\left|\,\psi_{\mathrm{n}}\,\right\rangle
\label{eq:cons expansion}
\end{equation}

From Eq.~\eqref{eq:cons expansion} it becomes clear immediately that any complete orthonormal system must adhere the \emph{completeness relation}

\begin{equation}
\sum\limits_{\mathrm{n}}\,\left|\,\psi_{\mathrm{n}}\,\right\rangle\left\langle\,\psi_{\mathrm{n}}\,\right|\,
=\,\mathds{1}\ \ .
\label{eq:completeness relation}
\end{equation}

If the basis vectors belong to a continuous spectrum $\left\{\,\left|\,\xi\,\right\rangle\,\right\}$, rather than a discrete one, the summation in the expansion (Eq.~\eqref{eq:cons expansion}) as well as in the completeness relation (Eq.~\eqref{eq:completeness relation}) need to be replaced by an integration, i.e.

\begin{eqnarray}
\left|\,\psi\,\right\rangle &=& \,\int d\xi\,\psi(\xi)\,\left|\,\xi\,\right\rangle\,=
		\,\int d\xi\,\left\langle\,\xi\,\right|\left.\psi\,\right\rangle\,\left|\,\xi\,\right\rangle
		\label{eq:continuous cons expansion}\\
\ \ &\mbox{and}& \int d\xi\,\left|\,\xi\,\right\rangle\left\langle\,\xi\,\right|\,=\,\mathds{1}\ \ ,
\end{eqnarray}
where $\psi(\xi)$ are the components of the state vector $\left|\,\psi\,\right\rangle$ with respect to the chosen continuous basis $\left\{\,\left|\,\xi\,\right\rangle\,\right\}$, while the ket $\left|\,\psi\,\right\rangle$ provides the basis independent notion of the state.\\

As stated before, all physical observables are represented by Hermitian operators (see Def.~\ref{def:hermitian operator}), whose eigenvalues correspond to possible measurement outcomes. Since generally the system will not be in an eigenstate of the observable measured, we can only predict mean values, so called \emph{expectation values}, of observables in a given state.

\begin{defi}\label{def:expectation value}
 	\begin{tabbing}
		\hspace*{3.5cm}\=\hspace*{3cm}\=\kill
			\>The \emph{expectation value} of an observable $\mathcal{O}$, represented\\
			\>by the Hermitian operator $\hat{\mathcal{O}}$, in the state $\left|\,\psi\,\right\rangle$ is given by\\[2mm]
			\> \> $\left\langle\right.\mathcal{O}\left.\right\rangle_{\psi}\,=
						\,\left\langle\,\psi\,\right|\,\hat{\mathcal{O}}\,\left|\,\psi\,\right\rangle$\ .
	\end{tabbing}
\end{defi}

		\subsubsection{Schr\"odinger Equation and Wave Functions}\label{subsec:schroedinger and wave functions}
		
After having established the mathematical formalism we now want to tie this in with the physical concepts. The equation of motion governing the dynamics of the Hilbert space vectors $\psi(t,\vec{x})$ is the \emph{Schr\"odinger equation}

\begin{equation}
i\,\hbar\,\frac{\partial}{\partial t}\psi(t,\vec{x})\,=\,H\,\psi(t,\vec{x})\ ,
\label{eq:schroedinger equation}
\end{equation}		
where $H$ is the \emph{Hamiltonian}, or Hamilton operator, given by
\begin{equation}
H\,=\,-\,\frac{\hbar^{2}}{2m}\Delta\,+\,V(\vec{x})\ \ .
\label{eq:hamiltonian}
\end{equation}		

The free\footnote{Since we do not study any interactions in this work this will be sufficient.} Schr\"odinger equation can be viewed as the non-relativistic energy-momentum relation, where the physical quantities energy $E$ and momentum $\vec{p}$, are replaced by operators $\hat{E}$ and $\hat{\vec{p}}$.

\begin{equation}
E\,\rightarrow\,\hat{E}\,=\,i\hbar\frac{\partial}{\partial t}\ \ , \ \ \
\vec{p}\,\rightarrow\,\hat{\vec{p}}\,=\,-i\hbar\,\vec{\nabla}\ .
\label{eq:first quantization}
\end{equation}

The solutions of the Schr\"odinger equation are the wave functions $\psi(t,\vec{x})$, which we can rewrite as the position representation of the state vector $\left|\,\psi(t)\,\right\rangle$ according to Eq.~\eqref{eq:continuous cons expansion},

\begin{equation}
\psi(t,x)\,=\,\left\langle\,x\right.\left|\,\psi(t)\,\right\rangle \ \ ,
\label{eq:wave function pos rep}
\end{equation}
where $\left|\,x\,\right\rangle$ are the eigentstates of the position operator\footnote{We must note here, that we have changed to the one dimensional case for the sake of simplicity, all arguments do however apply to the three dimensional case in equal measure, by integrating over the full space and including factors of $2\pi$ in the Fourier transforms.} $\hat{x}$.
\begin{equation}
\hat{x}\,\left|\,x\,\right\rangle\,=\,x\,\left|\,x\,\right\rangle
\label{eq:position eigenstates}
\end{equation}

It is sometimes convenient to choose a different basis, the momentum eigenstates, which, similar to Eq.~\eqref{eq:position eigenstates}, are determined by their eigenvalue equation.
\begin{equation}
\hat{p}\,\left|\,p\,\right\rangle\,=\,p\,\left|\,p\,\right\rangle
\label{eq:momentum eigenstates}
\end{equation}

To determine the corresponding wave function $\psi_{p}(x)$ in position representation, we apply the momentum operator as defined by Eq.~\eqref{eq:position eigenstates} and use the r.h.s. of Eq.~\eqref{eq:first quantization}. Solving the resulting differential equation, where the integration constant is fixed by the normalization condition, one easily gets
\begin{equation}
\left\langle\,x\right.\left|\,p\,\right\rangle\,=\,\psi_{\rm{p}}(x)\,=
\,\frac{1}{\sqrt{2\pi\hbar}}\,e^{\,ipx/\hbar}\ \ .
\label{eq:bra x ket p}
\end{equation}

We can then check the orthogonality of the position eigenfunctions.
\begin{eqnarray}\nonumber
    \left\langle\,x^{\,\prime}\right.\left|\,x\,\right\rangle 	&=&
	\,\left\langle\,x^{\,\prime}\,\right|\,\mathds{1}\,\left|\,x\,\right\rangle\,
	=\,\left\langle\,x^{\,\prime}\,\right|\,\int\!dp\,
    \left|\,p\,\right\rangle\left\langle\,p\,\right|\,\left|\,x\,\right\rangle\,
	=\,\int\!dp\,\left\langle\,x^{\,\prime}\right.\left|\,p\,\right\rangle
    \left\langle\,p\right.\left|\,x\,\right\rangle\,=\\ \nonumber
    \ &=& \,\int\!dp\ \psi_{\rm{p}}(x^{\,\prime}) \ \psi_{\rm{p}}^{*}(x)\,=\,
    \frac{1}{2\pi\hbar}\int\!dp\ \exp\left(\frac{ipx^{\,\prime}}{\hbar}\right)\ \exp\left(\frac{-ipx}{\hbar}\right)\,=\\
		\label{eq:position eigenstate orthogonality}
    \ &=& \,\frac{1}{2\pi\hbar}\int\!dp\ \exp\left(\frac{i}{\hbar}(x^{\,\prime}\,-\,x)p\right)\,=\,\delta(x^{\,\prime}-x)\,.\\ \nonumber
\end{eqnarray}

Analogously for the momentum states we get

\begin{equation}
\left\langle\,p^{\,\prime}\right.\left|\,p\,\right\rangle\,=\,\delta(p^{\,\prime}-p)\ \ .
\label{eq:momentum eigenstates orthogonality}
\end{equation}

Expanding a state vector $\left|\,\psi\,\right\rangle$ in the momentum eigenstate basis,

\begin{equation}
\left|\,\psi\,\right\rangle\,=
\,\int\!dp\,\left|\,p\,\right\rangle\left\langle\,p\right.\left|\,\psi\,\right\rangle\,=
\,\int\!dp\,\tilde{\psi}(p)\,\left|\,p\,\right\rangle\,,
\label{eq:momentum eigenstate expansion}
\end{equation}
and labeling the components of $\psi$ with respect to the momentum eigenstates by $\tilde{\psi}(p)$, the connection to the position representation $\psi(x)$ via Fourier transformation becomes apparent,

\begin{equation}
\psi(x)\,=\,\left\langle\,x\right.\left|\,\psi\,\right\rangle\,=
\,\int\!dp\,\tilde{\psi}(p)\,\left\langle\,x\right.\left|\,p\,\right\rangle\,=
\,\frac{1}{\sqrt{2\pi\hbar}}\,\int\!dp\,\tilde{\psi}(p)\,e^{\,ipx/\hbar}\ .
\label{eq:fourier transformation}
\end{equation}

The interpretation (see Eq.~\eqref{eq:probability interpretation}) of the wave function in position representation, describing a particle, can now be rephrased.

\begin{center}
\parbox{0.8\textwidth}{The probability to find the particle in an interval $\left[x,x+dx\right]$ is given by $\left|\psi(x)\right|^{2}dx$.}
\end{center}

In order for such a probability interpretation to be applied the wave functions need to be continuous, square integrable functions. Consequently position- and momentum eigenstates, i.e. plane waves in momentum- or position space respectively, cannot be regarded as physically realizable, though they are often used to simplify analysis.

		\subsubsection{Density Operators}\label{subsec:density operators}

Previously we have argued that the states of Hilbert space obey the superposition principle such that linear combinations of physical states are elements of Hilbert space themselves. This is not only mathematically justified, but also realized in nature in the form of coherent superpositions.\\

However, we have not yet accounted for the possibility of incoherent mixing of states, i.e. statistical mixtures of states. This cannot be successfully done with the state vectors we have used so far, since the states are only determined up to a global phase, i.e. $\left|\,\psi\,\right\rangle$ and $e^{i\alpha}\left|\,\psi\,\right\rangle$ represent the same physical situation. So from evaluating a probability distribution of measurements, for instance of the energy levels of a harmonic oscillator, performed on a given state, we cannot infer these phases and must therefore average over them.

\begin{equation}
\left\langle\right.e^{i\alpha}\left.\right\rangle\,=
\,\frac{1}{2\pi}\int\limits_{0}^{2\pi}\!d\alpha\,e^{i\alpha}\,=\,0
\label{eq:averaging phases}
\end{equation}

Although averaging of the phases yields a correct result for the probability distributions, since the probabilities depend only on the moduli squared of the complex amplitudes, where the phases cancel, we must conclude that the state vectors themselves cannot be used as an appropriate tool to describe statistical mixtures. We therefore introduce a new class of operators, so called \emph{density matrices}, to provide a more realistic description. For a state, described by a state vector $\left|\,\psi\,\right\rangle$, the density operator is simply given by the projector on the given state.\\

\begin{defi}\label{def:pure state density matrix}
 	\begin{tabbing}
		\hspace*{3.5cm}\=\hspace*{3cm}\=\kill
			\> The \emph{density matrix} $\rho$ for the ``pure" state $\left|\,\psi\,\right\rangle$\\
			\> is given by\\[1mm]
			\> \> $\rho\,:=\,\left|\,\psi\,\right\rangle\left\langle\,\psi\,\right|$ \ .
	\end{tabbing}
\end{defi}

The density matrix of a pure state thus uniquely defines a ray in Hilbert space, i.e. the state vector $\left|\,\psi\,\right\rangle$ up to a phase. From Def.~\ref{def:pure state density matrix} it is also imminent, that a pure state density operator satisfies the following properties:

\begin{eqnarray}	
& \rho^{2}\,=\,\rho & \mbox{projector} \label{eq:projective pure state}\\
	[2mm]
& \rho^{\dagger}\,=\,\rho & \mbox{hermiticity} \label{eq:density matrix hermitian}\\
	[2mm]
& \rho\,\geq\,0 & \mbox{positivity} \label{eq:density matrix positivity}		\\
	[2mm]
& Tr(\rho)\,=\,1 & \mbox{normalization}\ , \label{eq:density matrix normalization}
\end{eqnarray}
where the \emph{trace operation} is defined as
\begin{defi}\label{def:trace operation}
 	\begin{tabbing}
		\hspace*{3.5cm}\=\hspace*{2cm}\=\kill
			\> \> $ Tr(\mathcal{O})\,:=
			\,\sum\limits_{\rm{n}}\,\left\langle\,n\,\right|\,\mathcal{O}\,\left|\,n\,\right\rangle$ \\
			[2mm]
			\> for any complete orthonormal system $\left\{\,\left|\,n\,\right\rangle\,\right\}$.
	\end{tabbing}
\end{defi}

\begin{defi}\label{def:density matrix expectation value}
 	\begin{tabbing}
		\hspace*{3.5cm}\=\hspace*{4cm}\=\kill
			\>The \emph{expectation value} of an observable $\mathcal{O}$ in the \\
			\> state $\rho$ is given by\\
			\> \> $\left\langle\right.\mathcal{O}\left.\right\rangle_{\rho}\,=
						\,Tr(\mathcal{O}\rho)$
	\end{tabbing}
\end{defi}

In contrast to the pure states above, we can not introduce \emph{mixed states}, by choosing the density operator to be a convex sum of pure state density matrices.

\begin{defi}\label{def:mixed state density matrix}
 	\begin{tabbing}
		\hspace*{3.5cm}\=\hspace*{3cm}\=\kill
			\> The \emph{density matrix} $\rho$ for the ``mixed" state $\rho$\\
			\> is given by\\
			\> \> $\rho\,:=\,\sum\limits_{i}\,p_{\mathrm{\,i}}
						\left|\,\psi_{\mathrm{\,i}}\,\right\rangle\left\langle\,\psi_{\mathrm{\,i}}\,\right|$ \\[1mm]
			\> where $p_{\mathrm{\,i}}\in\mathds{R}_{+}$ and $\sum\limits_{i}\,p_{\mathrm{\,i}}\,=\,1\ $.
	\end{tabbing}
\end{defi}

Clearly, properties \eqref{eq:density matrix hermitian}, \eqref{eq:density matrix positivity}, and \eqref{eq:density matrix normalization} as well as our definition of the expectation value (Def.~\ref{def:density matrix expectation value}) hold for mixed density matrices as well as for pure ones. The projection property (Eq.~\eqref{eq:projective pure state}) however is no longer valid. Combining this insight with the normalization of the density operators, we can use this to define the \emph{mixedness} of a density matrix.

\begin{defi}\label{def:mixedness of density matrix}
 	\begin{tabbing}
		\hspace*{3.5cm}\=\hspace*{2.5cm}\=\kill
			\> The \emph{mixedness} of a density matrix $\rho$ is defined as\\[2mm]
			\> \> $M(\rho)\,:=\,1\,-\,Tr(\rho^{2})$ \ .\\
	\end{tabbing}
\end{defi}

The mixedness is bounded, i.e. $0\,\leq\,M(\rho)\leq\,1-\tfrac{1}{d}$, where $d$ is the dimension of the Hilbert space. It only vanishes for pure states and is strictly greater than zero for mixed states.\\

The density operators are elements of the so called \emph{Hilbert-Schmidt space}, where the Hilbert-Schmidt inner product $\left\langle\right..,.\left.\right\rangle_{HS}$ of two operators $A$ and $B$ is defined by the trace operation,

\begin{equation}
\left\langle\right.A,B\left.\right\rangle_{HS}\,:=\,Tr(A^{\dagger}B)\ \ .
\label{eq:HS inner product}
\end{equation}

Since the trace in Eq.~\eqref{eq:HS inner product} is defined only for a chosen basis in the Hilbert space, on which the density operators are linear operators, there is a natural association of Hilbert-Schmidt spaces with the corresponding Hilbert spaces. In fact, by defining the inner product in Eq.~\eqref{eq:HS inner product}, the Hilbert-Schmidt space effectively becomes a Hilbert space.
We can also assume, that all operators on finite dimensional Hilbert-Schmidt spaces, can be expressed as matrices, such that their components with respect to a chosen basis $\left\{\left|\,n\,\right\rangle\right\}$ are given by

\begin{equation}
\rho_{mn}\,=\,\left\langle\,m\,\right|\,\rho\,\left|\,n\,\right\rangle\ \ .
\label{eq:matrix representation on HS space}
\end{equation}

The decomposition of these density matrices $\rho$ is therefore not uniquely determined, since for different choices of decompositions into pure states $\left|\,\psi\,\right\rangle\left\langle\,\psi\,\right|$ (see Def.~\ref{def:mixed state density matrix}) the same matrix components can be generated. One particular choice to break down the  density matrix elements of a two dimensional quantum system, i.e. a \emph{qubit}, is the so called \emph{Bloch decomposition},

\begin{equation}
\rho_{\mathrm{qubit}}\,=\,\frac{1}{2}\left(\mathds{1}_{2}\,+\,\vec{a}\cdot\vec{\sigma}\right)\ ,\ \ |\vec{a}|\,\leq\,1\ \ \ ,
\label{eq:bloch decomposition}
\end{equation}
where $\vec{a}$ is a vector in $\mathbb{R}^{3}$ and $\vec{\sigma}$ is the vector of Pauli matrices, given by \eqref{eq:pauli matrices}. This decomposition is very descriptive for spin $\tfrac{1}{2}$ systems, where the vector $\vec{a}$ represents the spin orientation of the state $\rho$ in three dimensional space. If $\vec{a}$ is a unit vector, it lies on the so called \emph{Bloch sphere} and describes a pure state, while, conversely, if $|\vec{a}|<1$ the vector describes a mixed state lying inside of the sphere.

\newpage
	\subsection{Rotations \& Spin}\label{subsec:rotations and spin}

To cover all the dynamical variables of classical physics in quantum mechanics we need to introduce the
\emph{orbital angular momentum} as an operator in our theory. Clearly this is not difficult if we compose the
$i$-th component of the corresponding operator $\hat{\vec{L}}$ from the canonical position and momentum operators
(Eq.~\eqref{eq:position eigenstates} and r.h.s. of Eq.~\eqref{eq:first quantization}) as

\begin{equation}
\hat{L}^{\,i}\,=\,\varepsilon^{\,ijk}\,\hat{x}^{\,j}\,\hat{p}^{\,k} \ \ \ ,
\label{eq:orbital angular momentum i component}
\end{equation}
where $\varepsilon^{\,ijk}$ is the totally antisymmetric Levi-Civita symbol in three dimensions ($i,j,k=1,2,3$\ or alternatively\ $i,j,k=x,y,z$) and summation
convention is implied. It can be easily checked that the angular momentum operator satisfies the commutation relation

\begin{equation}
\left[\right.\hat{L}^{\,i},\hat{L}^{\,j}\left.\right]\,=
\,i\hbar\,\varepsilon^{\,ijk}\,\hat{L}^{\,k}\ \ \ ,
\label{eq:so3 lie algebra}
\end{equation}
which defines the Lie algebra $so(3)$ (see Sec.~\ref{subsec:lie groups and lie algebras}, in particular Eq.~\eqref{eq:so(3) structure constants}) of the three dimensional rotation group $SO(3)$. Furthermore, the operator $\hat{\vec{L}}$ thus constructed
is the generator of infinitesimal rotations on the Hilbert space of states, i.e. we can construct a unitary
operator $U(R(\vec{\varphi}\,))$,

\begin{equation}
U(R(\vec{\varphi}\,))\,=\,e^{-\tfrac{i}{\hbar}\vec{\varphi}\,\vec{L}}\ \ \ ,
\label{eq:angular momentum rotation generator}
\end{equation}
where $R(\vec{\varphi}\,)$ is the usual rotation matrix in three dimensions, i.e. its components are given by

\begin{equation}
R(\vec{\varphi}\,)_{\,ij}\,=\,
\cos\varphi\,\delta_{\,ij}\,+\,
(1-\cos\varphi)\,\frac{\varphi_{\,i}\,\varphi_{\,j}}{|\vec{\varphi}\,|^{2}}\,-\,
\sin\varphi\,\varepsilon_{\,ijk}\,\frac{\varphi_{\,k}}{|\vec{\varphi}\,|}
\label{eq:rotation in R3}
\end{equation}
such that for rotations about an angle $|\vec{\varphi}\,|=\varphi$ around an axis $\vec{\varphi}$,
$U(R(\vec{\varphi}\,))$ maps a state vector to that of the rotated coordinate system.

\begin{equation}
\phi(\vec{x})\,\longrightarrow\,U(R(\vec{\varphi}\,))\,\phi(\vec{x})\,=\,\psi(R^{-1}(\vec{\varphi}\,)\,\vec{x})
\label{eq:rotations on hilbert space}
\end{equation}

The Lie algebra \eqref{eq:so3 lie algebra} suggests that different components of the angular momentum
operator do not commute and therefore cannot have common eigenfunctions. The operators satisfying
this property and in addition commuting with the Hamiltonian \eqref{eq:hamiltonian}, thus supplying a complete
basis of eigenfunctions, are the squared orbital angular momentum operator\footnote{The hat symbol
$``\,\hat{\ }\,"$ will be suppressed from now on, applying it only where confusion could arise.}
$\vec{L}^{2}$ and an arbitrary component of $\vec{L}$, commonly taken to be $L^{z}$.

\begin{equation}
\left[\right.\vec{L}^{2},L^{z}\left.\right]\,=\,
\left[\right.H,L^{z}\left.\right]\,=\,
\left[\right.H,\vec{L}^{2}\left.\right]\,=\,0\ .
\label{eq:commutator L squared Lz}
\end{equation}

The corresponding eigenfunctions are the spherical harmonics $Y_{lm}$, which translate to our ket vectors as
$\left|\,l,m\,\right\rangle$.

\begin{equation}
\vec{L}^{2}\,\left|\,l,m\,\right\rangle\,=\,\hbar^{2}\,l(l+1)\,\left|\,l,m\,\right\rangle\ \ \ \mbox{and}\ \ \
L^{z}\,\left|\,l,m\,\right\rangle\,=\,\hbar\,m\,\left|\,l,m\,\right\rangle
\label{eq:spherical harmonics eigenvalue equation}
\end{equation}
where $l$ and $m$ are the azimuthal and magnetic quantum numbers respectively, and the former is $(2l+1)$-fold
degenerate, i.e. for each $l$ the magnetic quantum number can take values in integer steps in the interval
$[-l,+l]$.

		\subsubsection{Spin in Quantum Mechanics}\label{subsubsec:spin in quantum mechanics}

However, the introduction of orbital angular momentum alone could not account for the results of the famous
Stern-Gerlach experiment \cite{sterngerlach22}, showing that an additional \emph{intrinsic angular momentum},
i.e. \emph{spin}, is carried by elementary particles, the valence electrons of silver atoms in case of
\cite{sterngerlach22}, and that it is quantized, taking on two possible values.\\

As will be seen later on in Sec.~\ref{sec:Relativistic Description of Quantum Systems}, the union with
special relativity allows us to acknowledge the spin quantum number as a label for the irreducible representations
of the Poincar$\acute{e}$ group (\cite{ryderqft}), the most interesting cases of which are the spin $0$
representation - scalar particles, transforming under rotations according to Eq.~\eqref{eq:rotations on hilbert space},
 spin $1$ - vector particles, transforming like usual vector fields, i.e.

\begin{equation}
\vec{A}(\vec{x})\,\longrightarrow\,R(\vec{\varphi}\,)\,\vec{A}(R^{-1}(\vec{\varphi}\,)\,\vec{x})
\label{eq:rotation of vector field}
\end{equation}
and for our discussion most important, the spin $\tfrac{1}{2}$ representation, described (at this stage)
by two-component spinors $\psi(\vec{x})\in\mathbb{C}^{2}$, which, in a slight abuse of notation, we will
denote by the same symbol as the ordinary (scalar) wave function of Eq.~\eqref{eq:wave function pos rep}
such that for some chosen basis in $\mathbb{C}^{2}$ the spinor is of the form

\begin{equation}
\psi(\vec{x})\,=\,\begin{pmatrix} \psi_{\mathrm{1}}(\vec{x}) \\ \psi_{\mathrm{2}}(\vec{x})\end{pmatrix}
\ , \ \ \mbox{where}\ \ \
\int\limits_{-\infty}^{\infty}\!d^{3}x\,
\left(|\psi_{\mathrm{1}}(\vec{x})|^{2}\,+\,|\psi_{\mathrm{2}}(\vec{x})|^{2}\right)\,=\,1\ \ .
\label{eq:two component spinors}
\end{equation}

All previous definitions and analysis can then be applied to these spinors by simply choosing the (one-particle)
Hilbert space to be

\begin{equation}
\mathcal{H}\,=\,\left\{\,\mathcal{L}_{2}\otimes\mathbb{C}^{2},\,d\mu\,\right\}
\label{eq:spin one half hilbert space}
\end{equation}
where $d\mu$ is a suitable integration measure. The angular momentum operator corresponding to spin and
taking the place of Eq.~\eqref{eq:orbital angular momentum i component} is then given by

\begin{equation}
S^{\,i}\,=\,\frac{\hbar}{2}\,\sigma^{\,i}
\label{eq:spin operator i component}
\end{equation}
where $\sigma^{\,i}$ are the \emph{Pauli matrices}.

\begin{equation}
\sigma^{\,x}\,=\,\begin{pmatrix} \,0 & 1\, \\ \,1 & 0\,\end{pmatrix}\ ,\,
\sigma^{\,y}\,=\,\begin{pmatrix} \,0 & \!\!-i\, \\ \,i & \,0\,\end{pmatrix}\ ,\,
\sigma^{\,z}\,=\,\begin{pmatrix} \,1 & \,0 \\ \,0 & \!-1\end{pmatrix}
\label{eq:pauli matrices}
\end{equation}

Let us regard some important properties of the Pauli matrices. They are Hermitian
(Def.~\ref{def:hermitian operator}), traceless, and their square is the identity.

\begin{equation}
(\sigma^{\,i}\,)^{\dagger}\,=\,\sigma^{\,i}\ ,\ \ Tr(\sigma^{\,i}\,)\,=\,0\ , \ \ (\sigma^{\,i}\,)^{2}\,=\,\mathds{1}_{2}
\label{eq:pauli matrix properties}
\end{equation}

Furthermore they satisfy the commutation relation
\begin{equation}
\left[\right.\sigma^{\,i},\sigma^{\,j}\left.\right]\,=
\,2i\,\varepsilon^{\,ijk}\,\sigma^{\,k}\ \ \ ,
\label{eq:pauli matrices commutator}
\end{equation}
thereon together with Eq.~\eqref{eq:spin operator i component} implying that the
spin operators $S^{\,i}$ satisfy the same lie algebra \eqref{eq:so3 lie algebra} as the
angular momentum operators $L^{i}$, i.e.

\begin{equation}
\left[\right.S^{\,i},S^{\,j}\left.\right]\,=
\,i\hbar\,\varepsilon^{\,ijk}\,S^{\,k}\ \ \ ,
\label{eq:SU2 lie algebra}
\end{equation}
which will be explained in Sec.~\ref{subsubsec:SO3 and SU2} in detail. For now we can recognize that due to this
fact we can choose two commuting operators $\vec{S}^{\,2}$ and $S^{\,z}$, whose eigenfunctions will form
the basis states of choice for the one-particle Hilbert space $\mathbb{C}^{2}$ in complete analogy to
Eq.~\eqref{eq:spherical harmonics eigenvalue equation}.

\begin{equation}
\vec{S}^{\,2}\,\left|\,s,m_{s}\,\right\rangle\,=\,\hbar^{2}\,s(s+1)\,\left|\,s,m_{s}\,\right\rangle
\label{eq:spin squared eigenvalue equation}
\end{equation}
\begin{equation}
S^{\,z}\,\left|\,s,m_{s}\,\right\rangle\,=\,\hbar\,m_{s}\,\left|\,s,m_{s}\,\right\rangle
\label{eq:spin z eigenvalue equation}
\end{equation}

The degeneracy of the spin quantum number $s$ is $(2s+1)$, thus giving two possible values
for the magnetic spin quantum number $m_{s}$ for spin $\tfrac{1}{2}$ particles, and consequently
two orthogonal spin states for our particles, which are the eigenstates of the matrix
$\sigma^{\,z}$ (shown here for the basis choice of Eq.~\eqref{eq:pauli matrices}).

\begin{equation}
\left|\right.\tfrac{1}{2},\tfrac{1}{2}\left.\right\rangle\,=\,\left|\right.\up\left.\right\rangle\,=\,
\begin{pmatrix} 1 \\ 0 \end{pmatrix}\ \ ,\
\left|\right.\tfrac{1}{2},-\tfrac{1}{2}\left.\right\rangle\,=\,\left|\right.\down\left.\right\rangle\,=\,
\begin{pmatrix} 0 \\ 1 \end{pmatrix}
\label{eq:spin z eigenstates}
\end{equation}

In the context of this work, the eigenstates of Eq.~\eqref{eq:spin z eigenstates} are denoted by the symbols
$\up$ and $\down$ and not, as usual in quantum information theory, by $0$ and $1$ (see e.g.
\cite{nielsenchuangQI}).

		\subsubsection{Lie Groups \& Lie Algebras}\label{subsec:lie groups and lie algebras}

To reach the main goal of this work, i.e. describe and analyze quantum entanglement in a relativistic framework,
it is essential to study the symmetry group of special relativity, the Poincar$\acute{e}$ group, or its subgroup
the (homogeneous) Lorentz group and their representations on Hilbert space.\\

The cornerstone of this analysis was
laid by Wigner in 1939 in his seminal paper on the unitary representations of the Poincar$\acute{e}$ group
(\cite{wigner39}), the extent of which we cannot reflect here. Neither can we introduce the whole sizeable machinery
of Lie groups and Lie algebras necessary, to truly understand the generality of the concepts used in this work.
Nevertheless do we feel the need to give some basic definitions and results and redirect the reader to the
circumstantial literature available on (Lie) group theory, e.g. \cite{tunggrouptheory} or \cite{warnerliegroups}, and differential geometry
\footnote{For a detailed description of differential geometry, especially differentiable manifolds
(note Def.\ref{def:lie group}), tangent spaces and the Lie bracket (note Def.~\ref{def:lie algebra}) see
\cite{bertlmannanomalies} (chapters 2.3 and 2.6.) or \cite{waldGR} (chapter 2).} for more rigorous treatment
of the topic.\\

\begin{defi}\label{def:group}
	\begin{tabbing} \hspace*{3.5cm}\=\hspace*{2cm}\=\kill
			\> A \textbf{\emph{group}} is a set $G$, together with a map \\[2mm]
			\> \> $G \times G\,\rightarrow\,G$\\
            \> \> $(g_{1},g_{2})\,\mapsto\,g_{1}g_{2}$ \\[2mm]
            \> with the properties\\
            \> $\cdot$\ $g_{1}\,(g_{2}\,g_{3})\,=\,(g_{1}\,g_{2})\,g_{3}$\ \ \emph{(associativity)}\\[1mm]
            \> $\cdot$\ $\exists$ \emph{identity element} $e\in G$:\ \ $e\,g\,=\,g\,e\,=\,g$\\[1mm]
            \> $\cdot$\ $\exists$ \emph{inverse element} $g^{-1}\in G$:\ \ $g\,g^{-1}\,=\,g^{-1}\,g\,=\,e$\\[1mm]
            \> for all $g,g_{1},g_{2}\,\in\,G$.\\[3mm]
            \> Additionally, if \ $g\,h\,=\,h\,g\ \forall\,g,h\,\in\,G$, the group is\\
            \> called \emph{abelian} or \emph{commutative}.\\
	\end{tabbing}
\end{defi}
\begin{defi}\label{def:lie group}
	\begin{tabbing} \hspace*{3.5cm}\=\hspace*{1.5cm}\=\kill
			\> A \textbf{\emph{Lie group}} is a differentiable manifold endowed\\[1mm]
			\> with a group structure, such that the group operations \\[2mm]
      \> \> $(g_{1}\,,\,g_{2})\mapsto g_{1}\,g_{2}$ \ and\  $g\mapsto g^{-1}$\\[2mm]
      \> are differentiable $\forall\,g,g_{1},g_{2}\,\in\,G$.
	\end{tabbing}
\end{defi}
\begin{defi}\label{def:subgroup}
	\begin{tabbing} \hspace*{3.5cm}\=\hspace*{2cm}\=\kill
			\> A subset $H$ of a group $G$ is a \textbf{\emph{subgroup}} of $G$, if it is\\[1mm]
            \> closed under the group operations, i.e. $h_{1}\,h_{2}\in H$,\\[1mm]
            \> $\forall\,h_{1},h_{2}\in H$, and if $h\,\in\,H$, then $h^{-1}\,\in\,H$, especially\\[1mm]
            \> it contains the identity, $e\,\in\,H$.\\
	\end{tabbing}
\end{defi}

One particular interesting result of Lie group theory is the fact, that all closed subgroups of
Lie groups are again Lie groups. Similarly all products, quotients by closed
normal subgroups and universal covers of Lie groups are Lie groups themselves.

\begin{defi}\label{def:group homomorphism}
	\begin{tabbing} \hspace*{3.5cm}\=\hspace*{2cm}\=\kill
			\> A \textbf{\emph{group homomorphism}} between two groups $G$ and $H$\\[1mm]
            \> is a map \ $\phi :\,G\,\rightarrow\,H$\ , which respects the group\\[1mm]
            \> structure, $\phi(g_{1}\,g_{2})\,=\,\phi(g_{1})\,\phi(g_{2})$,\ \ $\forall\,g_{1},g_{2}\in G$.\\[3mm]
            \> If it is bijective the map $\phi$ is called an \emph{isomorphism}.\\
	\end{tabbing}
\end{defi}

The physical significance of Lie groups lies in their role as symmetry groups, in this case of quantum Theory, i.e.
the action of the group on the Hilbert space leaves the transition probabilities \eqref{eq:probability interpretation}
invariant. We therefore need to define how the group action is represented on a vector space (Hilbert space).

\begin{defi}\label{def:representation}
	\begin{tabbing} \hspace*{3.5cm}\=\hspace*{2cm}\=\kill
			\> A real (complex) \textbf{\emph{representation}} of a group $G$ is a\\[1mm]
            \> group homomorphism \ $\pi :\,G\,\rightarrow\,GL(V)$\ , from $G$ to\\[1mm]
            \> the group of linear transformations on the real (complex)\\[1mm]
            \> vector space $V$.\\
	\end{tabbing}
\end{defi}

Alternatively it can be expressed as a \emph{left-action} of the group $G$ on the space $V$,
i.e. a map $G \times V\,\rightarrow\,V$, $(g,\psi)\,\mapsto\,g\,\psi$, which respects the group law.
From now on we will identify the vector space $V$ in Def.~\ref{def:representation} with the Hilbert space
$\mathcal{H}$ at hand as well as associate the linear operators\footnote{It is also assumed implicitly that
the left-action on Hilbert space is a continuous map, which would require us to replace the group of linear
operators $GL(\mathcal{H})$ in Def.~\ref{def:representation} with the group of bounded linear operators
$B(\mathcal{H})$.} of Def.~\ref{def:linear operator} with elements of $GL(\mathcal{H})$, $A\in GL(\mathcal{H})$.\\

To accommodate the symmetries of the Hilbert space the notion of representation is still too broad, since
the transformations of the state vectors which leave invariant the transition probabilities are those
represented by (anti-)unitary operators alone. We therefore define

\begin{defi}\label{def:unitary representation}
	\begin{tabbing} \hspace*{3.5cm}\=\hspace*{2cm}\=\kill
			\> A \textbf{\emph{unitary representation}} is a group homomorphism\\[2mm]
            \>  \> $\pi :\,G\,\rightarrow\,U(\mathcal{H})$\ \\[2mm]
            \> from $G$ to the group of unitary operators, $U^{\dagger}=U^{-1}$,\\[1mm]
            \> on the Hilbert space $\mathcal{H}$.\\
	\end{tabbing}
\end{defi}
To ensure that we consider a representation suitable for the space of interest, and not one
more complicated than necessary, we need to find the so called \emph{irreducible representations}.
\begin{defi}\label{def:invariant subspace}
	\begin{tabbing} \hspace*{3.5cm}\=\hspace*{2cm}\=\kill
			\> Consider a representation $\pi$ of $G$ on $V$. A subspace\\[1mm]
            \> $W$ of $V$ is said to be \textbf{\emph{invariant}} under the action of $\pi$,\\[1mm]
            \> if $\ \pi(g)w\in W\ \ \forall\,w\in W$ and $\forall\,g\in G$.\\
	\end{tabbing}
\end{defi}
\begin{defi}\label{def:irreducible rep}
	\begin{tabbing} \hspace*{3.5cm}\=\hspace*{2cm}\=\kill
			\> A representation is called \textbf{\emph{irreducible}} if it has no \\[1mm]
            \> non-trivial, invariant subspaces, i.e. only the empty set\\[1mm]
            \> $W=\emptyset$, and the vector space $W=V$ itself are invariant\\[1mm]
            \> subspaces.\\[3mm]
            \> A representation $\pi$ which can be written as the direct\\[1mm]
            \> sum of irreducible representations $\pi_{1},\pi_{2},\ldots$, i.e. \\[1mm]
            \> $\pi\,=\,\pi_{1}\oplus\pi_{2}\oplus\ldots$, is called \emph{completely reducible}.\\
	\end{tabbing}
\end{defi}

A useful property of \emph{finite-dimensional unitary representations} (on Hilbert space) is their attribute always
to be completely reducible. This can be easily seen by considering any subspace $W$ of $\mathcal{H}$, invariant
under the unitary representation $\pi$, and its orthogonal complement $W^{\bot}$. For all $w\in W$ and
$v\in W^{\bot}$ we have
\begin{equation}
\left\langle\,v,\pi(g)w\,\right\rangle\,=\,
\left\langle\,\pi(g)^{\dagger}v,w\,\right\rangle\,=\,0\ \ ,
\label{eq:finite dim groups completely reducible unitar reps}
\end{equation}
proving that $\pi(g)v\in W^{\bot}\ \forall\,v\in W^{\bot}$, and since $\mathcal{H}=W\oplus W^{\bot}$ the
representation $\pi$ is completely reducible\footnote{If $W$ or $W^{\bot}$ have further invariant subspaces, the procedure can be applied again on these spaces.}. For \emph{compact groups} this property can even be extended, as
\emph{every unitary representation} $\pi$ of a compact group is completely reducible, i.e. there exists a unitary map
\begin{equation}
U:\ \mathcal{H}\,\rightarrow\,\bigoplus\limits_{k}\mathcal{H}_{k}
\label{eq:compact groups completely reducible unitar reps 1}
\end{equation}
and irreducible representations $\pi_{k}$ on $\mathcal{H}_{k}$, such that
\begin{equation}
\pi_{U}(g)\,=\,U\,\pi(g)\,U^{\dagger}\,=\,\bigoplus\limits_{k}\pi_{k}(g)
\label{eq:compact groups completely reducible unitar reps 2}
\end{equation}

Furthermore, even if the representation $\pi$ given initially is not unitary, it is always possible to
gain a unitary representation, if $\pi$ is a finite-dimensional representation of a compact group, by
redefining the inner product on $\mathcal{H}$ appropriately, which is known as ``Weyl's unitarity trick".
In case of matrix Lie groups, the structure of Eq.~\eqref{eq:compact groups completely reducible unitar
reps 2} corresponds to a block-diagonal completely reducible unitary representation $\pi_{U}$, where each
block forms a (unitary) irreducible representation $\pi_{k}$, where $k=1,2,\ldots$.

\begin{equation}
\pi_{U}(g)\,=\,\begin{pmatrix}
\,\framebox[11mm][c]{$\pi_{1}(g)$} & & \\
 &\!\framebox[11mm][c]{$\pi_{2}(g)$} & \\
 & & \ddots\\
\end{pmatrix}
\label{eq:compl reducible matrix rep}
\end{equation}

Conversely it is generally not true that infinite-dimensional representations are completely reducible.
Over and above this fact non-compact groups (such as the Lorentz group, see Sec.~\ref{subsec:lorentz
and poincare group}) might not even have finite-dimensional unitary representations at all.\\

Finally, groups might furnish representations on Hilbert space which differ only by change of basis
of $\mathcal{H}$, and therefore are physically equivalent.

\begin{defi}\label{def:equivalent reps}
	\begin{tabbing} \hspace*{3.5cm}\=\hspace*{2cm}\=\kill
			\> Two representations $\pi_{1}$ and $\pi_{2}$ are called (unitary)\\[1mm]
            \> \textbf{\emph{equivalent}} if there exists a unitary map $S:\,\mathcal{H}\,\rightarrow\,\mathcal{H}$\\[2mm]
            \> such that \ $\pi_{1}\,=\,S\,\pi_{2}\,S^{\,\dagger}\ \ \forall\,g\in G$\ .\\
	\end{tabbing}
\end{defi}

The major step to simplify finding the representations of a Lie group $G$ is recognizing the connection of
the Lie group to its Lie algebra $\mathfrak{g}$, since by finding the representation of $\mathfrak{g}$ one can
easily obtain a representation of $G$. As mentioned earlier we will not present the full background of differentiable
manifolds in this work but will make remarks as to the underlying geometric structure where this provides further insight.

\begin{defi}\label{def:lie algebra}
	\begin{tabbing} \hspace*{3.5cm}\=\hspace*{2cm}\=\kill
			\> A \textbf{\emph{Lie algebra}} $\mathfrak{g}$ is a vector space over a field $\mathbb{K}$,\\[1mm]
            \> usually $\mathbb{R}$ or $\mathbb{C}$, which has a bilinear map (over $\mathbb{K}$)\\[2mm]
            \> \> $\left[\, .\,,\, .\, \right]:\,\mathfrak{g}\,\times\,\mathfrak{g}\,\rightarrow\,\mathfrak{g}$\\[2mm]
            \> defined on it, which is \emph{antisymmetric}\\[2mm]
            \> \> $\left[\,X\, ,\,Y\,\right]\,=\,-\,\left[\,Y\, ,\,X\,\right]$ \\[2mm]
            \> and satisfies the \emph{Jacobi identity} $\forall X,Y,Z\in\mathfrak{g}$\\[2mm]
            \>$\left[\,X\, ,\,\left[\,Y\, ,\,Z\,\right]\,\right]\,+\,
                    \left[\,Y\, ,\,\left[\,Z\, ,\,X\,\right]\,\right]\,+\,
                     \left[\,Z\, ,\,\left[\,X\, ,\,Y\,\right]\,\right]\,=\,0$\ .\\[2mm]
	\end{tabbing}
\end{defi}

As before with Lie groups (Def.~\ref{def:subgroup}-\ref{def:representation}) we can define the relations
between Lie algebras in terms of subalgebras, homomorphisms and representations.

\begin{defi}\label{def:subalgebra}
	\begin{tabbing} \hspace*{3.5cm}\=\hspace*{2cm}\=\kill
			\> A subspace $\mathfrak{h}$ of a Lie algebra $\mathfrak{g}$ is called
                \textbf{\emph{subalgebra}}\\[1mm]
            \> of $\mathfrak{g}$, if it is closed under the Lie bracket $\left[\, .\,,\, .\, \right]$, i.e.\\[2mm]
            \> $\left[\,h_{1}\,,\,h_{2}\,\right]\in\mathfrak{h}\ \ \forall\,h_{1},h_{2}\in\mathfrak{h}$\ .\\
	\end{tabbing}
\end{defi}

\begin{defi}\label{def:lie algebra homomorphism}
	\begin{tabbing} \hspace*{3.5cm}\=\hspace*{2cm}\=\kill
			\> A linear map \ $\phi :\,\mathfrak{g}\,\rightarrow\,\mathfrak{h}$, such that\\[2mm]
            \> $\phi\left(\left[\,X\,,\,Y\,\right]\right)\,=\,\left[\,\phi\left(X\right)\,,\,\phi\left(Y\right)\,\right]\ \ \forall\,X,Y\in\mathfrak{g}$\ ,\\[2mm]
            \> is called a \textbf{\emph{Lie algebra homomorphism}}.\\
	\end{tabbing}
\end{defi}

\begin{defi}\label{def:lie algebra representation}
	\begin{tabbing} \hspace*{3.5cm}\=\hspace*{2cm}\=\kill
			\> A \textbf{\emph{representation}} (over the field $\mathbb{K}$) of a Lie algebra $\mathfrak{g}$\\[1mm]
            \> is a Lie algebra homomorphism $\pi$\\[2mm]
            \> \> $\pi :\,\mathfrak{g}\,\rightarrow\, gl(n,\mathbb{K})$
	\end{tabbing}
\end{defi}
where $gl(n,\mathbb{K})$ is the space of $n\times n$ matrices over the field $\mathbb{K}$.
\begin{equation}
gl(n,\mathbb{K})\,:=\,Mat(n\times n,\,\mathbb{K})
\label{eq:gl(n,K)}
\end{equation}
Since a Lie algebra is a vector space, a basis $\left\{X_{i}\right\}$ can be chosen in $\mathfrak{g}$ such that
the Lie bracket (introduced in Def.~\ref{def:lie algebra}) of two basis elements can uniquely be written as
\begin{equation}
\left[\,X^{\,i}\,,\,X^{\,j}\,\right]\,=\,f^{\,ij}_{\ \ k}\,X^{\,k}
\label{eq:structure constants}
\end{equation}
where summation convention is implied and the numbers $f^{\,ij}_{\ \ k}$ corresponding to the chosen basis are
the (antisymmetric) \emph{structure constants} of the Lie algebra, which is completely determined in turn by its
structure constants. One therefore often uses equations of the form of Eq.~\eqref{eq:structure constants}, i.e.
a choice of basis together with the resulting structure constants, as the defining property of the Lie algebra.\\

We now want to formulate the intimate connection of Lie algebras and Lie groups. As stated in Def.~\ref{def:lie group} a Lie group has the structure of a differentiable manifold. It is straightforward to define functions and their directional derivatives on such manifolds, see e.g. \cite{bertlmannanomalies}, and it turns out that all vectors on a manifold $M$ can be defined as tangent vectors to curves on $M$. Consequently, we define

\begin{defi}\label{def:tangent space}
	\begin{tabbing} \hspace*{3.5cm}\=\hspace*{2cm}\=\kill
			\> The \textbf{\emph{tangent space}} $T_{p}(M)$ to the manifold $M$ at the\\[1mm]
            \> point $p$ is the space of all possible tangent vectors at $p$.\\
	\end{tabbing}
\end{defi}

If a (smooth) map $\phi$ between two manifolds $M$ and $N$ is given, then this naturally induces a map between their tangent spaces $T_{p}(M)$ and $T_{q}(N)$, the \emph{differential map} $\phi_{*}$\footnote{The differential map is sometimes also called tangent map for obvious reasons and alternative notations include $d\phi$ and $\phi^{*}$. In the latter case the pullback, which we don't discuss in this work, is denoted by $\phi_{*}$.},
\begin{equation}
\phi_{*}:\,T_{p}(M)\,\rightarrow\,T_{\phi(p)}(N)\ \ \ ,
\label{eq:differential map}
\end{equation}
 mapping the vector fields $X\in T_{p}(M)$ to vector fields $\phi_{*}X\in T_{\phi(p)}(N)$ by evaluating the vector fields $X$ over functions on $M$, composed in the manner $\psi\circ\phi:\,M\rightarrow\mathbb{R}$ where $\psi:N\rightarrow\mathbb{R}$,
\begin{equation}
(\phi_{*}X)(\psi)\,=\,X(\psi\circ\phi)
\label{eq:differential map of vector field}
\end{equation}

The important relation between a Lie group and its algebra then is the following.\\

\begin{theorem}\label{thm:lie group to algebra homomorphism}
    \begin{tabbing} \hspace*{3.3cm}\=\hspace*{2cm}\=\kill
			\> If $\phi$ is a group homomorphism between two Lie groups $G$\\[1mm]
            \> and $H$, then the differential map $\phi_{*}$ is a Lie algebra\\[1mm]
	        \> homomorphism between the corresponding Lie algebras $\mathfrak{g}$\\[1mm]
            \> and $\mathfrak{h}$, respectively.\\
    \end{tabbing}
\end{theorem}
\begin{theorem}\label{thm:lie algebra to group homomorphism}
    \begin{tabbing} \hspace*{3.3cm}\=\hspace*{2cm}\=\kill
			\> Reversely, starting with a Lie algebra homomorphism \\[1mm]
      \> $\varphi:\,\mathfrak{g}\rightarrow\mathfrak{h}$, there exists a unique Lie group homomorphism\\[1mm]
	    \> $\phi:G\rightarrow H$, determined by the differential map $\varphi=\phi_{*}$,\\[1mm]
      \> if $G$ is simply connected.\\
    \end{tabbing}
\end{theorem}

The proof of these statements is rather complicated and can be found in \cite{warnerliegroups} (Theorems 3.14, 3.16, and 3.27). One important conceptual consequence of the connection of a Lie group homomorphism to a Lie algebra homomorphism via the differential map is the geometric interpretation, that a Lie algebra is the tangent space at the identity of the corresponding Lie group and their dimensions coincide, $\dim(G)=\dim(\mathfrak{g})$. If furthermore
$\mathfrak{h}$ is a Lie subalgebra of $\mathfrak{g}$, then $\mathfrak{h}$ is the Lie algebra of a Lie subgroup
$H$ of $G$. Moreover the following insight, explained in more detail and proven in \cite{warnerliegroups} (Theorem 3.25), will prove to be of great importance to our discussion.

\begin{theorem}\label{thm:universal covering}
    \begin{tabbing} \hspace*{3.3cm}\=\hspace*{2cm}\=\kill
            \> Every connected Lie group $G$ can be covered by a simply\\[1mm]
            \> connected Lie group, its \emph{universal cover} $\tilde{G}$, such that\\[1mm]
            \> $\tilde{G}$ is mapped to $G$ by a surjective group homomorphism,\\[1mm]
            \> which is an isomorphism close to the identity. \\[2mm]
            \> Consequently their Lie algebras $\mathfrak{g}$ and $\tilde{\mathfrak{g}}$ as well\\[1mm]
            \> as their dimensions are identical, $\dim(G)=\dim(\tilde{G})$.\\
    \end{tabbing}
\end{theorem}

So far all the definitions and statements made were very abstract and apply to a broad class of possible realizations. For the physical applications, however, the most important class of Lie groups is that of the matrix groups, i.e. those acting as finite- or infinite-dimensional matrices on a linear vector space, the most general group of which is $GL(n,\mathbb{K})$, the $n\times n$ invertible matrices over the field $\mathbb{K}$.

\begin{equation}
GL(n,\mathbb{K})\,:=\,\left\{\,A\in Mat(n\times\,n, \mathbb{K})\left.\right|\,\det(A)\neq\,0\right\}
\label{eq:GL(n,K)}
\end{equation}

This is an $n^{2}$-dimensional vector space and it is easy to see that the matrices of $GL(n,\mathbb{K})$ automatically supply their own representation on an $n$-dimensional vector space. The Lie algebra belonging to $GL(n,\mathbb{K})$ is simply the set of all $n\times n$ matrices $gl(n,\mathbb{K})$ (see Eq.~\eqref{eq:gl(n,K)}) where the Lie bracket of two matrices $A$ and $B$ is given by their commutator,

\begin{equation}
\left[\,A\,,\,B\,\right]\,=\,A\,B\,-\,B\,A\ \ \ ,
\label{eq:commutator}
\end{equation}
which can in principle be proven easily by showing that the commutator of two left-invariant vector fields again is a left-invariant vector field (see Sec.~3.5 and Sec.~3.10 of Ref.~\cite{warnerliegroups}). Alternatively, we prefer to introduce the \emph{exponential map} instead, which establishes the map from a Lie algebra to its Lie group, given the so called Baker-Campbell-Hausdorff-formula, which we will find more useful in the continuing discussion. Consider therefore
the matrix exponential $e^{X}$ of the matrix $X$, which we had already used without further justification earlier \eqref{eq:rotations on hilbert space}, defined by the Taylor series of the usual exponential function where the arguments are replaced by matrix products,

\begin{equation}
e^{X}\,=\,\exp(X)\,=\,\sum\limits_{n=0}^{\infty}\,\frac{X^{n}}{n!}\ \ \ ,
\label{eq:matrix exponential}
\end{equation}
where the property
\begin{equation}
\det(e^{X})\,=\,e^{Tr(X)}\ \ ,
\label{eq:determinant of matrix exponential}
\end{equation}
is satisfied\footnote{For a diagonalizable matrix this can be seen easily, since the trace is just the sum of the eigenvalues, the exponential of which is just the product of the eigenvalues, i.e. the determinant. The generalization to non-diagonalizable matrices is straightforward.}. Now take an element $X$ of $gl(n,\mathbb{K})$ and $\alpha\in\mathbb{R}$, then the map

\begin{eqnarray}
e^{\alpha X}:\,\mathbb{R} &\rightarrow & GL(n,{\mathbb{K}}) \nonumber \\
\alpha &\mapsto & e^{\alpha X}
\label{eq:one parameter subgroup}
\end{eqnarray}
is a smooth curve $\gamma(\alpha)$ in the space $\mathbb{K}^{n^{2}}$ (of which $GL(n,{\mathbb{K}})$ is an open subset) and $\gamma(\alpha)$ respects the group structure, $\gamma(\alpha+\beta)=\gamma(\alpha)\,\gamma(\beta)\ \forall\,\alpha,\beta\in\mathbb{R}$, i.e. it is a \emph{one-parameter subgroup}. It therefore is a group homomorphism (see Def.~\ref{def:group homomorphism}) and by Theorem \ref{thm:lie group to algebra homomorphism} the corresponding Lie algebra homomorphism is given by the tangent map $\gamma(\alpha)_{*}$.

\begin{equation}
\gamma(\alpha)_{*}\,=\,\frac{d}{d\alpha}\,e^{\alpha X}
\label{eq:tangent map of matrix exponential}
\end{equation}

Evaluating this at the origin, $\left.\frac{d}{d\alpha}e^{\alpha X}\right|_{\alpha=0}\,=\,X$, we get again $X\ \forall X$, showing that the the tangent space at the identity is in fact $gl(n,\mathbb{K})$ and we have a Lie algebra homomorphism. Then, using Theorem \ref{thm:lie algebra to group homomorphism}, it is easy to see, that the Lie group homomorphism of Eq.~\eqref{eq:one parameter subgroup} is uniquely determined because $\mathbb{R}$ is simply connected.\\

This result is very strong indeed, since we now have a simple way of obtaining Lie group representations, once the
representations of the corresponding Lie algebra are determined. For matrix Lie groups this is particularly easy, since they provide their own representation. This is guaranteed by the \emph{Baker-Campbell-Hausdorff} formula,

\begin{equation}
e^{X}e^{Y}=e^{\left(X\,+\,Y\,+\,\tfrac{1}{2}\left[X,Y\right]
\,+\,\tfrac{1}{12}\left[X,\left[X,Y\right]\right]
\,-\,\tfrac{1}{12}\left[Y,\left[X,Y\right]\right]\,+\,\cdots \right)}
\label{eq:baker campbell hausdorff}
\end{equation}
which ensures that the group is closed under the exponential map, since the Lie algebra is closed under the Lie bracket, represented here by the commutator.\\

Thus once we have found a basis of a Lie algebra $\mathfrak{g}$, i.e. a set of $n=\dim{\mathfrak{h}}$ matrices satisfying the commutation relation Eq.~\eqref{eq:structure constants} (for the appropriate structure constants), we immediately have a representation of the Lie group $G$. Moreover it is useful to work with the simply connected universal covering group $\tilde{G}$ (if the considered group $G$ is connected), since all results are
easily translated to the group $G$ if necessary (remember Theorem \ref{thm:lie algebra to group homomorphism}).\\

As a last preliminary assertion, which will be fundamentally important in Sec.~\ref{sec:composite systems}, let us study how we can compose  a new representation out of representations already gained by the tensor product. Suppose therefore that $\pi_{A}$ and $\pi_{B}$ are representations of a Lie group $G$ on Hilbert spaces $\mathcal{H}_{A}$ and $\mathcal{H}_{B}$ respectively, then we can define the representation $\pi$ of $G$ on the tensor product space $\mathcal{H}_{A}\otimes\mathcal{H}_{B}$ by

\begin{eqnarray} \nonumber
\pi:\ G & \rightarrow & GL(\mathcal{H}_{A}\otimes\mathcal{H}_{B})\,=\,GL(\mathcal{H}_{A})\otimes
    GL(\mathcal{H}_{B})\\
    g   & \mapsto & \pi_{A}(g)\otimes \pi_{B}(g)
    \label{eq:tensor product lie group rep}
\end{eqnarray}
and translating this to the Lie algebra by applying the Leibniz rule we get
\begin{eqnarray} \nonumber
\pi_{*}:\ \mathfrak{g} & \rightarrow & gl(\mathcal{H}_{A}\otimes\mathcal{H}_{B})\,=\,gl(\mathcal{H}_{A})\otimes
    gl(\mathcal{H}_{B})\\
    X   & \mapsto & \pi_{A}(X)\otimes\mathds{1}_{B}\,+\,\mathds{1}_{A}\otimes\pi_{B}(X)
    \label{eq:tensor product lie algebra rep}
\end{eqnarray}

This tensor product structure will typically appear when we consider composite quantum systems and entanglement in Chapter \ref{chap:entanglement and nonlocality}.

        \subsubsection{The Connection of SO(3) and SU(2)}\label{subsubsec:SO3 and SU2}

Let us now implement the abstract concepts of Sec.~\ref{subsec:lie groups and lie algebras} in a more practical approach by studying the 3-dimensional rotation group $SO(3)$, defined as the group of unimodular, orthogonal matrices over the field $\mathbb{R}$,

\begin{equation}
SO(3)\,:=\,\left\{R\in\,GL(3,\mathbb{R})\,|\,R^{\,T}\!R=RR^{\,T}=\mathds{1},\,\det(R)=1\right\}
\label{eq:SO(3)}
\end{equation}

and its representations on Hilbert space. So if we consider
a rotation as a transformation on a 3-dimensional coordinate system on $\mathbb{R}^{3}$ it can be generally expressed as in Eq.~\eqref{eq:rotation in R3}, clearly this is a representation of the rotation group $SO(3)$ on $\mathbb{R}^{3}$, where an observer, whose coordinate system is rotated by $R(\vec{\varphi}\,)$, would find that the vectors $\vec{x}^{\,\prime}$ in his system are rotated by $R^{-1}(\vec{\varphi}\,)$. The Lie algebra $so(3)$ belonging to $SO(3)$ is then given by all traceless (see Eq.~\eqref{eq:determinant of matrix exponential}), antisymmetric matrices of $gl(n,\mathbb{R})$,

\begin{equation}
so(3)\,:=\,\left\{A\in\,gl(3,\mathbb{R})\,|\,A^{\,T}=-A\,,\,Tr(A)=0\right\}
\label{eq:so(3)real}
\end{equation}
however, it is much more convenient to ``complexify" the Lie algebra, $\mathfrak{g}\rightarrow i\mathfrak{g}$, such that the generators of the group become Hermitian instead of antisymmetric and we have to include a factor $i$ in the matrix exponential as well as in the structure constants. We can thus alternatively write the Lie algebra of $SO(3)$ as
\begin{equation}
so(3)\,=\,\left\{T\in\,gl(3,\mathbb{C})\,|\,T^{\,\dagger}=T\,,\,Tr(T)=0\right\}
\label{eq:so(3)complexified}
\end{equation}

Let us then choose an appropriate basis of $so(3)$,

\begin{equation}
T^{\,x}\,=\,\begin{pmatrix} \,0 & 0 & \,0 \\
                                   \,0 & 0 & -i \\
                                   \,0 & i & \,0 \end{pmatrix}\, ,\ \
T^{\,y}\,=\,\begin{pmatrix}  \,0 & 0 & i\, \\
                                    \,0 & 0 & 0\, \\
                                     -i & 0 & 0\, \end{pmatrix}\, ,\ \
T^{\,z}\,=\,\begin{pmatrix}  0 & -i & 0\, \\
                                    i & \,0 & 0\, \\
                                    0 & \,0 & 0\, \end{pmatrix} \, ,\ \
\label{eq:so(3) basis}
\end{equation}
or in a more compact notation
\begin{equation}
(T^{\,k})_{\,ij}\,=\,-i\,\varepsilon_{\,ijk}\ \ \ ,
\label{eq:so(3) basis compact}
\end{equation}
where the commutation relation for the structure constants (compare Eq.~\eqref{eq:so3 lie algebra} and Eq.~\eqref{eq:structure constants}) takes the form
\begin{equation}
\left[\,T^{\,i}\,,\,T^{\,j}\,\right]\,=\,i\,\varepsilon^{\,ijk}\,T^{\,k}\ \ \ \  ,\
(i,j,k\,=\,x,y,z),
\label{eq:so(3) structure constants}
\end{equation}
and the rotation matrices $R(\vec{\varphi}\,)$ of Eq.~\eqref{eq:rotation in R3} are then given by the (complexified) exponential map (Eq.~\eqref{eq:matrix exponential}),
\begin{equation}
R(\vec{\varphi}\,)\,=\,e^{-i\,\vec{\varphi}\overrightarrow{T}}\ \in SO(3)\ \ .
\label{eq:exponential rep of 3d rotation}
\end{equation}

Interestingly, the rotation group $SO(3)$, although connected, is not simply connected. This can be seen by viewing the group manifold as a ball of radius $\pi$ in $\mathbb{R}^{3}$, where every point represents a rotation around an axis defined by the vector from the origin to the point and an angle given by the length of that vector. In addition, since rotations around $\pi$ and $-\pi$ give the same results, the points on the surface need to be identified with their antipodal counterparts. It is then obvious, that curves connecting two such antipodal points can never be shrunk to a point without breaking the curve.\\

Let us now study how the spinors of $\mathbb{C}^{2}$ transform under rotations in $\mathbb{R}^{3}$. Needless to say that we expect our Hilbert space description to be invariant under simple rotations of the observer, which means that the matrices representing the rotations on the Hilbert space need to be unitary, unimodular $2\times 2$ matrices, i.e. elements of $SU(2)$,

\begin{equation}
SU(2)\,:=\,\left\{U\in\,GL(2,\mathbb{C})\,|\,U^{\,\dagger}U=UU^{\,\dagger}=\mathds{1},\,\det(U)=1\right\}\ .
\label{eq:SU(2)}
\end{equation}

The Lie algebra of $SU(2)$ therefore must consist of traceless, Hermitian $2\times 2$ matrices,

\begin{equation}
su(2)\,:=\,\left\{X\in\,gl(2,\mathbb{C})\,|\,X^{\,\dagger}=X,\,Tr(X)=0\right\}\ \ \ ,
\label{eq:su(2)}
\end{equation}
which we also have already encountered, since the Pauli matrices (Eq.~\eqref{eq:pauli matrices}) form a
basis of this vector space, i.e. every $X\in su(2)$ can be written as a linear combination of them,

\begin{equation}
X\,=\,\sum\limits_{i}\,\alpha_{\,i}\,\sigma^{\,i}\,=\,\vec{\alpha}\cdot\vec{\sigma}\,=\,
\begin{pmatrix} \alpha_{\mathrm{z}} & \alpha_{\mathrm{x}}-i\alpha_{\mathrm{y}} \\
                \alpha_{\mathrm{x}}+i\alpha_{\mathrm{y}} & -\alpha_{\mathrm{z}} \end{pmatrix}\ \ ,
\label{eq:pauli matrices as su(2) basis}
\end{equation}
and their commutator (Eq.~\eqref{eq:pauli matrices commutator}), or the commutator
of the spin matrices $S_{\mathrm{i}}$, acts as Lie bracket supplying the structure constants.

\begin{equation}
\left[\right.\frac{\sigma^{\,i}}{2},\frac{\sigma^{\,j}}{2}\left.\right]\,=
\,i\,\varepsilon^{\,\mathrm{ijk}}\,\frac{\sigma^{\,k}}{2}\ \ \ \ \
(i,j,k\,=\,x,y,z)
\label{eq:pauli matrices over 2 commutator}
\end{equation}

Then every element of $SU(2)$ is expressible as

\begin{equation}
U(\vec{\alpha}\,)\,=\,e^{-i\,\vec{\alpha}\frac{\vec{\sigma}}{2}}\ \in SU(2),\ \mbox{where}\ \vec{\alpha}\in\mathbb{R}^{3}\ .
\label{eq:SU(2) exponential rep of 3d rotation}
\end{equation}

We can now see the similarity of the Lie algebras $su(2)$ and $so(3)$, they are both three dimensional vector spaces, consist of traceless, Hermitian matrices, and their elements satisfy the same commutation relations, which means that replacing $T^{\,i}$ with $\frac{\sigma^{\,i}}{2}$ is a Lie algebra homomorphism and because $SU(2)$ is simply connected, there is a unique Lie group homomorphism from $SU(2)$ to $SO(3)$ by Theorem \ref{thm:lie algebra to group homomorphism}.\\

The fact that $SU(2)$ is simply connected can be argued as follows. The group manifold is a three dimensional sphere in $\mathbb{R}^{4}$, since the requirements of unimodularity and unitarity leave only three independent parameters, see e.g. \cite{sexlurbantke}. A curve connecting antipodal points corresponding to $\alpha=|\vec{\alpha}|=0$ and $\alpha=2\pi$ will be a closed curve for $SO(3)$, since these points are identified, while it is only closed in $SU(2)$ if one adds a second part also connecting those two points on the three-sphere. The resulting closed curve can then easily be deformed and shrunk to a point.\\

From a physical point of view, the group homomorphism from $SU(2)$ to $SO(3)$ can be argued by noting, that the
expectation value of the spin operator $\vec{S}$ (Eq.~\eqref{eq:spin operator i component}) transforms like a vector under rotations.

\begin{equation}
R\,\left\langle\,\psi\right|\,\vec{S}\left.\psi\,\right\rangle\,=\,
\left\langle\,U\psi\right|\,\vec{S}\left.U\psi\,\right\rangle\,
\label{eq:spin operator exp value transformation}
\end{equation}

From the positive definiteness of the scalar product, we can then infer that independent of $\left|\psi\right\rangle$, the Pauli matrices must transform as

\begin{equation}
U^{\,\dagger}\,\sigma^{\,i}\,U\,=\,R_{\,ij}\,\sigma^{\,j}\ \ \Rightarrow\ \
U\,\sigma^{\,i}\,U^{\,\dagger}\,=\,\sigma^{\,j}\,R_{\,ji} \ \ ,
\label{eq:pauli matrices transformation}
\end{equation}
and using Eq.~\eqref{eq:pauli matrices as su(2) basis}, where we set
\begin{equation}
X\,=\,\vec{x}\cdot\vec{\sigma}\ ,\ \ \vec{x}\in\mathbb{R}^{3}\ ,
\label{eq:3dim vector as matrix}
\end{equation}
we end up with the expression
\begin{equation}
U\,X\,U^{\,\dagger}\,=\,\vec{\sigma}(R\,\vec{x})\ .
\label{eq:su(2) matrix transformation}
\end{equation}

The l.h.s. of Eq.~\eqref{eq:su(2) matrix transformation} corresponds to a vector $\vec{x}^{\,\prime}$, such that

\begin{equation}
\vec{x}^{\,\prime}\cdot\vec{\sigma}\,=\,X^{\prime}\,=\,U\,X\,U^{\,\dagger}\ \ \ ,
\label{eq:rotated vector as matrix}
\end{equation}
and the norm $|\vec{x}^{\,\prime}|$ coincides with that of $\vec{x}$ since $\det(X^{\,\prime})=\det(X)=-|\vec{x}|^{2}$. Since by setting $U=\pm\mathds{1}$ one arrives at
the identity element of $SO(3)$ and $SU(2)$ is connected, we can conclude that we have found a suitable mapping
between those two groups, which can be checked to be a group homomorphism with kernel $\left\{\mathds{1},-\mathds{1}\right\}$. It is obvious from Eq.~\eqref{eq:su(2) matrix transformation} that $U$ and $-U$ give the same rotation in $SO(3)$ and the map is therefore not bijective, but still surjective. Finally we can establish the fact, that $SU(2)$ is the universal (double) covering group (Theorem \ref{thm:universal covering}) of $SO(3)$,

\begin{equation}
SU(2)/\left\{\mathds{1},-\mathds{1}\right\}\,\cong\,SO(3)\ .
\label{eq:SU(2) covering SO(3)}
\end{equation}

The $\mathbb{C}^{\,2}$ spinors, transforming under the $SU(2)$ matrices, are the irreducible representations of the rotation group to the weight $s=\tfrac{1}{2}$ (see \eqref{eq:spin squared eigenvalue equation}). The spinors introduced in \eqref{eq:two component spinors}, on the other hand, transform more complicated under the rotation group, they transform as spin-$\tfrac{1}{2}$ fields, which means they transform as the tensor product of the spinor representation $D_{g}$ (to the weight $s=\tfrac{1}{2}$) \eqref{eq:spin z eigenstates}, and the representation $T_{g}$ \eqref{eq:rotations on hilbert space} in the (Hilbert) space of scalar fields. Let us formerly write this (irreducible) representation of the rotation group, the spin $\tfrac{1}{2}$ fields, as
\begin{equation}
{}_{D}T_{g}\,=\,D_{g}\,\otimes\,T_{g}\ .
\label{eq:spin one half field representation}
\end{equation}

The generators of the corresponding transformations are the angular momentum operators
\begin{equation}
J^{\,i}\,=\,L^{\,i}\,+\,S^{\,i}\ ,\ \ \mathrm{i}\,=\,1,2,3\ ,
\label{eq:total angular momentum}
\end{equation}
where $L^{\,i}$, and $S^{\,i}$ are the orbital angular momentum- and spin operators, given by \eqref{eq:orbital angular momentum i component} and \eqref{eq:spin operator i component} respectively.

\addtocontents{toc}{\vspace{-1ex}}
\newpage\section{Entanglement \& Nonlocality}\label{chap:entanglement and nonlocality}

This chapter is devoted to the study of non-classical correlations between two (or more) quantum systems,  known as \emph{entanglement}. The phenomenon arises from the quantum mechanical description of composite systems and the \emph{superposition principle}, which in turn originates from the linearity of the Schr\"odinger equation \eqref{eq:schroedinger equation}. It is thus a feature inherent to all quantum systems, that can be decomposed into two or more (nontrivial) subsystems, even if they all describe the same particle, but its consequences become more incomprehensible and fascinating for subsystems which are far apart from each other.\\

We therefore study the description of composite systems, and the entanglement between them, for pure bipartite and pure multipartite states in Sec.~\ref{subsec:pure bipartite entanglement} and Sec.~\ref{subsec:pure multipartite entanglement} respectively, and continue with the description of mixed state entanglement in Sec.~\ref{subsec:entanglement of mixed states} and its quantification in Sec.~\ref{subsec:entanglement measures for mixed states}, before we present the \emph{EPR - Paradox} and (a version of) \emph{Bell's Theorem} in Sec.~\ref{sec:EPR Bell}, which underlines, as well as utilizes the concept of entanglement to investigate the fundamental understanding of reality. An extensive mathematical review on quantum entanglement can be found in \cite{horodeckisentanglement07}, while textbook approaches from a more physical perspective are given in \cite{audretschentangledsystems}, \cite{mintertcarvalhokusbuchleitner05}, and \cite{nielsenchuangQI}.
	
		\subsection{Composite Quantum Systems \& Entanglement}\label{sec:composite systems}

Let us start out with two different quantum systems $A$ and $B$ and their respective Hilbert spaces $\mathcal{H}_{A}$ and $\mathcal{H}_{B}$, to acquire a composite quantum system, one simply forms the tensor product of the Hilbert spaces of the constituent systems $A$ and $B$.
\begin{equation}
\mathcal{H}_{AB}\,=\,\mathcal{H}_{A}\,\otimes\,\mathcal{H}_{B}
\label{eq:composite quantum system hilbert space}
\end{equation}

The dimension of the product space is the product of the dimensions of the constituent spaces, i.e.

\begin{equation}
\dim(\mathcal{H}_{AB})\,=\,\dim(\mathcal{H}_{A})\cdot\dim(\mathcal{H}_{B})\ \ \ ,
\label{eq:product space dimension}
\end{equation}

and in a similar fashion we can construct states $\in\mathcal{H}_{AB}$ by forming the tensor product of any of the states $\in\mathcal{H}_{A}$ with states $\in\mathcal{H}_{B}$. All the previously discussed operations and analysis apply, since the inner product on the product space can be defined by the inner products of the spaces $\mathcal{H}_{A}$ and $\mathcal{H}_{B}$.
\begin{equation}
\langle\,.\,,\,.\,\rangle_{AB}\,=\,\langle\,.\,,\,.\,\rangle_{A}\,\cdot \langle\,.\,,\,.\,\rangle_{B}\,
\label{eq:product space inner product}
\end{equation}

Furthermore, given symmetry groups of the Hilbert spaces $\mathcal{H}_{A}$ and $\mathcal{H}_{B}$ and the respective representation thereon, then Eq.~\eqref{eq:tensor product lie group rep} and Eq.~\eqref{eq:tensor product lie algebra rep} yield the appropriate representations on the product space.\\

However, a general state of the product space need not be a product of states of the individual systems since the superposition principle allows for linear combinations of these product states. The descriptive power in terms of
transition probabilities, i.e. inner products, expectation values, operators, representations of symmetry groups and the likes remains undiminished in the process. The problems arise solely in our comprehension of spatially separated physical systems, which we would intuitively assume to have properties, i.e. well defined states, independently of each other. States for which this assumption is true are called \emph{separable states}. Those states, on the other hand, for which this assumption fails are called \emph{entangled states}.\\
\begin{center}\parbox{0.80\textwidth}{The state of a composite quantum system is said to be \emph{entangled}, if the total system is in a well defined state, while the subsystems are not.}\end{center}\ \\

            \subsubsection{Pure Bipartite Entanglement}\label{subsec:pure bipartite entanglement}

For the state vectors in Hilbert space the distinction between separable and entangled states can be easily defined in a rigorous way, although we need to specify the number of subsystems considered.

\begin{defi}\label{def:pure bipartite state}
	\begin{tabbing} \hspace*{3.5cm}\=\hspace*{2cm}\=\kill
			\> A pure state $|\psi^{\,AB}\rangle$ is called \textbf{\emph{bipartite}}, if it has the form\\[2mm]
            \> \> $|\psi^{\,AB}\rangle\, = \,\sum\limits_{\mathrm{i,\,j}=1}^{\mathrm{n,m}} \,c_{\mathrm{ij}}\, |\psi_{\mathrm{i}}^{\,A}\rangle\otimes|\psi_{\mathrm{j}}^{\,B}\rangle$\\[2mm]
            \> where $\left\{|\psi_{\mathrm{i}}^{\,A}\rangle\right\}$, $\left\{|\psi_{\mathrm{j}}^{\,B}\rangle\right\}$ form a basis in $\mathcal{H}_{A}$ and $\mathcal{H}_{B}$ \\[2mm]
            \> respectively, $\dim(\mathcal{H}_{A})=n$, $\dim(\mathcal{H}_{B})=m$, and the\\[1mm]
            \> complex coefficients $c_{\mathrm{ij}}$ satisfy $\sum\limits_{\mathrm{i,\,j}=1}^{\mathrm{n,m}} \,|c_{\mathrm{ij}}|^{2}=1$.\\
	\end{tabbing}
\end{defi}

\begin{defi}\label{def:pure bipartite separable and entangled state}
	\begin{tabbing} \hspace*{3.5cm}\=\hspace*{2cm}\=\kill
			\> A bipartite pure state $|\psi^{\,AB}\rangle\in\mathcal{H}_{AB}$ is called \textbf{\emph{separable}}\\[2mm]
            \> (with respect to the decomposition of $\mathcal{H}_{AB}$ into $\mathcal{H}_{A}$\\[2mm]
            \>  and $\mathcal{H}_{B}$), if it can be written as \ \ $|\psi^{\,AB}\rangle\,=\,|\psi^{\,A}\rangle\otimes|\psi^{\,B}\rangle$\ ,\\[2mm]
            \> where $|\psi^{\,A}\rangle\in\mathcal{H}_{A}$ and $|\psi^{\,B}\rangle\in\mathcal{H}_{B}$.\\[3mm]
            \> A (bipartite) state is called \textbf{\emph{entangled}}, if it is not\\[1mm]
            \> separable.
	\end{tabbing}
\end{defi}

To describe the constituent systems individually it is convenient to formulate the state vectors $|\psi^{\,AB}\rangle$ in terms of their projectors, i.e. we switch to the density matrix formalism introduced in Sec.~\ref{subsec:density operators}, and write

\begin{equation}
\rho^{\,AB}\,=\,|\psi^{\,AB}\rangle\langle\psi^{\,AB}|\ \ \ .
\label{eq:pure bipartite density operator}
\end{equation}

The subsystems $A$ and $B$ are then described by their respective \emph{reduced density matrices} $\rho^{\,A}$ and $\rho^{\,B}$, defined by the partial trace of the total system over the remaining subsystem,

\begin{equation}
\rho^{\,A}\,=\,Tr_{B}(\rho^{\,AB})\,=\,\sum\limits_{\mathrm{j}=1}^{\mathrm{m}}\,
\langle\psi_{\mathrm{j}}^{\,B}|\,\rho^{\,AB}\,|\psi_{\mathrm{j}}^{\,B}\rangle\ \ \ ,
\label{eq:partial trace over B}
\end{equation}
\begin{equation}
\rho^{\,B}\,=\,Tr_{A}(\rho^{\,AB})\,=\,\sum\limits_{\mathrm{i}=1}^{\mathrm{n}}\,
\langle\psi_{\mathrm{i}}^{\,A}|\,\rho^{\,AB}\,|\psi_{\mathrm{i}}^{\,A}\rangle\ \ \ ,
\label{eq:partial trace over A}
\end{equation}
where $\left\{|\psi_{\mathrm{i}}^{\,A}\rangle\right\}$ and $\left\{|\psi_{\mathrm{j}}^{\,B}\rangle\right\}$
are complete orthonormal bases of $\mathcal{H}_{A}$ and $\mathcal{H}_{B}$, such as in Def.~\ref{def:pure bipartite state}. It can be immediately seen then, that for separable states (Def.~\ref{def:pure bipartite separable and entangled state}), the reduced density operators for the individual subsystems coincide with the projectors on the states $|\psi^{\,A}\rangle$ and $|\psi^{\,B}\rangle$ and are thus pure states themselves.\\

In this case measurements performed on the subsystems are always independent of each other. States of that form are therefore said to be \emph{uncorrelated}, which can be formulated operationally. If for every observable of the form $\mathcal{O}_{A}\otimes\mathcal{O}_{B}$ on $\mathcal{H}_{AB}$, where $\mathcal{O}_{A}$ and $\mathcal{O}_{B}$ are (Hermitian) operators on the subspaces $\mathcal{H}_{A}$ and $\mathcal{H}_{B}$ respectively, the expectation value in the state $\rho^{\,AB}$ coincides with the product of expectation values of the reduced density matrices,
\begin{equation}
\langle\,\mathcal{O}_{A}\otimes\mathcal{O}_{B}\,\rangle_{\rho^{\,AB}}\,=\,
\langle\,\mathcal{O}_{A}\,\rangle_{\rho^{\,A}}\cdot\langle\,\mathcal{O}_{B}\,\rangle_{\rho^{\,B}}\ \ \forall\,\mathcal{O}_{A}, \mathcal{O}_{B}\ \ \ ,
\label{eq:uncorrelated}
\end{equation}
the state is uncorrelated. For pure states the set of correlated states is identical to that of entangled states, in other words pure states can be either uncorrelated or exhibit quantum mechanical correlations, i.e. be entangled. The reduced density matrices of entangled states do not contain all the information about the total system, which is why Eq.~\eqref{eq:uncorrelated} does not hold. This is closely related to the fact that the reduced density matrices of entangled pure states are mixed. Let us investigate this statement using the B\emph{ell states} $|\psi^{\,\pm}\rangle$,$|\phi^{\,\pm}\rangle$ as a palpable example. The Bell states are the most obvious instance of (maximally) entangled bipartite \emph{qubit} states, i.e. describing a quantum system composed of two subsystems with two degrees of freedom each.
\begin{eqnarray}
|\psi^{\,\pm}\rangle &=&
        \frac{1}{\sqrt{2}}\left(\,|\up\,\rangle\,|\down\,\rangle\,\pm\,|\down\,\rangle\,|\up\,\rangle\,\right)
        \label{eq:bell psi states}\\
|\phi^{\,\pm}\rangle &=&
        \frac{1}{\sqrt{2}}\left(\,|\up\,\rangle\,|\up\,\rangle\,\pm\,|\down\,\rangle\,|\down\,\rangle\,\right)
        \label{eq:bell phi states}
\end{eqnarray}

The reduced density matrices of $\rho^{\,\pm}=|\psi^{\,\pm}\rangle\langle\psi^{\,\pm}|$ and $\omega^{\,\pm}=|\phi^{\,\pm}\rangle\langle\phi^{\,\pm}|$ are all proportional to the identity, $\rho^{\,A}=\rho^{\,B}=\tfrac{1}{2}\mathds{1}$, and consequently, maximally mixed, $M(\rho^{\,A})=M(\rho^{\,B})=\tfrac{1}{2}$ (see Def.~\ref{def:mixedness of density matrix}), which we will use to define entanglement.

\begin{defi}\label{def:maximal entanglement}
	\begin{tabbing} \hspace*{3.5cm}\=\hspace*{2cm}\=\kill
			\> A pure state $\rho$ is called \textbf{\emph{maximally entangled}}, if all\\[2mm]
            \> reduced density matrices of the subsystems are \emph{maximally}\\[2mm]
            \> \emph{mixed}.
	\end{tabbing}
\end{defi}

This definition also contains the main point of interest about entanglement. The subsystems of an entangled state are somehow correlated in a way, that they cannot be assigned properties independently of the other subsystem(s), i.e. their properties are not well defined. The total system, in contrast, is in a well defined quantum state, which is the interpretation we already stated in the beginning of this chapter. By ignoring (or simply having no access to) certain subsystems, which mathematically corresponds to the partial trace operation, the state of the considered subsystem will become well defined, but also mixed, if the total state is entangled.
To capitalize on this connection, we introduce an important theorem of linear algebra, the \emph{Schmidt-decomposition theorem} (see \cite{schmidt1907}\cite{audretschentangledsystems}).

\begin{theorem}\label{thm:schmidt decomposition theorem}
    \begin{tabbing} \hspace*{3.3cm}\=\hspace*{2cm}\=\kill
            \> For every pure bipartite state $|\,\psi^{\,AB}\rangle$ there exist\\[1mm]
            \> orthonormal bases $\left\{|\,\chi_{\,\mathrm{i}}^{\,A}\,\rangle\in\mathcal{H}_{A}\right\}$ and $\left\{|\,\chi_{\,\mathrm{i}}^{\,B}\,\rangle\in\mathcal{H}_{B}\right\}$,\\[1mm]
            \> called the \emph{Schmidt-bases}, such that \\[2mm]
            \> \> $|\,\psi^{\,AB}\rangle\,=\,\sum\limits_{\mathrm{i}=1}^{\mathrm{k}}\,\sqrt{p_{\,\mathrm{i}}}\,
                    |\,\chi_{\,\mathrm{i}}^{\,A}\,\rangle\otimes|\,\chi_{\,\mathrm{i}}^{\,B}\,\rangle$\ \ ,\\[2mm]
            \> where $p_{\,\mathrm{i}}\geq0$ are real numbers satisfying $\sum\limits_{\mathrm{i}}p_{\,\mathrm{i}}=1$\ , and\\[1mm]
            \> $k\leq\min(\dim(\mathcal{H}_{A}),\dim(\mathcal{H}_{B}))$. The smallest $k=k_{\mathrm{min}}$ is\\[2mm]
            \> called the Schmidt-rank.
    \end{tabbing}
\end{theorem}

\paragraph{Proof:}\ \ To prove this let us rewrite a pure, bipartite state, which was previously specified in Def.~\ref{def:pure bipartite state}, in terms of a generally non-orthonormal basis $\left\{|\,\omega_{\,\mathrm{i}}^{\,B}\,\rangle\right\}$ in subsystem $B$, where

\begin{equation}
|\,\omega_{\,\mathrm{i}}^{\,B}\,\rangle\,=\,\sum\limits_{\mathrm{j}=1}^{\mathrm{m}}\,c_{\,\mathrm{ij}}\,|\,\psi_{\,\mathrm{j}}^{\,B}\,\rangle
\label{eq:relative states for bob}
\end{equation}
and a basis of eigenstates $\left\{|\,\psi_{\,\mathrm{j}}^{\,A}\,\rangle\right\}$ of the reduced density operator $\rho^{\,A}$ in subsystem $A$.

\begin{equation}
\rho^{\,A}\,=\,\sum\limits_{\mathrm{i}=1}^{\mathrm{n}}\,p_{\,\mathrm{i}}\,
|\,\psi_{\,\mathrm{i}}^{\,A}\,\rangle\langle\,\psi_{\,\mathrm{i}}^{\,A}\,|
\label{eq:red density eigenstates for alice}
\end{equation}
where $p_{\,\mathrm{i}}\in\mathbb{R}_{+}$, and $\sum\limits_{\mathrm{i}=1}^{n}\,p_{\,\mathrm{i}}\,=\,1$. We can then rearrange our basis in $A$, such that there is some number $k$, for which $p_{\,\mathrm{i}}>0$, if $1\leq i\leq k$, and  $p_{\,\mathrm{i}}=0$, if $k+1\leq i\leq n$. Our bipartite pure state then has the form

\begin{equation}
|\,\psi^{\,AB}\rangle\,=
\,\sum\limits_{\mathrm{i}=1}^{\mathrm{n}}\,|\,\psi_{\,\mathrm{i}}^{\,A}\,\rangle\otimes|\,\omega_{\,\mathrm{i}}^{\,B}\,\rangle\ \ ,
\label{eq:pure bipartite state for schmidt decomp proof}
\end{equation}
and the corresponding density matrix is given by

\begin{equation}
\rho^{\,AB}\,=\,|\,\psi^{\,AB}\rangle\langle\,\psi^{\,AB}\,|=
\,\sum\limits_{\mathrm{i,l}=1}^{\mathrm{n}}\,|\,\psi_{\,\mathrm{i}}^{\,A}\,\rangle\langle\,\psi_{\,\mathrm{l}}^{\,A}\,|\otimes|\,\omega_{\,\mathrm{i}}^{\,B}\,\rangle\langle\,\omega_{\,\mathrm{l}}^{\,B}\,|\ \ .
\label{eq:pure bipartite state density matrix for schmidt decomp proof}
\end{equation}

We can then calculate the reduced density matrix $\rho^{\,A}$ for Alice.

\begin{eqnarray}
\rho^{\,A} &=& 	Tr_{B}(\rho^{\,AB})\,=\,
	\,\sum\limits_{\mathrm{i,l}=1}^{\mathrm{n}}\,
	|\,\psi_{\,\mathrm{i}}^{\,A}\,\rangle\langle\,\psi_{\,\mathrm{l}}^{\,A}\,|\,
	Tr\left(\,|\,\omega_{\,\mathrm{i}}^{\,B}\,\rangle\langle\,\omega_{\,\mathrm{l}}^{\,B}\,|\,\right)\nonumber\\
 &=& \,\sum\limits_{\mathrm{i,l}=1}^{\mathrm{n}}\,
	|\,\psi_{\,\mathrm{i}}^{\,A}\,\rangle\langle\,\psi_{\,\mathrm{l}}^{\,A}\,|\,
	\sum\limits_{\mathrm{k}=1}^{\mathrm{m}}\,\langle\,\psi_{\,\mathrm{k}}^{\,B}\,|
	\,\omega_{\,\mathrm{i}}^{\,B}\,\rangle\langle\,\omega_{\,\mathrm{l}}^{\,B}\,
	|\,\psi_{\,\mathrm{k}}^{\,B}\,\rangle\,= \nonumber\\
 &=&\,\sum\limits_{\mathrm{i,l}=1}^{\mathrm{n}}\,
	|\,\psi_{\,\mathrm{i}}^{\,A}\,\rangle\langle\,\psi_{\,\mathrm{l}}^{\,A}\,|\,
	\sum\limits_{\mathrm{k}=1}^{\mathrm{m}}\,\langle\,\omega_{\,\mathrm{l}}^{\,B}\,
	|\,\psi_{\,\mathrm{k}}^{\,B}\,\rangle\langle\,\psi_{\,\mathrm{k}}^{\,B}\,|
	\,\omega_{\,\mathrm{i}}^{\,B}\,\rangle\,= \nonumber\\
 &=&\,\sum\limits_{\mathrm{i,l}=1}^{\mathrm{n}}\,
	|\,\psi_{\,\mathrm{i}}^{\,A}\,\rangle\langle\,\psi_{\,\mathrm{l}}^{\,A}\,|\,
	\langle\,\omega_{\,\mathrm{l}}^{\,B}\,|\,\omega_{\,\mathrm{i}}^{\,B}\,\rangle
 \label{eq:proof of schmidt decomposition}
\end{eqnarray}

Comparing this result with Eq.~\eqref{eq:red density eigenstates for alice} it becomes clear, that the state vectors $\ |\,\omega_{\,\mathrm{i}}^{\,B}\,\rangle/\sqrt{p_{\,\mathrm{i}}}$, $\ i=1,\ldots,k\ $, are orthonormal. Since we can repeat the same procedure with interchanged roles of systems $A$ and $B$ we arrive at $\ k\leq\min(\dim(\mathcal{H}_{A}),\dim(\mathcal{H}_{B}))$.\qed

As a direct consequence of Theorem~\ref{thm:schmidt decomposition theorem} the Schmidt-bases, $\left\{|\,\chi_{\,\mathrm{i}}^{\,A}\,\rangle\right\}$ and $\left\{|\,\chi_{\,\mathrm{i}}^{\,B}\,\rangle\right\}$, are the eigenbases of the reduced density operators $\rho^{\,A}$ and $\rho^{\,B}$ respectively and their eigenvalues coincide. The reduced density matrices of a pure, bipartite state therefore also give the same value for all functions of their eigenvalues, such as the \emph{von-Neumann entropy} introduced in Sec.~\ref{subsec:quantification of pure state entanglement}. The value of the \emph{Schmidt-rank} $k$ of a given state at once discloses wether or not a state is separable, which it is only for $k=1$. Also, for a pure bipartite state, the Schmidt-decomposition theorem tells us, that it is enough to look at the reduced density operator of one subsystem, not only to detect, but also to quantify the entanglement present.\\

   \subsubsection{Pure Multipartite Entanglement}\label{subsec:pure multipartite entanglement}

For pure, multipartite states $|\psi^{\,AB\cdots N}\rangle\in\mathcal{H}_{AB\cdots N}$, we can define separability in total analogy to Def.~\ref{def:pure bipartite separable and entangled state}.

\begin{defi}\label{def:pure multipartite separable state}
	\begin{tabbing} \hspace*{3.5cm}\=\hspace*{1.5cm}\=\kill
			\>A pure, $n$-partite state $\,|\,\psi^{\,AB\cdots N}\rangle\,\in\,\mathcal{H}_{AB\cdots N}$ is called
              \\[2mm]
            \> \textbf{\emph{fully}} \emph{($n$-partite)} \textbf{\emph{separable}}, if it can be written as\\[2mm]
            \> \> $|\psi^{\,AB\cdots N}\rangle\,=\,|\psi^{\,A}\rangle\otimes|\psi^{\,B}\rangle\otimes\cdots\otimes
                |\psi^{\,N}\rangle$\ ,\\[2mm]
            \> where $|\psi^{\,A}\rangle\in\mathcal{H}_{A}$, $|\psi^{\,B}\rangle\in\mathcal{H}_{B}$, $\ldots$,
                $|\psi^{\,N}\rangle\in\mathcal{H}_{N}$.\\
	\end{tabbing}
\end{defi}

While the generalization of the separability definition to $n$-partite systems is straightforward, it is not as simple to check wether or not a given pure state is separable in the multipartite case, since these do not universally admit a \emph{generalized Schmidt-decomposition} (see e. g. \cite{peres95}), i.e. a general $n$-partite state $|\,\psi^{\,AB\cdots N}\,\rangle\,\in\mathcal{H}_{AB\cdots N}$, consisting of subsystems $A$,$B,\ldots,N$, can only be written as a single sum,
\begin{equation}
|\,\psi^{\,AB\cdots N}\,\rangle\,=\,\sum\limits_{\mu}\,\sqrt{p_{\,\mu}}\,
|\,\chi_{\,\mu}^{\,A}\,\rangle\,\otimes\,|\,\chi_{\,\mu}^{\,B}\,\rangle\,\otimes\,\cdots\,\otimes
|\,\chi_{\,\mu}^{\,N}\,\rangle\ \ \ ,
\label{eq:generalized Schmidt decomposition}
\end{equation}
if the reduced density matrices of all bipartite partitions of $\mathcal{H}_{AB\cdots N}$ are separable, i.e. they can be written as convex sum of projection operators\footnote{A more detailed description of separability and entanglement of general density matrices will be given in Sec.~\ref{subsec:entanglement of mixed states}.}. This can be seen by applying the same procedure as in the proof of Theorem~\ref{thm:schmidt decomposition theorem} to the tripartite case, choosing the bases $\left\{|\,\psi_{\,\mathrm{j}}^{\,A}\,\rangle\right\}$, $\left\{|\,\psi_{\,\mathrm{j}}^{\,B}\,\rangle\right\}$ as before and additionally a basis $\left\{|\,\psi_{\,\mathrm{j}}^{\,C}\,\rangle\right\}$ in $\mathcal{H}_{\,C}$, but instead of Eq.~\eqref{eq:relative states for bob}, we introduce a relative state for subsystems $B$ and $C$ combined.

\begin{equation}
|\,\omega_{\,\mathrm{i}}^{\,BC}\,\rangle\,=\,\sum\limits_{\mathrm{j,k}=1}\,c_{\,\mathrm{ijk}}
\,|\,\psi_{\,\mathrm{j}}^{\,B}\,\rangle
\,|\,\psi_{\,\mathrm{k}}^{\,C}\,\rangle
\label{eq:relative states for bob and charlie}
\end{equation}

Calculating the reduced density matrix of system $A$ as in Eq.~\eqref{eq:proof of schmidt decomposition} we find that this expression will only agree with the choice of $\left\{|\,\psi_{\,\mathrm{j}}^{\,A}\,\rangle\right\}$ (Eq.~\eqref{eq:red density eigenstates for alice}), if the coefficients $c_{\,\mathrm{ijk}}$ satisfy
\begin{equation}
\sum\limits_{\mathrm{j,k}}\,c_{\,\mathrm{ijk}}(c_{\,\mathrm{ljk}})^{*}\,=\,\delta_{\,\mathrm{il}}\,p_{\mathrm{i}}\ \ \ .
\label{eq:generalized schmidt coefficients restriction}
\end{equation}
This requirement is only met, if the reduced density matrix of the combined systems $B$ and $C$ can be written as
\begin{equation}
Tr_{A}\left(\,|\,\psi^{\,AB\cdots N}\,\rangle\langle\,\psi^{\,AB\cdots N}\,|\,\right)\,=\,
\sum\limits_{\mathrm{i,j,k}}\,c_{\,\mathrm{ijk}}(c_{\,\mathrm{ijk}})^{*}\,
|\,\psi_{\,\mathrm{j}}^{\,B}\,\rangle\langle\,\psi_{\,\mathrm{j}}^{\,B}\,|\,\otimes\,
|\,\psi_{\,\mathrm{k}}^{\,C}\,\rangle\langle\,\psi_{\,\mathrm{k}}^{\,C}\,|\ \ ,
\label{eq:reduced density for bob and charlie}
\end{equation}
which is a separable state\footnote{Compare $\sum\limits_{\mathrm{i}}\,c_{\,\mathrm{ijk}}(c_{\,\mathrm{ijk}})^{*}\,=:\,p_{\,\mathrm{jk}}$ to Def.~\ref{def:mixed separable and entangled state} and set $\sum\limits_{\mathrm{j,k}}\,p_{\,\mathrm{jk}}=\sum\limits_{\mathrm{i}}\,p_{\,\mathrm{i}}$.}. The generalization to the $n$-partite case is straightforward. It is therefore generally necessary to check the purity of all the reduced density matrices of a given pure state.\\

The definition of multipartite entanglement, in contrast to that of multipartite separability (Def.~\ref{def:pure multipartite separable state}), is slightly more intricate, since the entanglement can be confined to certain subsystems and need not generally correlate all subspaces.

\begin{defi}\label{def:pure multipartite entangled state}
	\begin{tabbing} \hspace*{3.5cm}\=\hspace*{1.5cm}\=\kill
            \> A pure, $n$-partite state $\,|\,\psi^{\,AB\cdots N}\rangle\,$ is called (genuinely)\\[2mm]
            \> \textbf{\emph{$n$-partite entangled}}, if the reduced density matrices\\[2mm]
            \> of \emph{all} bipartite partitions are mixed.\\
	\end{tabbing}
\end{defi}

Clearly, as in Def.~\ref{def:maximal entanglement}, the entanglement in the $n$-partite case is called maximal if the reduced density matrices of all bipartite partitions are maximally mixed. Although this seems to be a tedious task, the Schmidt-decomposition theorem still leaves an advantage, since for every bipartite partition, we still only have to calculate the mixedness of one of the reduced density matrices, regardless of the dimensions of the subspaces involved.

	\subsubsection{Quantification of Pure State Entanglement}\label{subsec:quantification of pure state entanglement}

In Sec.~\ref{subsec:pure multipartite entanglement} we found that it is necessary to consider bipartite partitions of the state space to investigate the entanglement contained in the total system. Therefore we restrict our analysis of the detection and quantification of entanglement to (pure) bipartite systems, represented by a density matrix $\rho^{\,AB}$, although the methods presented for pure states can be naturally extended to larger numbers of subsystems. As previously argued, the \emph{mixedness} (see Def.~\eqref{def:mixedness of density matrix}) of the reduced density matrices, \eqref{eq:partial trace over B} or \eqref{eq:partial trace over A}, is then a sufficient measure of entanglement of the total system, also called the \emph{linear entropy} $S_{L}$ of $\rho^{\,A}$ or $\rho^{\,B}$ respectively.

\begin{defi}\label{def:linear entropy}
	\begin{tabbing} \hspace*{3.5cm}\=\hspace*{1.5cm}\=\kill
            \> The \textbf{\emph{linear entropy}} $S_{L}(\rho)$ of a density matrix $\rho$ is\\[2mm]
            \> defined as \ \ \ \ $S_{L}(\rho)\,:=\,1\,-\,Tr(\rho^{2})$\ \ .\\
	\end{tabbing}
\end{defi}

The linear entropy can alternatively be defined with an additional normalization factor $\frac{d}{d-1}$, such that it ranges from $0$ to $1$, instead of $0$ to $1-\tfrac{1}{d}$, but we refrain from doing so in this work. It is also invariant under unitary transformations on Hilbert space.\\

The task of finding an appropriate \emph{pure state entanglement measure} certainly is completed, i.e. the linear entropy is a \emph{real, non-negative function}, which can be seen easily, since a density matrix is Hermitian \eqref{eq:density matrix hermitian} and positive semidefinite \eqref{eq:density matrix positivity}, and therefore only has real, non-negative eigenvalues, it \emph{vanishes for separable states}, since the reduced density matrices of pure, separable states are pure, and it does not increase under \emph{local operations and classical communication} (LOCC), i.e. operations performed only locally in the subspaces A and B, possibly using classical communication. It is difficult to express the most general LOCC operation, which makes a direct proof hard, but there exists a natural inclusion of the LOCC operations in the set of \emph{separable operations}, see \cite{vedralplenio98}\cite{horodeckisentanglement07}, for which the monotonicity property is generally easier to show. An example for such a (one-way forward) LOCC operation, is the unitary ``twirling" (see \cite{horodeckisentanglement07}) operation $\tau(\rho)$ used in \cite{bennettetal96}\footnote{This LOCC operation is called ``random bilateral rotation" in Ref.~\cite{bennettetal96}.}.

\begin{equation}
\tau(\rho)\,=\,\int\!dU\ U\otimes U\rho\ U^{\,\dagger}\!\otimes U^{\,\dagger}
\label{eq:twirling}
\end{equation}

However, there are other, more general, concepts of entropy, the \emph{von \!Neumann entropy}, and the \emph{R$\acute{e}$nyi $\alpha$-entropy}, which provide deeper insight into the correlations of the entangled states. Also it turns out, that these entropy measures are more suited to characterize the entanglement present. In classical information theory, the fundamental entropy measure is \emph{Shannon's entropy} $H(\{p_{\,\mathrm{i}}\})$.

\begin{defi}\label{def:shannon entropy}
	\begin{tabbing} \hspace*{3.5cm}\=\hspace*{1.5cm}\=\kill
            \> The \textbf{\emph{Shannon entropy}} $H(\{p_{\,\mathrm{i}}\})$ of a probability\\[3mm]
            \> distribution $P$ is given by \ \  $H(\{p_{\,\mathrm{i}}\})\,=\,-\,\sum\limits_{i}\,p_{\,\mathrm{i}}\,\log(p_{\,\mathrm{i}})$\,,\\[2mm]
            \> where the logarithm is taken to the base $2$ and $\sum\limits_{i}p_{\,\mathrm{i}}=1$\,.
	\end{tabbing}
\end{defi}

It describes operationally the average number of bits one needs to transmit a message with a classical statistical source, associated to the random variables $p_{\,\mathrm{i}}$, without knowing the values of said variables. The natural quantum mechanical extension of this quantity is the \emph{von Neumann entropy} $S(\rho)$.

\begin{defi}\label{def:von Neumann entropy}
	\begin{tabbing} \hspace*{3.3cm}\=\hspace*{1.5cm}\=\kill
            \> The \textbf{\emph{von \!Neumann entropy}} $S(\rho)$ of a density matrix $\rho$\\[3mm]
            \> is given by \ \  $S(\rho)\,=\,-\,Tr(\rho\,\log(\rho))\,=\,-\,\sum\limits_{i}\,p_{\,\mathrm{i}}\,\log(p_{\,\mathrm{i}})$ \ ,\\[2mm]
            \> where $p_{\,\mathrm{i}}$ are the eigenvalues of $\rho$.
	\end{tabbing}
\end{defi}

Similar to the Shannon entropy, the von Neumann entropy has an operational interpretation if the logarithm is taken to the base $2$, then, according to Schumacher \cite{schumacher95} ``\emph{...the von Neumann entropy $S$ of the density operator describing an ensemble of pure quantum signal states is equal to the number of spin-$\tfrac{1}{2}$ systems (``quantum bits" or ``qubits") necessary to represent the signal faithfully.}"\\

One can check that the von Neumann entropy vanishes for pure states, this means, in particular, for the reduced density matrices of a pure separable state. Furthermore the logarithm in Def.~\ref{def:von Neumann entropy} is sometimes chosen to the base $d$, the dimension of the Hilbert space, since it is then bounded by $1$, which it reaches for the maximally mixed state.

\begin{equation}
0\,\leq\,S(\rho)\,\leq\,1
\label{eq:bounds of von Neumann entropy}
\end{equation}
It is a positive (semi-definit) functional,
\begin{equation}
S(\rho)\,\geq\,0\ \ ,
\label{eq:positivity of von Neumann entropy}
\end{equation}
and is invariant under unitary transformations $U$,
\begin{equation}
S(U\rho\,U^{\,\dagger})\,=\,S(\rho)\ \ .
\label{eq:unitary invariance of von Neumann entropy}
\end{equation}
Also the von\! Neumann entropy of a bipartite state $\rho^{\,AB}$ satisfies the \emph{subadditivity} property, i.e. it is always smaller (or equal to) the sum of the entropies of the reductions $\rho^{\,A}$ and $\rho^{\,B}$, which provides an upper bound, while a lower bound is given by the modulus of their difference \cite{arakilieb70}.

\begin{equation}
|S(\rho^{\,A})\,-\,S(\rho^{\,B})|\,\leq\,S(\rho^{\,AB})\,\leq\,S(\rho^{\,A})\,+\,S(\rho^{\,B})
\label{eq:araki lieb inequality}
\end{equation}

The most obvious pure state entanglement measure, which can be constructed from the von\! Neumann entropy is called \emph{entropy of entanglement}, defined as follows.

\begin{defi}\label{def:entropy of entanglement}
	\begin{tabbing} \hspace*{3.3cm}\=\hspace*{1.5cm}\=\kill
            \> The \textbf{\emph{entropy of entanglement}} $\mathcal{E}(\rho)$ of a bipartite pure\\[2mm]
            \> state $\rho$ is defined as the von\! Neumann entropy of either of\\[2mm]
            \> the two reduced density matrices $\rho^{\,A}$ or $\rho^{\,B}$,\\[3mm]
            \> \> \ \ \ $\mathcal{E}(\rho)\,=\,S(\rho^{\,A})=\,S(\rho^{\,B})$\ . \\
	\end{tabbing}
\end{defi}

The connection to the linear entropy can be easily found, it is established on the grounds that the linear entropy $S_{L}(\rho)$ is the first order approximation of the von\! Neumann entropy, when the logarithm is taken to be the natural logarithm.

\begin{equation}
-\,Tr(\rho\,\ln\rho)\,\approx\,-\,Tr(\rho\,(\rho\,-\,1))\,=\,Tr(\rho)\,-\,Tr(\rho^{2})\,=\,1\,-\,Tr(\rho^{2})
\label{eq:connection of von Neumann and linear entropy}
\end{equation}

Finally, let us introduce the R$\acute{e}$nyi $\alpha$-entropy $S_{\alpha}(\rho)$ (see  \cite{renyi61} and \cite{horodecki-rpm96}) by defining:

\begin{defi}\label{def:alpha entropy}
	\begin{tabbing} \hspace*{3.3cm}\=\hspace*{1.5cm}\=\kill
            \> The \textbf{\emph{R$\mathbf{\acute{e}}$nyi $\alpha$ - entropy}} $S_{\alpha}(\rho)$ of a density matrix $\rho$\\[3mm]
            \> is given by \ \ $S_{\alpha}(\rho)\,=\,\frac{1}{1-\alpha}\,\log\,Tr(\rho^{\alpha})$\ , $\alpha\geq0$, $\alpha\neq1$\ . \\ \end{tabbing}
\end{defi}

In the limit of $\alpha\rightarrow1$ we obtain the von\! Neumann entropy (with $\log$ taken to be $\ln$), which ca be seen by applying the rule of de l'H$\mathrm{\hat{o}}$spital.

\begin{equation}
\lim_{\alpha\,\to\,1}\,S_{\alpha}(\rho)\,=\,\lim_{\alpha\,\to\,1}\,\frac{\frac{d}{d\alpha}Tr(\rho^{\alpha})}{\frac{d}{d\alpha}(1-\alpha)}\,=\,-\,\lim_{\alpha\,\to\,1}\,Tr(\rho^{\alpha}\,\ln\rho)\,=\,S(\rho)
\label{eq:limit of alpha to one for renyi entropy}
\end{equation}

For pure states these $\alpha$ - entropies also detect entanglement, i.e. for pure bipartite states $\rho^{\,AB}$ with reduced density matrices $\rho^{\,A}$ and $\rho^{\,B}$ the following inequalities are violated if, and only if, they are entangled.

\begin{equation}
S_{\alpha}(\rho^{\,AB})\,\geq\,S_{\alpha}(\rho^{\,A})\ ,\ \ S_{\alpha}(\rho^{\,AB})\,\geq\,S_{\alpha}(\rho^{\,B})
\label{eq:alpha entropy inequalities}
\end{equation}

This is obvious from the fact that the entropies of the pure states vanish identically, while those of their reductions are nonzero if the state is entangled, but we will come back to these inequalities in Sec.~\ref{subsec:entanglement measures for mixed states}.

            \subsubsection{Entanglement of Mixed States}\label{subsec:entanglement of mixed states}

As discussed earlier in Sec.~\ref{subsec:density operators}, statistical mixtures of quantum states need to be described by density operators, representing more realistic sources of entangled states. Describing the entanglement contained in a general mixed state, however, presents two difficulties not occurring with pure states.\\

The first difference to pure states arises from the statistical mixture of composite system. Consider an ensemble of pure, bipartite states $\rho_{\,\mathrm{i}}^{\,AB}$, which in addition are all separable. Clearly, since the tensor product structure is preserved, these states can be written as $\rho_{\,\mathrm{i}}^{\,AB}=\rho_{\,\mathrm{i}}^{\,A}\otimes\rho_{\,\mathrm{i}}^{\,B}$. Since the states of the ensemble are all separable, we do not expect any quantum correlations to occur, when convex sums of these states are considered, but it is then possible that the subsystems $A$ and $B$ of the ensemble are \emph{classically correlated}, e.g.

\begin{equation}
\rho^{\,AB}\,=\,\frac{1}{2}\,\left(\,|\up\,\rangle\langle\,\up|\otimes|\down\,\rangle\langle\,\down|\,+\,
|\down\,\rangle\langle\,\down|\otimes|\up\,\rangle\langle\,\up|\,\right)\ \ ,
\label{eq:classical correlation example}
\end{equation}
where measurements of the spins along the quantization axis will give perfectly anti-correlated results, i.e. spin up in one subsystem and spin down in the other, while measurements along directions perpendicular to that axis will show no correlation whatsoever. This can be easily checked by calculating the expectation values of the observable $\sigma_{\mathrm{z}}\otimes\sigma_{\mathrm{z}}$ in Eq.~\eqref{eq:uncorrelated}, which gives $-1$ for the total state, but vanishes for the reduced density matrices. We therefore acknowledge, mixed states that are separable can be classically correlated.

\begin{defi}\label{def:mixed separable and entangled state}
	\begin{tabbing} \hspace*{3.5cm}\=\hspace*{2cm}\=\kill
			\> A state $\rho^{\,AB}$ is called \textbf{\emph{separable}} (w.r.t. the partition\\[2mm]
            \> into subsystems $A$ and $B$), if it can be written as \\[3mm]
            \> \> $\rho^{\,AB}\,=\,\sum\limits_{\mathrm{i}}\,p_{\,\mathrm{i}}\,
            \rho_{\,\mathrm{i}}^{\,A}\otimes\rho_{\,\mathrm{i}}^{\,B}$\ ,\\[3mm]
            \> where $\sum\limits_{\mathrm{i}}\,p_{\,\mathrm{i}}=1$ and $p_{\,\mathrm{i}}\geq0$.\\[3mm]
            \> A (bipartite) state is called \textbf{\emph{entangled}}, if it is not\\[1mm]
            \> separable.
	\end{tabbing}
\end{defi}

These definitions now naturally extend to $n$-partite states, which are separable only if they can be written as the convex sum of $n$-fold product states,

\begin{equation}
\rho^{\,AB\cdots N}\,=\,\sum\limits_{\mathrm{i}}\,p_{\,\mathrm{i}}\
\rho_{\,\mathrm{i}}^{\,A}\otimes\rho_{\,\mathrm{i}}^{\,B}\otimes\cdots\otimes\rho_{\,\mathrm{i}}^{\,N} \ \ \
\label{eq:mixed multipartite separability}
\end{equation}
and are (at least partially) entangled otherwise.
\begin{equation}
\rho^{\,AB\cdots N}\,\neq\,\sum\limits_{\mathrm{i}}\,p_{\,\mathrm{i}}\
\rho_{\,\mathrm{i}}^{\,A}\otimes\rho_{\,\mathrm{i}}^{\,B}\otimes\cdots\otimes\rho_{\,\mathrm{i}}^{\,N} \ \ \
\label{eq:mixed multipartite entanglement}
\end{equation}

However, it is now much more complicated to determine, wether or not a given mixed state is entangled. The reason for this comprises the other difference to pure states. The mixedness of the subsystem's reduced density matrices does not give conclusive information about the entanglement of the total state, if the latter is mixed. Going back to the example given in Eq.~\eqref{eq:classical correlation example}, it can be easily seen, that the reduced density matrices of both subsystems are maximally mixed, i.e. $\rho^{\,A}=\tfrac{1}{2}\mathds{1}_{2}$, even though the total state \eqref{eq:classical correlation example} is not maximally mixed, and, most importantly, separable.

            \subsubsection{Quantification of Mixed State Entanglement}\label{subsec:entanglement measures for mixed states}

In contrast to pure states, it is much more difficult generally to tell wether or not a given mixed state is entangled or not, and to quantify the entanglement present if it is. We briefly want to mention some of the many different approaches to this problem but refer the reader to the substantial literature on entanglement, such as \cite{horodeckisentanglement07} measures for mixed states for more in depth analysis. The proposed entanglement measures fall into two categories, \emph{operational separability criteria}, such as e.g. the \emph{CHSH-criterion} (see Sec.~\ref{subsec:bell theorem}), \emph{entropy inequalities}, \emph{entanglement cost}, \emph{entanglement of distillation}\footnote{Entanglement cost and entanglement of distillation are only operational if a particular LOCC operation is chosen, see \eqref{eq:entanglement cost}, and \eqref{eq:entanglement of distillation}.}, the \emph{PPT-criterion}, or the \emph{logarithmic negativity}, as well as \emph{non-operational separability criteria}, e.g. \emph{entanglement of formation}, \emph{entanglement witnesses}, or \emph{relative entropy of entanglement}.\\

\pa{Entropy Measures:}\ \ \\

The most obvious starting point to look for mixed state entanglement measures, is the exploitation of the entropy of entanglement, i.e. the von\! Neumann entropy (Def.~\ref{def:von Neumann entropy}). Previously we have always started out with a pure composite system, for which we calculated the entropy. If we proceed with an ensemble of pure states $\{(p_{\,\mathrm{i}},\rho_{\,\mathrm{i}})\}$, we find that the von\! Neumann entropy reduces to the Shannon entropy (Def.~\ref{def:shannon entropy}) of the probability distribution $\{p_{\,\mathrm{i}}\}$. For an ensemble of mixed density matrices, with supports on orthogonal subspaces, the equality

\begin{equation}
S(\sum\limits_{\mathrm{i}}\,p_{\,\mathrm{i}}\,\rho_{\,\mathrm{i}})\,=\,H(\{p_{\,\mathrm{i}}\})\,+\,
\sum\limits_{\mathrm{i}}\,p_{\,\mathrm{i}}\,S(\rho_{\,\mathrm{i}})\ \ ,
\label{eq:entropy theorem}
\end{equation}
holds, which can be easily seen by first diagonalizing the density matrices, i.e.
\begin{equation}
\rho_{\,\mathrm{i}}\,=\,\sum\limits_{\mathrm{j}}\,\lambda_{\mathrm{i}}^{\,\,\mathrm{j}}\,
|\,e_{\mathrm{i}}^{\,\,\mathrm{j}}\,\rangle\langle\,e_{\mathrm{i}}^{\,\,\mathrm{j}}\,|\ \ ,
\label{eq:ensemble density matrix diagonalization}
\end{equation}
and using the property $\log ab=\log a\,+\,\log b$, and the normalization of the density matrices,  $\sum\limits_{\mathrm{j}}\,\lambda_{\mathrm{i}}^{\,\,\mathrm{j}}=1$ .
\begin{eqnarray}
S(\sum\limits_{\mathrm{i}}\,p_{\,\mathrm{i}}\,\rho_{\mathrm{i}}) &=& -\,\sum\limits_{\mathrm{i,j}}\,p_{\,\mathrm{i}}\,\lambda_{\mathrm{i}}^{\,\,\mathrm{j}}\,\log\,p_{\,\mathrm{i}}\,\lambda_{\mathrm{i}}^{\,\,\mathrm{j}}\,= \nonumber \\[2mm]
&=& -\,\sum\limits_{\mathrm{i}}\,p_{\,\mathrm{i}}\,
(\log\,p_{\,\mathrm{i}})\,\sum\limits_{\mathrm{j}}\,\lambda_{\mathrm{i}}^{\,\,\mathrm{j}}\,-\,\sum\limits_{\mathrm{i}}\,p_{\,\mathrm{i}}\,\sum\limits_{\mathrm{j}}\,\lambda_{\mathrm{i}}^{\,\,\mathrm{j}}\,\log\,\lambda_{\mathrm{i}}^{\,\,\mathrm{j}}\,= \nonumber\\[2mm]
&=& H(\{p_{\,\mathrm{i}}\})\,+\,
\sum\limits_{\mathrm{i}}\,p_{\,\mathrm{i}}\,S(\rho_{\,\mathrm{i}})
\label{eq:proof of entropy theorem}
\end{eqnarray}

Since the Shannon entropy is non-negative, in this case, trivially we can formulate the inequality
\begin{equation}
S(\sum\limits_{\mathrm{i}}\,p_{\,\mathrm{i}}\,\rho_{\mathrm{i}})\,\geq\,
\sum\limits_{\mathrm{i}}\,p_{\,\mathrm{i}}\,S(\rho_{\,\mathrm{i}})\ .
\label{eq:concavity of entropy}
\end{equation}
This property is more generally true for all decompositions of density matrices and is called \emph{concavity}. We are now in the position to formulate the following theorem for the $\alpha$\emph{-entropy inequalities} \eqref{eq:alpha entropy inequalities}, initially introduced for pure states, which hold for arbitrarily mixed separable states, see \cite{horodecki-rpm96} and \cite{horodecki-rm96}.
\begin{theorem}\label{thm:alpha entropy inequalities}
    \begin{tabbing} \hspace*{3.1cm}\=\hspace*{1cm}\=\kill
            \> The $\alpha$-entropy inequalities for mixed bipartite states $\rho^{\,AB}$,\\[2mm]
            \> \> $S_{\alpha}(\rho^{\,AB})\,\geq\,S_{\alpha}(\rho^{\,A})\ ,\ \ S_{\alpha}(\rho^{\,AB})\,\geq\,S_{\alpha}(\rho^{\,B})$ , \\[2mm]
            \> are violated only if the state $\rho^{\,AB}$ is entangled.\\
    \end{tabbing}
\end{theorem}
We will present a proof for the special case of the von\! Neumann entropy (Def.~\ref{def:von Neumann entropy}) here. Proofs for $\alpha=0,1,2,$ and $\infty$ can be found in  Ref.~\cite{terhal02}. Theorem \ref{thm:alpha entropy inequalities} only provides a necessary separability condition, but not a sufficient one, which means that non-violation of the inequality does not give a conclusive result.\\

\paragraph{Proof:}\ \ To prove this let us first calculate the entropy of a bipartite uncorrelated state $\rho=\rho^{\,A}\otimes\rho^{\,B}$ using Eq.~\eqref{eq:proof of entropy theorem}.

\begin{eqnarray}
S(\rho^{\,A}\otimes\rho^{\,B}) &=& S(\sum\limits_{\mathrm{i}}\,p_{\,\mathrm{i}}\,\rho^{\,A}\otimes\,|\,\mathrm{i}\,\rangle\langle\,\mathrm{i}\,|)\,=\nonumber\\
&=&
H(\{p_{\,\mathrm{i}}\})\,+\,\sum\limits_{\mathrm{i}}\,p_{\,\mathrm{i}}\,S(\rho^{\,A}\otimes\,|\,\mathrm{i}\,\rangle\langle\,\mathrm{i}\,|)\,=\nonumber\\
&=& S(\rho^{\,B})\,+\,\sum\limits_{\mathrm{i}}\,p_{\,\mathrm{i}}\,S(\rho^{\,A})\,=\,
S(\rho^{\,A})\,+\,S(\rho^{\,B})
\label{eq:entropy of uncorrelated state}
\end{eqnarray}
We can then use Eq.~\eqref{eq:proof of entropy theorem} and Eq.~\eqref{eq:entropy of uncorrelated state} to calculate the entropy of a bipartite separable state $\rho^{\,AB}$ (Def.~\ref{def:mixed separable and entangled state}).
\begin{eqnarray}
S(\rho^{\,AB}) &=& S(\sum\limits_{\mathrm{i}}\,p_{\,\mathrm{i}}\,
		\rho_{\,\mathrm{i}}^{\,A}\otimes\rho_{\,\mathrm{i}}^{\,B})\,=\nonumber\\
&=& H(\{p_{\,\mathrm{i}}\})\,+\,\sum\limits_{\mathrm{i}}\,p_{\,\mathrm{i}}\,
		S(\rho_{\,\mathrm{i}}^{\,A}\otimes\rho_{\,\mathrm{i}}^{\,B}))\,=\nonumber\\
&=& H(\{p_{\,\mathrm{i}}\})\,+\,
		\sum\limits_{\mathrm{i}}\,p_{\,\mathrm{i}}\,S(\rho_{\,\mathrm{i}}^{\,A})\,+\,
		\sum\limits_{\mathrm{i}}\,p_{\,\mathrm{i}}\,S(\rho_{\,\mathrm{i}}^{\,B})
\label{eq:entropy of separable state}
\end{eqnarray}
The entropies of the reduced density operators $\rho^{\,A}$ and $\rho^{\,B}$ on the other hand are
\begin{equation}
S(\rho^{\,A})\,=\,S(\sum\limits_{\mathrm{i}}\,p_{\,\mathrm{i}}\,\rho_{\,\mathrm{i}}^{\,A})\,=\,
H(\{p_{\,\mathrm{i}}\})\,+\,\sum\limits_{\mathrm{i}}\,p_{\,\mathrm{i}}\,S(\rho_{\,\mathrm{i}}^{\,A})\ ,
\label{eq:separable reduction  A entropy}
\end{equation}
\begin{equation}
\mbox{and}\ \ S(\rho^{\,B})\,=\,
H(\{p_{\,\mathrm{i}}\})\,+\,\sum\limits_{\mathrm{i}}\,p_{\,\mathrm{i}}\,S(\rho_{\,\mathrm{i}}^{\,B})\ ,
\label{eq:separable reduction  B entropy}
\end{equation}
which due to the non-negativity of the entropies are always smaller or equal than the entropy of the total separable state $\rho^{\,AB}$.\qed

It should be pointed out here, however, that the inequalities for $\alpha=2$ provide a stronger separability condition than the CHSH-criterion (Theorem~\ref{thm:chsh criterion}), which means that some entangled states that satisfy the CHSH-criterion violate the the entropy inequality for $\alpha=2$. Also there exists another entropy measure based on the idea of relative entropy $S(\rho||\sigma)$,
\begin{equation}
S(\rho||\sigma)=Tr\rho\log\rho-Tr\rho\log\sigma\ \ ,
\label{eq:relative entropy}
\end{equation}
of two states $\rho$ and $\sigma$, characterizing how close states are to the set of separable states, called \emph{relative entropy of entanglement}, which can be found in Ref.~\cite{vedralplenio98}.

\pa{Entanglement of Formation \& Concurrence:}\ \\

A non-operational way to use the entropy of entanglement $\mathcal{E}$ (Def.~\ref{def:entropy of entanglement}) not only to detect, but also to quantify mixed state entanglement is the so called \emph{Entanglement of Formation} $E_{\mathrm{F}}(\rho)$ \cite{bennettetal96b}, a so called \emph{convex roof} construction, defined as
\begin{equation}
E_{\mathrm{F}}(\rho)\,:=\,\inf_{\{(p_{\,\mathrm{i}},\rho_{\,\mathrm{i}})\}}\,\sum\limits_{\mathrm{i}}\,p_{\,\mathrm{i}}\,\mathcal{E}(\rho_{\,\mathrm{i}})\ \ ,
\label{eq:entanglement of formation}
\end{equation}
where the infimum is evaluated over all ensembles $\{(p_{\,\mathrm{i}},\rho_{\,\mathrm{i}})\}$ of pure states $\rho_{\,\mathrm{i}}$ and probability distributions $\{p_{\,\mathrm{i}}\}$ realizing $\rho$. This is generally very difficult to calculate, since the decompositions of a general mixed state into ensembles of pure states is not unique and the infimum is to take over all possible decompositions. Only in the special case of two qubits is an operational method of calculating the entanglement of formation known, for a proof and more details see \cite{bennettetal96b} and \cite{wootters98}. It can then be expressed as a function of the \emph{concurrence} $C(\rho)$ as,
\begin{equation}
E_{\mathrm{F}}(\rho)\,=\,E_{\mathrm{F}}(C(\rho))\,=\,h\left(\frac{1+\sqrt{1-C^{2}(\rho)}}{2}\right)\ \ ,
\label{eq:entanglement of formation for 2 qubits}
\end{equation}
where $h(p)=H(\{p,1-p\})$, $H$ is the Shannon entropy (Def.~\ref{def:shannon entropy}), and the concurrence is given by
\begin{equation}
C(\rho)\,=\,\max\{\,0,\sqrt{\lambda_{1}}-\sqrt{\lambda_{2}}-\sqrt{\lambda_{3}}-\sqrt{\lambda_{4}}\,\}\ \ .
\label{eq:concurrence}
\end{equation}
The values $\lambda_{\mathrm{i}}$ are the eigenvalues of the matrix $\rho(\sigma_{\mathrm{y}}\otimes\sigma_{\mathrm{y}})\rho^{*}(\sigma_{\mathrm{y}}\otimes\sigma_{\mathrm{y}})$, where the asterisk indicates complex conjugation, in decreasing order.\\

\pa{Entanglement Cost \& Entanglement of Distillation:}\ \\

The question remains, how a physical interpretation can be given to the entanglement of formation, since there is, a priori, no physical reason, why the ensemble of pure states minimizing the convex sum of entropies is singled out, while all the ensembles are equivalent for the quantum mechanical description of the total state. Such a connection can be established by the so called \emph{Entanglement Cost} $E_{\mathrm{C}}(\rho)$, see e.g. \cite{bruss02},
\begin{equation}
E_{\mathrm{C}}(\rho)\,=\,\inf_{\{LOCC\}}\,\lim_{n_{\rho}\,\to\,\infty}\,\frac{m^{\,\mathrm{in}}}{n^{\,\mathrm{out}}_{\rho}}\ \ ,
\label{eq:entanglement cost}
\end{equation}
which quantifies how many copies $m^{\,\mathrm{in}}$ of maximally entangled 2-qubit states $|\,\phi^{+}\,\rangle$ are needed to prepare the state $\rho$ applying LOCC operations only. It can then be shown that the entanglement cost is related the entanglement of formation by
\begin{equation}
E_{\mathrm{C}}(\rho)\,=\,\lim_{n\,\to\,\infty}\,\frac{1}{n}\,E_{\mathrm{F}}(\rho^{\otimes n})\ \ ,
\label{eq:entanglement cost and entanglement of formation}
\end{equation}
where ``$\rho^{\otimes n}$" symbolizes the $n$-fold tensor product of $\rho$ with itself. A proof of Eq.~\eqref{eq:entanglement cost and entanglement of formation} can be found in Ref.~\cite{haydenhorodeckiterhal01}. Obviously the entanglement of formation, the concurrence and the entanglement cost reduce to the entropy of entanglement for pure states.\\

Alternatively to asking how difficult it is to prepare a certain quantum state in terms of maximally entangled states, one can also define entanglement measures by the number $m^{\,\mathrm{out}}$ of maximally entangled states $|\,\phi^{+}\,\rangle$, which can be distilled by (LOCC) purification protocols from the original state $\rho$, which comprises the \emph{Entanglement of Distillation} $E_{\mathrm{D}}(\rho)$,
\begin{equation}
E_{\mathrm{D}}(\rho)\,=\,\sup_{\{LOCC\}}\,\lim_{n_{\rho}\,\to\,\infty}\,\frac{m^{\,\mathrm{out}}}{n^{\,\mathrm{in}}_{\rho}}\ \ ,
\label{eq:entanglement of distillation}
\end{equation}
see Ref.~\cite{bennettetal96},\cite{bennettetal96b}, and \cite{bruss02}. Generally the entanglement of Distillation provides a lower bound for other entanglement measures, such as the relative entropy of entanglement (see Ref.~\cite{vedralplenio98}\cite{vedral02}, and \cite{vollbrechtwerner01}) and logarithmic negativity \cite{vidalwerner02}\cite{plenio05}, while the entanglement cost is an upper bound.
\begin{equation}
E_{\mathrm{D}}(\rho)\,\leq\,E(\rho)\,\leq\,E_{\mathrm{C}}(\rho)
\label{eq:bounds on entanglement measures}
\end{equation}

\pa{Entanglement Witnesses:}\ \\

A different way of looking at entanglement is by analyzing the geometry of the state space. Since the set of separable states is convex, one can construct a hyperplane separating a given entangled state from this set, which is formulated in the Entanglement Witness Theorem \ref{thm:entanglement witness theorem}, introduced in \cite{horodecki-mpr96}.
\begin{theorem}\label{thm:entanglement witness theorem}
    \begin{tabbing} \hspace*{3.1cm}\=\hspace*{1cm}\=\kill
            \> A state $\rho$ is entangled if, and only if, there $\exists$ a Hermitian\\[2mm]
            \> operator $W$, an \textbf{\emph{entanglement witness}}, such that \\[2mm]
            \> \> $\langle\,\rho,\,W\,\rangle\,<\,0$\ \  and \ \ $\langle\,\sigma,\,W\,\rangle\,\geq\,0$\\[2mm]
            \> for all separable states $\sigma$.\\
    \end{tabbing}
\end{theorem}
This theorem clearly does not supply an operational criterion for the detection of entanglement, but it has proven to be useful in characterizing different forms of entanglement, see e.g. \cite{bertlmannkrammer08}. Also shown in Ref.~\cite{horodecki-mpr96} is the connection to the so called \emph{Positive Map Theorem} (PMT).
\begin{theorem}\label{thm:positive map theorem}
    \begin{tabbing} \hspace*{3.1cm}\=\hspace*{2.5cm}\=\kill
            \> A bipartite state $\rho$ is separable if, and only if\\[2mm]
            \> \> $(\mathds{1}\,\otimes\,\mathcal{P})\rho\,\geq\,0$\\[2mm]
            \> for all positive operators $\mathcal{P}$.\\
    \end{tabbing}
\end{theorem}
For this theorem to have any useful result we need to find operators, which are positive, but not completely positive\footnote{A map $\mathcal{P}_{C}$ is called completely positive if it is positive and extensions $\mathcal{P}_{C}\otimes\mathds{1}_{\mathrm{d}}$ to larger (Hilbert) spaces are also positive $\forall\,d$.}, in order to detect entanglement. One such operator is the transposition $T$, which leads us to a necessary separability criterion for bipartite states, which becomes sufficient if the dimension d of the product space does not exceed $d=6$. It is called the \emph{Peres - Horodecki criterion} (\cite{peres96},\cite{horodecki-mpr96}), or \emph{PPT - criterion}, and is even stronger than the entropy inequalities of Theorem \ref{thm:alpha entropy inequalities}.

\begin{theorem}\label{thm:PPT theorem}
    \begin{tabbing} \hspace*{3cm}\=\hspace*{2.5cm}\=\kill
            \> A bipartite state $\rho\,\in\,\mathcal{H}^{2}\otimes\mathcal{H}^{2},\ \mathcal{H}^{2}\otimes\mathcal{H}^{3},\,$ or $\mathcal{H}^{2}\otimes\mathcal{H}^{3}$ is\\[2mm]
            \> separable if, and only if, its partial transposition is positive.\\[2mm]
            \> \> $\rho^{\,T_{\mathrm{B}}}\,=\,(\mathds{1}\otimes T)\rho\,\geq\,0$\\
    \end{tabbing}
\end{theorem}

Another separability criterion, which can be formulated as an entanglement witness, is the CHSH-criterion (Theorem \ref{thm:chsh criterion}), which we introduce in Sec.~\ref{subsec:bell theorem}.

\newpage
    \subsection{The EPR - Paradox \& Bell's Theorem}\label{sec:EPR Bell}	
The most intriguing importance of entanglement lies at the heart of the Gedankenexperiment, presented by Albert Einstein and his colleagues, Boris Podolsky, and Nathan Rosen, in their famous 1935 paper \cite{EPR35}, titled ``\emph{Can Quantum-Mechanical Description of Physical Reality Be Considered Complete?}", which throughout the literature and in the following is referred to as the \emph{EPR - Paradox}. It illustrates how certain requirements, \emph{reality}, \emph{locality} and \emph{completeness}, which we intuitively would expect to be fulfilled by a physical theory, cannot all be met by quantum theory at the same time, due to the existence of entangled states.\\

Despite the far-reaching conceptual consequences of the paradox, it was mostly ignored, or deemed purely philosophical, by the physical community as a result of the reply \cite{bohr1935},  and the authority, of Niels Bohr, although the subject of the non-classical correlations between separated physical systems was not touched there in any way. Only much later were the paradox, and the quantum mechanical correlations inherent to it, appreciated by David Bohm and Yakir Aharonov, who reformulated the situation with spin variables instead of position and momentum.\\

The underlying physical content of the apparently philosophical debate was then revealed in 1964, when John Stewart Bell found a way to probe the EPR - paradox in a real experiment, see \cite{bell64}. We will present the EPR - Paradox and its implications, in the version proposed by Bohm and Aharonov \cite{bohmaharonov57}, in Sec.~\ref{subsec:EPR}, followed by the resolution by Bell in Sec.~\ref{subsec:bell theorem}. More information about this topic can be found in Ref.~\cite{bellspeakable}.

    					\subsubsection{The Einstein - Podolsky - Rosen Paradox}\label{subsec:EPR}
    					
In order to express their discomfort with the probabilistic character of quantum theory, Einstein, Podolsky, and Rosen put forward three\footnote{The argumentation in \cite{EPR35} does not treat all three assumptions equally, in contrast to completeness and reality, the locality requirement is not questioned in their work, but since the argument also fundamentally depends on it, we will include it, as commonly done, as a main requirement.} requirements in their work \cite{EPR35}, which should be met by any satisfactory physical theory. These are

\begin{enumerate}
\item{\begin{tabbing}\hspace*{3.5cm}\=\kill
			 \textbf{\emph{Completeness:}}\> Every element of reality must be assigned to a\\[1mm]
			 \> corresponding element in the physical theory, in\\[1mm]
			 \> order for the theory to be complete.
			\end{tabbing}}
\item{\begin{tabbing}\hspace*{2.5cm}\=\kill
			 \textbf{\emph{Realism:}}
			 \> If the value of a physical quantity can be predicted with\\[1mm]
			 \> certainty, i.e. probability equal to unity, without in any\\[1mm]
			 \> way perturbing the system, the quantity corresponds to\\[1mm]
			 \> an element of physical reality.\\
			\end{tabbing}}
\item{\begin{tabbing}\hspace*{2.5cm}\=\kill
			 \textbf{\emph{Locality:}}
			 \> There is no instantaneous interaction at a distance, i.e.\\[1mm]
			 \> neither between the individual distant subsystems, nor\\[1mm]
			 \> between subsystems and distant measurement devices.\\
			\end{tabbing}}
\end{enumerate}

Assume now that a system of total spin $0$ dissociates into two subsystems Alice (A), and Bob (B), each of spin $\tfrac{1}{2}$, which do not interact after some initial time, such that the joint spin state for the spatially separated subsystems is given by the antisymmetric Bell state
\begin{equation}
|\,\psi^{\,-}\,\rangle \,=\,
\frac{1}{\sqrt{2}}\left(\,|\up\,\rangle\,|\down\,\rangle\,-\,|\down\,\rangle\,|\up\,\rangle\,\right)\ \ \ .
\label{eq:bell psi minus state}
\end{equation}

Suppose then, that a spin measurement along an arbitrary direction, without loss of generality, the $z$-direction, is performed on system A. The result of this measurement will be either $+\tfrac{\hbar}{2}$, i.e. spin up, or $-\tfrac{\hbar}{2}$, i.e. spin down, regardless which direction is chosen. Since the spin operators for perpendicular directions do not commute, see Eqs.~\eqref{eq:pauli matrices commutator}\eqref{eq:SU2 lie algebra}, by Heisenberg's uncertainty principle (see e.g. \cite{griffithsQM}, pages 108-110), the exact values of different spin components can not be simultaneously known, i.e. they cannot be simultaneously predicted with certainty, and cannot therefore be parts of the physical reality at the same time. But it is still possible to argue, that by measuring the spin of system A along a particular direction, the measurement apparatus interacts with the system, and subsequently the spin along perpendicular directions is somehow disturbed and cannot therefore be predicted accurately.\\

However, regardless which measurement direction was chosen for system A and which result, up or down, was obtained, the quantum mechanical description predicts that a measurement along the same direction on system B will give the opposite result, down or up, every time. Assuming the locality condition to hold, subsystem B has not been disturbed by the measurement performed on A and therefore, by the argumentation of EPR, the spin along the chosen direction of system B can be said to have physical reality. This statement causes conceptual problems, the so called \emph{EPR - paradox}. Since the choice of measurement direction was arbitrary, we could have chosen a different direction, perpendicular to the former. Following the same line of argument, the spin of B along the alternative direction is an element of physical reality as well. As no interaction between the subsystems has taken place, system B has no information about the chosen direction. Consequently, all possible spin directions of system B should be elements of (a simultaneous) physical reality. This clearly is in contradiction with the uncertainty principle. Moreover, quantum theory does not contain any element, that a priori corresponds to the physical reality of the spin directions of either of the two subsystems of the entangled state \eqref{eq:bell psi minus state}.\\

According to EPR, such a situation would be interpreted as an \emph{incompleteness} of quantum mechanics, since its intrinsic probabilistic character cannot explain the apparent randomness, with which a certain measurement result is realized in nature. Certainly quantum theory does not provide the corresponding description of the elements of reality mentioned in the above example. In \cite{EPR35} EPR thus conclude that ``\emph{...the wave function does not provide a complete description of physical reality,...}" although they admit they have ``\emph{...left open the question of wether or not such a description exists.}" (p.780). Dwelling on the possibility of more complete models, different interpretations of quantum theory, so called Hidden-Variable-Theories (HVT), can be constructed, although they underly severe constraints, such as the theorems by Bell \cite{bell64} or Kochen and Specker \cite{kochenspecker67}.\\

Certainly it is also possible to abandon or redefine the assumptions of \emph{locality} and/or \emph{realism} in order to avoid the paradox. Allowing for instantaneous interactions between distant subsystems is however starkly contrasting to the concept of causality and the theory of special relativity. Nonetheless, inherently non-local, deterministic hidden-variable models, have been proposed, e.g. by David Bohm, see \cite{hollandQTofM}.\\

The last option is to drop the assumption of realism (in the sense of EPR) or restrict it such that ``\emph{...two or more physical quantities can be regarded as simultaneous elements of reality only when they can be simultaneously measured or predicted.}" (\cite{EPR35}, p.780). This restriction would imply that the reality of physical properties of a system generally depends on the choice of measurement, performed on distant systems, which EPR reject by stating ``\emph{No reasonable definition of reality could be expected to permit this.}" (\cite{EPR35}, p.780).\\

This last possibility is strengthened by the observation first made by Kochen and Specker in \cite{kochenspecker67}, whose arguments were later on simplified by Peres \cite{peres91} and Mermin \cite{mermin90}, that it is in fact contradictory to assume that definitive measurement outcomes for all possible observables of a composite system could be pre-assigned (e.g. by hidden variables) to that system even for commuting observables. Subsequently it is not meaningful to picture physical properties of quantum systems as elements of reality, which are ``uncovered" by a measurement, but rather, that the hidden variable model must include the \emph{context} of the measurement.

    					\subsubsection{Bell's Theorem}\label{subsec:bell theorem}

The major breakthrough in the debate about the EPR - paradox and the viability of Hidden-Variable theories was accomplished by John S. Bell in 1964, see \cite{bell64}, with the formulation of the conflict in terms of an experimentally testable inequality, in which the assumptions made by EPR in \cite{EPR35} enter. The concept of such a \emph{Bell inequality} has been reformulated, extended and generalized in the subsequent years, at first to a more easily testable version, the \emph{CHSH - Inequality}, named after Clauser, Horne, Shimony, and Holt, see \cite{chsh69}, and later on to pure systems of higher dimensions, \cite{popescurohrlich1992}, mixed states, see \cite{werner89}, and \cite{popescu95}, as well as geometric interpretations in Hilbert space, \cite{bertlmann-narnhofer-thirring02}. We will in the following present the Bell inequality in the CHSH form, but we will use the term ``Bell inequality" synonymously for all different types of such expressions, while we refer to the original inequality, derived in \cite{bell64}, as Bell's inequality.\\

Regardless of which form of Bell inequality is used, the essence of Bell's proposal can be formulated in the following theorem.

\begin{theorem}\label{thm:bell theorem}
    \begin{tabbing} \hspace*{3.1cm}\=\hspace*{2cm}\=\kill
            \> In certain experiments all local realistic theories (LRT) are\\[1mm]
            \> incompatible with the predictions of quantum mechanics.\\
    \end{tabbing}
\end{theorem}

To formulate this mathematically, let us go back to the situation discussed in Sec.~\ref{subsec:EPR}, Alice and Bob performing spin measurements on the entangled 2-qubit state $|\,\psi^{\,-}\,\rangle$ in Eq.~\eqref{eq:bell psi minus state}. The spin
measurement along an arbitrary direction, represented by the unit vector $\vec{a}$, is described by the operator $\vec{a}\cdot\vec{\sigma}$, but let us stay more general for now and assume that both Alice and Bob measure some observable $A(\vec{a},\lambda)$ and $B(\vec{b},\lambda)$ respectively, where the possible outcomes both also depend on some internal hidden parameter $\lambda$. This comprises the reality assumption of EPR. It is not necessary to assume that $\lambda$ is a single parameter, but it can be an arbitrarily large set of numbers, or even functions, containing the information, which includes the possibility for the observables to depend on different such parameters.\\

The possible measurement results for Alice and Bob are then

\begin{equation}
A(\vec{a},\lambda)\,=\,\pm\,1,\,0\ ,\ \ B(\vec{b},\lambda)\,=\,\pm\,1,\,0\ \ \ ,
\label{eq:possible measurement results in bell inequality}
\end{equation}
where $\pm\,1$ represents $\up$ and $\down$ respectively, while $0$ corresponds to no detection. The locality assumption only requires, that neither result depends on the measuring direction of the other side,
\begin{equation}
A(\vec{a},\vec{b}\!\!\!\textbf{/},\lambda)\ ,\ \ B(\vec{a}\!\!\!\textbf{/},\vec{b},\lambda)\ \ .
\label{eq:locality assumption in bell inequality}
\end{equation}
Furthermore, there must be some normalized distribution function $\rho(\lambda)$,
\begin{equation}
\int\!d\lambda\,\rho(\lambda)\,=\,1\ \ \ ,
\label{eq:normalization of distribution function}
\end{equation}
determining the outcome for a given $\lambda$. The expectation value of a combined measurement
on both Alice's and Bob's side then is just
\begin{equation}
E(\vec{a},\,\vec{b})\,=\,\int\!d\lambda\,\rho(\lambda)\,A(\vec{a},\,\lambda)\,B(\vec{b},\,\lambda)\ \ .
\label{eq:bell expvalue}
\end{equation}

Let us then consider a certain combination of such expectation values for measurements in different directions, but in the same state.

\begin{eqnarray}
E(\,\vec{a},\vec{b}\,)\,-\,E(\,\vec{a},\vec{b}^{\,\prime}\,) &=& \int\!d\lambda\,\rho(\lambda)\,
		\left(A(\vec{a},\lambda)\,B(\vec{b},\lambda)\,-\,
						A(\vec{a},\lambda)\,B(\vec{b}^{\,\prime},\lambda)\right)\,= \nonumber \\
						[2mm]
&=&	\int\!d\lambda\,\rho(\lambda)\,\underbrace{A(\vec{a},\lambda)\,B(\vec{b},\lambda)}
		_{|\ \,|\,\leq\,1}	
		\left(1-A(\vec{a}^{\,\prime},\lambda)\,B(\vec{b}^{\,\prime},\lambda)\right)\,-
		\nonumber \\
&-&	\int\!d\lambda\,\rho(\lambda)\,\underbrace{A(\vec{a},\lambda)\,B(\vec{b}^{\,\prime},\lambda)}
		_{|\ \,|\,\leq\,1}	
		\left(1-A(\vec{a}^{\,\prime},\lambda)\,B(\vec{b},\lambda)\right)\nonumber \\
& & \label{eq:chsh derivation 1}		
\end{eqnarray}
where we just added and subtracted the same term to rewrite the equation. Noting that the
moduli of the products of $A$ and $B$ on the r.h.s. must be less or equal than $1$, we can make the estimation,

\begin{eqnarray}
\left|\,E(\,\vec{a},\vec{b}\,)\,-\,E(\,\vec{a},\vec{b}^{\,\prime}\,)\,\right| &\leq & 	
		\left|\,\int\!d\lambda\,\rho(\lambda)\,\left(
		1-A(\vec{a}^{\,\prime},\lambda)\,B(\vec{b}^{\,\prime},\lambda)\right)\,\right|\,+
		\nonumber \\
		[2mm]
&+&	\left|\,\int\!d\lambda\,\rho(\lambda)\,\left(
		1-A(\vec{a}^{\,\prime},\lambda)\,B(\vec{b},\lambda)\right)\,\right|\ \ \ .
		\label{eq:chsh derivation 2}		
\end{eqnarray}

Using Eqs.~\eqref{eq:normalization of distribution function} and \eqref{eq:bell expvalue} we get
\begin{equation}
\left|\,E(\vec{a},\vec{b})\,-\,E(\vec{a},\vec{b}^{\,\prime})\,\right|\,\leq\,
2\,-\,\left|\,E(\vec{a}^{\,\prime},\vec{b}^{\,\prime})\,+\,E(\vec{a}^{\,\prime},\vec{b})\,\right|\ \ .
\label{eq:chsh derivation 3}
\end{equation}

Introducing the so called \emph{Bell-parameter} $S(\vec{a},\,\vec{a}^{\,\prime},\,\vec{b},\,\vec{b}^{\,\prime})$,
\begin{equation}
S(\vec{a},\,\vec{a}^{\,\prime},\,\vec{b},\,\vec{b}^{\,\prime})\,:=\,
E(\vec{a},\vec{b})\,-\,E(\vec{a},\vec{b}^{\,\prime})\,+\,
E(\vec{a}^{\,\prime},\vec{b}^{\,\prime})\,+\,E(\vec{a}^{\,\prime},\vec{b})
\label{eq:bell parameter}
\end{equation}
we can rewrite Eq.~\eqref{eq:chsh derivation 3} using the triangle inequality as
\begin{equation}
|\,S(\vec{a},\,\vec{a}^{\,\prime},\,\vec{b},\,\vec{b}^{\,\prime})\,|\,\leq\,2\ \ \ ,
\label{eq:chsh inequality}
\end{equation}
which is the desired \emph{Bell inequality}, known as the \emph{CHSH inequality}. Every local realistic theory in the sense of EPR, has to satisfy this inequality. Let us now check wether or not quantum mechanics satisfies the inequality. To this end we need the quantum mechanical expectation value $E(\vec{a},\vec{b})$ for the state \eqref{eq:bell psi minus state}, which is
\begin{equation}
E(\vec{a},\vec{b})\,=\, \left\langle\,\psi^{-}\,\right|\,\vec{a}\cdot\vec{\sigma}\,\otimes\,
\vec{b}\cdot\vec{\sigma}\,\left|\,\psi^{-}\,\right\rangle\,=\,-\,\vec{a}\,\vec{b}\,=
\,-\cos(\theta(a,b))\ \ ,
\label{eq:QM bell exp value calculation}
\end{equation}
where $\theta(a,b)$ is the angle between the unit vectors $\vec{a}$ and $\vec{b}$. Choosing these four directions to lie in the same plane in steps of $45$°, such that
\begin{equation}
\theta(a-b)\,=\,\theta(a^{\,\prime}-b^{\,\prime})\,=\,\theta(a^{\,\prime}-b)\,=
\,\frac{\pi}{4}\ ,\ \ \theta(a-b^{\,\prime})\,=\,\frac{3\pi}{4}\ \ ,
\label{eq:bell angles}
\end{equation}
the Bell parameter as predicted by quantum mechanics yields
\begin{equation}
|S_{\rm{QM}}|\,=\,|-\,\frac{\sqrt{2}}{2}\,-\,\frac{\sqrt{2}}{2}\,
-\,\frac{\sqrt{2}}{2}\,-\,\frac{\sqrt{2}}{2}|\,=\,2\sqrt{2}\,>\,2\ \ \ ,
\label{eq:QM bell parameter}
\end{equation}
in clear violation of the CHSH inequality \eqref{eq:chsh inequality}. It must therefore be concluded, that
a local realistic theory can never fully reproduce the predictions of quantum mechanics, which proves Theorem \ref{thm:bell theorem}.\qed

The Bell inequality palpably separates possible local realistic models (i.e. HVT) from quantum mechanics in a way, that can be distinguished in experiment. The overwhelming majority of experiments so far performed, e.g. by Weihs, Jennewein, Simon, Weinfurter, and Zeilinger \cite{weihsetal98}, violated Bell inequalities and thus suggested, that local realistic models cannot account for all phenomena in nature. The experimental verification was first achieved by Freedman and Clauser \cite{freedmanclauser72} in 1972, and demonstrated under different (more restrictive) circumstances and for different variants of Bell inequalities in 1981 and 1982 by Aspect, Grangier, Roger, and Dalibard, see \cite{aspectetal81},\cite{aspectetal82a}, and \cite{aspectetal82b}.\\

It is also possible to reformulate the Bell parameter \eqref{eq:bell parameter} as the expectation value of the \emph{Bell observable} $B$, which can be written as
\begin{equation}
B\,=\,\vec{a}\cdot\vec{\sigma}\,\otimes\,(\vec{b}\,+\,\vec{b}^{\,\prime})\cdot\vec{\sigma}\,
+\,\,\vec{a}^{\,\prime}\cdot\vec{\sigma}\,\otimes\,(\vec{b}\,-\,\vec{b}^{\,\prime})\cdot\vec{\sigma}\ \ .
\label{eq:bell observable}
\end{equation}
This observable can then be used (\cite{terhal00},\cite{horodecki-rpm95}) to construct an entanglement witness to detect entanglement such as in Theorem \ref{thm:entanglement witness theorem}.
\begin{theorem}\label{thm:chsh criterion}
    \begin{tabbing} \hspace*{3.1cm}\=\hspace*{3cm}\=\kill
            \> A bipartite qubit state $\rho$ violates the CHSH inequality\\[2mm]
            \> \> $\langle\,\rho,\,2\mathds{1}\,-\,B\,\rangle\,\geq\,0$ , \\[2mm]
            \> where $B$ is given by \eqref{eq:bell observable}, only if it is entangled.\\
    \end{tabbing}
\end{theorem}
As with the entanglement witness theorem before, it might not be easy to find the exact witness, which in this case means finding appropriate directions $\vec{a},\,\vec{a}^{\,\prime},\,\vec{b},\,\vec{b}^{\,\prime}$. This problem can be circumvented by a theorem, introduced in \cite{horodecki-rpm95}.
\begin{theorem}\label{thm:chsh criterion 2}
    \begin{tabbing} \hspace*{3.1cm}\=\hspace*{3cm}\=\kill
            \> A $2$-qubit state $\rho$ violates the CHSH inequality for some $B$\\[2mm]
            \> if, and only if, $\ M(\rho)\,>\,1$ \ , where $M(\rho)\,=\,\lambda_{\mathrm{max1}}+\lambda_{\mathrm{max2}}$\\[2mm]
            \> is the sum of the largest two eigenvalues of $\ U_{\rho}=(T_{\rho})^{\mathrm{T}}T_{\rho}$\\[2mm]
            \> and $T_{\rho}$ is the matrix with components $(T_{\rho})_{\mathrm{ij}}=Tr\,\rho\,\sigma_{\mathrm{j}}\otimes\sigma_{\mathrm{j}}$\ .
    \end{tabbing}
\end{theorem}
Here $\sigma_{\mathrm{i}}$ are the usual Pauli matrices \eqref{eq:pauli matrices}. It is thus very simple to find out wether a given state violates the CHCH-inequality, although one does not automatically know the corresponding measurement directions.\\

Often states violating some Bell inequality are called nonlocal. Although the term ``nonlocality" is often used in discussions about entanglement, neither Bell's, nor any other theorem, or experiments thereon, suggest, that the quantum correlations of entanglement need to involve any super-luminal signalling. Despite this fact, the terms entanglement and nonlocality are frequently used in the same context. We have already stated in Sec.~\ref{subsec:entanglement measures for mixed states} that there are stronger entanglement criteria than the violation of the CHSH-inequality, which leaves the question wether or not entanglement and nonlocality are essentially the same thing or not, i.e. if in principle no entangled state can be explained in terms of a local realistic model, unresolved. We do want to note here however, that this might be due to the (projective) measurements used in the CHSH-inequality, and that generalized Bell inequalities, using positive operator-valued measurements (POVM), can be constructed, see Ref.~\cite{popescu95}.\\

Having quantified the entanglement of quantum states and the apparent connection to the Bell Theorem, we can interpret this in such a way, that although local realistic models might be applicable to some situations, clearly, they cannot provide a more complete understanding of nature than quantum mechanics. The non-classical correlations found in entanglement seem to be the crossroads, where one of the two, or both, assumptions, locality and realism, must be abandoned, when constructing a physical model.

\addtocontents{toc}{\vspace{-1ex}}
\newpage\section{Relativistic Description of Physical \mbox{Systems}}\label{chap:relativistic quantum}

We have so far studied the structure of (non-relativistic) quantum mechanics and (some of) the representations of the 3-dimensional rotation group $SO(3)$ therein, e.g. the spinors introduced in Eq.~\eqref{eq:two component spinors} already are a unitary irreducible representation (see Def.~\ref{def:unitary representation},\ref{def:irreducible rep}) of the rotation group, fully characterized by the eigenvalue\footnote{We have set $\hbar=c=1$ .} $s(s+1)$ of $\vec{S}^{\,2}$ in Eq.~\eqref{eq:spin squared eigenvalue equation}. The rotation group is not the only symmetry group of physical relevance, in fact, there are two other groups, the (proper, orthochronous) Lorentz group $\mathcal{L}_{+}^{\up}$, containing $SO(3)$ as a subgroup, and the (proper, orthochronous) Poincar$\acute{e}$ group $\mathcal{P}_{+}^{\up}$, of which both $\mathcal{L}_{+}^{\up}$ and $SO(3)$ are subgroups, whose representations are fundamentally important.\\

This chapter is structured as follows: In Sec.~\ref{sec:special relativity} the main ideas of (the classical theory of) special relativity are briefly introduced, first in terms of vector space geometry related to physical observers in spacetime in Sec.~\ref{subsec:minkowski and lorentz trafo}, then followed by a more abstract formulation in the language of group theory in Sec.~\ref{subsec:lorentz and poincare group} and Sec.~\ref{subsec:lorentz group and SL(2,c)}. In the second part, Sec.~\ref{sec:Relativistic Description of Quantum Systems}, of this chapter, we will discuss how quantum objects, especially spin-$\tfrac{1}{2}$ particles, can be described relativistically as (irreducible) representations of the aforementioned groups $\mathcal{L}_{+}^{\up}$ (Sec.~\ref{subsec:reps of lorentz group}), and $\mathcal{P}_{+}^{\up}$ (Sec.~\ref{subsec:reps of poincare group}). Finally we will see how unitarity of these representations arises in Sec.~\ref{subsec:wigners little group}.

	\subsection{Special Relativity}\label{sec:special relativity}
	
The theory of special relativity is a well understood, and satisfyingly verified physical theory. We do not aim here to present it in its full extent, but we do want to establish some basic notions, which are frequently used throughout this work. A review on special relativity, including relativistic mechanics and especially focusing on its application to quantum theory and particle physics, can be found in Ref.~\cite{sexlurbantke}, while an account from the perspective of general relativity is given in Ref.~\cite{waldGR}.

		\subsubsection{Minkowski Spacetime \& Lorentz Transformations}\label{subsec:minkowski and lorentz trafo}

At the foundation of relativity lies a seemingly simple observation: Their seems to be no preferred inertial frame of reference in nature. We will take this observation, called the (special) \emph{principle of relativity}, to be the basic postulate of the theory.

\begin{center}\parbox{0.80\textwidth}{\textbf{\emph{Principle of Relativity:}} The laws of nature, and consequently their mathematical formulation, are invariant under a change of the inertial reference frame, i.e. they assume their usual form for every inertial observer.}\end{center}\ \\

To formulate this statement in terms of a mathematical model, we need to define the space, or in this case spacetime, which serves as a ``playground", which is the \emph{Minkowski spacetime}.

\begin{defi}\label{def:minkowski spacetime}
	\begin{tabbing} \hspace*{3.3cm}\=\hspace*{1.5cm}\=\kill
            \> The \textbf{\emph{Minkowski spacetime}} is a four dimensional affine\\[2mm]
            \> space $\mathbb{A}^{4}$, over a (real, Lorentzian) vector space, equipped\\[2mm]
            \> with a non-degenerate, symmetric bilinear form $\eta(\,.\,,\,.\,)$,\\[2mm]
            \> called \emph{\textbf{Minkowski metric}}, of signature $(+\,-\,-\,-)$.\\
	\end{tabbing}
\end{defi}
The principle of relativity can then be stated as follows:

\begin{defi}\label{def:inertial frames}
	\begin{tabbing} \hspace*{3.3cm}\=\hspace*{1.5cm}\=\kill
            \> A coordinate system of Minkowski spacetime, for which\\[2mm]
            \> the physical laws take on their usual form, is called an\\[2mm]
            \> \emph{\textbf{inertial frame}} (of reference).\\
	\end{tabbing}
\end{defi}
An important fact of special relativity is the existence of an invariant velocity, the \emph{speed of light} $c$, which has the same value for all observers. We choose units, such that its value is $c=1$. In the concept of (Minkowski) spacetime, space and time are now treated on equal footing and we therefore use coordinates $\{x^{\,\mu}\}=\{x^{0},x^{\mathrm{i}}\}$, where Greek indices run from $0$ to $3$, with $x^{0}$ being the time coordinate, and $\vec{x}=(x^{1},x^{2},x^{3})^{\mathrm{T}}$ is the 3-vector of spatial coordinates $\{x^{\mathrm{i}}\}$, for which we will use Latin indices going from $1$ to $3$. Together we have
\begin{equation}
x^{\,\mu}\,=\,\begin{pmatrix} x^{0} \\ x^{1} \\ x^{2} \\ x^{3} \end{pmatrix}\,=\,
\begin{pmatrix} \,x^{0} \\ \!\vec{x} \end{pmatrix} \ .
\label{eq:coordinates in minkowski space}
\end{equation}
Points $q$ in the Minkowski spacetime\footnote{We will from now on refer to the construction of Minkowski spacetime as ``spacetime", or ``Minkowski space".} will be called \emph{events}, which are formally represented by coordinates $\{x^{\,\mu}\}$ with respect to an inertial frame in the vector space $(\mathbb{R}^{4},\eta(\,.\,,\,.\,))$.\\

The question remains how we are to describe certain inertial observers in Minkowski space. This is done by choosing an event \textit{o} in $\mathbb{A}^{4}$ as origin\footnote{With a choice of coordinate system in spacetime a choice of origin in the affine space will be implied as well from now on.} and an orthonormal basis $\{e_{0},e_{1},e_{2},e_{3}\}$ of the vector space, where $\eta(e_{0},e_{0})=1$ and $\eta(e_{\mathrm{i}},e_{\mathrm{j}})=-\delta_{\mathrm{ij}}$. The timelike basis vector $e_{0}$ is clearly singled out, its orthogonal complement, i.e. the space spanned by $\{e_{1},e_{2},e_{3}\}$, is the \mbox{(hyper-)} plane of simultaneity of this particular inertial observer. The basis vector $e_{0}$ therefore describes an inertial observer up to the choice of a spatial basis $\{e_{1},e_{2},e_{3}\}$, which corresponds to the freedom of rotations in $\mathbb{R}^{3}$. The timelike basis vector is then given a new symbol\footnote{We will switch between notations, sometimes using the symbol $x^{\,\mu}$ for the abstract vector $x$, and sometimes for the $\mu$-th component of the vector $x$. This is however always clear from the context.}, e.g. $v$, and called the \emph{4-velocity} of the observer $X$ with components, with respect to any observer $X^{\,\prime}$ with 4-velocity $w$, given by
\begin{equation}
v\,=\,\gamma(\vec{v}\,)\,\begin{pmatrix} 1 \\ \vec{v} \end{pmatrix}\ ,\ \ \mbox{where}\ \ \gamma(\vec{v})\,=\,\eta(v,w)
\label{eq:four velocity}
\end{equation}
and $\vec{v}$, where $|\vec{v}|<1$, is the 3-velocity of $X$ with respect to $X^{\,\prime}$. The \mbox{\emph{4-momentum}} $p$ (of a particle) is then constructed by multiplying with its mass $m$,
\begin{equation}
p\,=\,m\gamma(\vec{v}\,)\,\begin{pmatrix} 1 \\ \vec{v} \end{pmatrix}\,=\,
\begin{pmatrix} E \\ \vec{p} \end{pmatrix}\ \ ,
\label{eq:four momentum}
\end{equation}
where $p^{2}=m^{2}$ and the energy of the particle for the observer $X^{\,\prime}$ is given by
\begin{equation}
E\,=\,\sqrt{m^{2}\,+\,\vec{p}^{\:2}}\ .
\label{eq:relativistic energy}
\end{equation}

From the invariance of the speed of light follows directly the invariance of the line element $ds^{2}$, given by
\begin{equation}
ds^{2}\,=\,\eta_{\,\mu\nu}\,dx^{\,\mu}dx^{\,\nu}\,=\,(dx^{0})^{2}\,-\,(d\vec{x}\,)^{2}\ \ ,
\label{eq:minkowski line element}
\end{equation}
which then can be used to classify vectors $w$ in spacetime, also called $4$-vectors, or pairs of two events $p$ and $q$ connected by those $4$-vectors, $\overline{pq}=w$, by their norm $w^{2}=\eta(w,w)=w_{\mu}w^{\,\mu}$, as being
\begin{eqnarray}
w^{2}\,>\,0 &\leftrightarrow& \mbox{timelike (separated)}, \label{timelike}\\
w^{2}\,=\,0 &\leftrightarrow& \mbox{null (separated)},\ \mbox{or}\label{null}\\
w^{2}\,<\,0 &\leftrightarrow& \mbox{spacelike (separated)}.\label{spacelike}
\end{eqnarray}

\begin{defi}\label{def:lorentz and poincare trafo}
	\begin{tabbing} \hspace*{3.3cm}\=\hspace*{1.5cm}\=\kill
            \> Coordinate transformations in Minkowski space, which\\[2mm]
            \> leave the laws of physics unchanged, i.e. transformations\\[2mm]
            \> between inertial frames, are called \emph{\textbf{Lorentz}}\\[2mm]
            \> \textbf{\emph{transformations}}, if they leave the origin fixed,\\[2mm]
            \> or \textbf{\emph{Poincar$\mathbf{\acute{e}}$ transformations}} otherwise.\\
	\end{tabbing}
\end{defi}
 Pure (spatial) rotations $R$ are obviously possible Lorentz transformations, which then are of the form
\begin{equation}
R\,=\,\begin{pmatrix}\,1\, & \\
 &\!\framebox[20mm][c]{
 $\begin{matrix}  &  & \\
                 & R(\vec{\varphi}\,) & \\
                 &  & \\ \end{matrix}$}  \\
\end{pmatrix}
\label{eq:rotations in minkowski space}
\end{equation}
where $R(\vec{\varphi}\,)\in\,SO(3)$ is a three dimensional rotation matrix, with components given by \eqref{eq:rotation in R3}. Other important Lorentz transformations are the so called \emph{Lorentz boosts}, which are the transformations relating two inertial frames, moving with constant velocity respect to each other. Assume an observer at rest in the inertial frame $S$ with coordinates $\{x^{\,\mu}\}$, and a second observer, at rest in a frame $S^{\,\prime}$ with coordinates $\{x^{\,\mu\,\prime}\}$, which is moving in the $x^{1}$ direction with velocity $\vec{v}=(v_{\mathrm{x}},0,0)^{\mathrm{T}}$ with respect to $S$. The transformation taking the coordinates $\{x^{\,\mu}\}$ to $\{x^{\,\mu\,\prime}\}$, which we will denote by $L_{\mathrm{x}}(\vec{v}\,)$, is given by
\begin{equation}
L_{\mathrm{x}}(\vec{v}\,)\,=\,\begin{pmatrix}   \!\,\gamma      &   \!\!\!-\gamma v_{\mathrm{x}}   &   0\    &   0\,\\
                        \!-\gamma v_{\mathrm{x}}   &   \!\!\!\,\gamma      &   0\    &   0\,\\
                        \!\,0           &   \!\!\!\,0           &   1\    &   0\,\\
                        \!\,0           &   \!\!\!\,0           &   0\    &   1\,\\
\end{pmatrix}\ ,
\label{eq:boost in x direction}
\end{equation}
where
\begin{equation}
\gamma\,=\,\gamma(\vec{v}\,)\,=\,\frac{1}{\sqrt{1-\vec{v}^{\,2}}}
\label{eq:gamma factor}
\end{equation}
is the so called $\gamma$ - factor. The components of a matrix representing a Lorentz boost $L(\vec{v}\,)$ in a general direction $\tfrac{\vec{v}}{|\vec{v}|}$ with velocity $\vec{v}$ are given by
\begin{eqnarray}
L(\vec{v}\,)^{\,\mathrm{i}}_{\ \ \mathrm{j}} &=& \delta^{\,\mathrm{i}}_{\ \ \mathrm{j}}\,+\,(\gamma(\vec{v})\,-\,1)\,\frac{v^{\,\mathrm{i}}v_{\,\mathrm{j}}}{|\vec{v}|^{\,2}}\nonumber\\
L(\vec{v}\,)^{\,\mathrm{i}}_{\ \ \mathrm{0}} &=& L(\vec{v}\,)^{\,\mathrm{0}}_{\ \ \mathrm{i}}\,=\,-\,\gamma(\vec{v})\,v^{\,\mathrm{i}}\label{eq:general lorentz boost}\\[3mm]
L(\vec{v}\,)^{\,\mathrm{0}}_{\ \ \mathrm{0}} &=& \gamma(\vec{v})\ \ .\nonumber
\end{eqnarray}
Instead of expressing the Lorentz boost using the velocity, one can also choose to express it in terms of the \emph{rapidity} $u$, which in contrast to the velocity is an additive quantity with respect to the composition of several boosts, and which is defined by the relations
\begin{equation}
\cosh\,u\,=\,\gamma(\vec{v}\,)\ ,\ \ \tanh\,u\,=\,|\vec{v}\,|\ ,\ \ \sinh\,u\,=\,\gamma(\vec{v}\,)\,|\vec{v}\,|\ .
\label{eq:rapidity}
\end{equation}
Also every Lorentz boost can be expressed by a combination of two rotations and a boost in a chosen direction, e.g. the $x$-direction, as
\begin{equation}
L(\vec{v}\,)\,=\,R_{\mathrm{x}}(\vec{v}\,)\,L_{\mathrm{x}}(|\vec{v}\,|)\,R_{\mathrm{x}}^{-1}(\vec{v}\,)\ \ ,
\label{eq:general boost as rotations and boost}
\end{equation}
where $R_{\mathrm{x}}(\vec{v}\,)$ is a rotation\footnote{The argument $\vec{v}$ does not imply, that $\vec{v}$ is the axis of rotation as in Eq.~\eqref{eq:rotation in R3}, which we have indicated also by using an index $x$ on $R_{\mathrm{x}}$ to distinguish it from $R(\vec{\varphi}\,)$.} taking the $x$-axis to the direction of $\vec{v}$ and $L_{\mathrm{x}}(|\vec{v}\,|)$ is given by Eq.~\eqref{eq:boost in x direction}. Generally a Lorentz transformation will be a combination of rotations \eqref{eq:rotations in minkowski space} and general boosts \eqref{eq:general lorentz boost} with components
\begin{eqnarray}
L(\vec{v},\vec{\varphi}\,)^{\,\mathrm{i}}_{\ \ \mathrm{j}} &=& R(\vec{\varphi}\,)^{\,\mathrm{i}}_{\ \ \mathrm{j}}\,+\,(\gamma(\vec{v})\,-\,1)\,\frac{(R(\vec{\varphi\,})v)^{\,\mathrm{i}}v_{\,\mathrm{j}}}{|\vec{v}|^{\,2}}\nonumber\\
L(\vec{v},\vec{\varphi}\,)^{\,\mathrm{i}}_{\ \ \mathrm{0}} &=& -\,\gamma(\vec{v})\,(R(\vec{\varphi\,})v)^{\,\mathrm{i}}\ ,\ \ L(\vec{v},\vec{\varphi}\,)^{\,\mathrm{0}}_{\ \ \mathrm{i}}\,=\,-\,\gamma(\vec{v})\,v^{\,\mathrm{i}}\label{eq:general lorentz trafo}\\[3mm]
L(\vec{v},\vec{\varphi}\,)^{\,\mathrm{0}}_{\ \ \mathrm{0}} &=& \gamma(\vec{v})\ \ .\nonumber
\end{eqnarray}

A \emph{Poincar$\acute{e}$} transformation $T(L,a)$ then combines this transformation with a (spacetime) \emph{translation},  $x^{\,\mu}\rightarrow x^{\,\mu}+a^{\,\mu}$, where $a^{\,\mu}$ is some constant vector in Minkowski space.
\begin{equation}
T(L,a)^{\,\mu}_{\ \ \nu}\,x^{\,\nu}\,=\,L^{\,\mu}_{\ \ \nu}\,x^{\,\nu}\,+\,a^{\mu}
\label{eq:poincare trafo}
\end{equation}

		\subsubsection{Lorentz - \& Poincar$\mathbf{\acute{e}}$ - group}\label{subsec:lorentz and poincare group}

It can be checked that the transformations listed above, rotations \eqref{eq:rotations in minkowski space}, boosts \eqref{eq:general lorentz boost}, and combinations thereof \eqref{eq:general lorentz trafo}, leave the line element \eqref{eq:minkowski line element} and thus the Minkowski metric $\eta$ invariant. This statement can be formulated in a coordinate independent way as
\begin{equation}
L^{T}\eta\,L\,=\,\eta\ \ ,
\label{eq:lorentz invariance of metric}
\end{equation}
i.e. the Lorentz transformations $L$ are completely characterized by this property, which lets us define the Lorentz group $\mathcal{L}$,
\begin{equation}
\mathcal{L}\,:=\,\left\{\,L\,\in\,GL(4,\mathbb{R})\,|\,L^{T}\eta\,L\,=\,\eta\,\right\}\ \ .
\label{eq:lorentz group}
\end{equation}
The group properties, see Def.~\ref{def:group}, are easily checked. The succession of two Lorentz transformations is again a Lorentz transformation,
\begin{equation}
(L\,L^{\prime})^{T}\eta\,L\,L^{\prime}\,=\,L^{\prime\,T}L^{T}\eta\,L\,L^{\prime}\,=\,L^{\prime\,T}\eta\,L^{\prime}\,=\,\eta\ \ ,
\label{eq:lorentz group closed}
\end{equation}
the group multiplication is associative, which is always the case for matrix groups, the identity element is $\mathds{1}_{4}$, trivially satisfying \eqref{eq:lorentz invariance of metric}, and to every element $L\in\mathcal{L}$ exists an inverse element $L^{-1}\in\mathcal{L}$, which we see by acting on \eqref{eq:lorentz invariance of metric} with $(L^{T})^{-1}$ from the left and $L^{-1}$ from the right.
\begin{eqnarray}
(L^{T})^{-1}L^{T}\eta\,L\,L^{-1} &=& (L^{T})^{-1}\eta\,L^{-1} \nonumber \\[2mm]
\eta &=& (L^{-1})^{T}\eta\,L^{-1}
\label{eq:inverse element of lorentz group}
\end{eqnarray}
Although the elements of $\mathcal{L}$ are in general neither orthogonal (only for pure rotations) or symmetric (only for pure boosts), we can further classify the elements of the Lorentz group by two conditions, the sign of $L^{\,0}_{\ \ 0}$ and their determinant, by noting that
\begin{prop}\label{prop:sign and det of lorentz trafo}
	\begin{tabbing} \hspace*{3.7cm}\=\hspace*{1.5cm}\=\kill
            \> Every element of the Lorentz group \eqref{eq:lorentz group} satisfies\\[2mm]
            \> \> $(L^{\,0}_{\ \ 0})^{2}\,\geq\,1$ and $(\det\,L)^{2}\,=\,1$ .\\
	\end{tabbing}
\end{prop}
\pa{Proof:}\ \\
First writing \eqref{eq:lorentz invariance of metric} in components we get
\begin{equation}
L^{\,\mu}_{\ \ \nu}\,\eta_{\,\mu\rho}\,L^{\,\rho}_{\ \ \sigma}\,=\,\eta_{\,\nu\sigma}\ \ ,
\label{eq:lorentz invariance of metric in components}
\end{equation}
of which we express the component $\eta_{\,00}=1$ of the r.h.s.
\begin{equation}
L^{\,\mu}_{\ \ 0}\,\eta_{\,\mu\rho}\,L^{\,\rho}_{\ \ 0}\,=\,
(L^{\,0}_{\ \ 0})^{2}\,-\,\sum\limits_{\mathrm{i}}\,(L^{\,\mathrm{i}}_{\ \ 0})^{2}\,=\,\eta_{\,00}\,=\,1\ \ ,
\label{eq:proof of L zerozero squared geq one}
\end{equation}
which proves the first part of Prop.~\ref{prop:sign and det of lorentz trafo}, since $(L^{\,\mathrm{i}}_{\ \ 0})^{2}\geq0\ \forall\,i$. The second part is easily shown by taking the determinant of \eqref{eq:lorentz invariance of metric} and noting that the determinant of a transposed matrix is equal to the determinant of the matrix itself.
\begin{equation}
\det(L^{T}\eta\,L)\,=\,(\det\,L)^{2}\,\det\eta\,=\,\det\eta
\label{eq:proof of det L squared is one}
\end{equation}
\ \qed

Gathering what we know about the structure of the Lorentz group, we find that it is a subgroup of $GL(4,\mathbb{R})$, which makes it a Lie group. Since a Lie group is a differentiable manifold and we can describe $\mathcal{L}$ by $6$ independent parameters, encoded in $\vec{v}$ and $\vec{\varphi}$ of \eqref{eq:general lorentz trafo}, we conclude that the Lorentz group forms a 6 - dimensional submanifold of $\mathbb{R}^{16}$. This can also be seen from \eqref{eq:lorentz invariance of metric in components}, which comprises 10 (pseudo-) orthogonality relations for the 16 components of $L^{\,\mu}_{\ \ \nu}$. Geometrically this can be interpreted as the equations for 10 hypersurfaces in a 16-dimensional Euclidean space\footnote{The formalism of differential geometry allows one to formulate such properties without referring to the $\mathbb{R}^{16}$ or any other embedding structure.}. The resulting structure is neither connected, $\mathcal{L}$ decomposes into $4$ disconnected, but internally connected, pieces, determined by the sign of the zero-zero-component, $L^{\,0}_{\ \ 0}\geq1$ or $L^{\,0}_{\ \ 0}\leq-1$, and their determinant, $\det\,L=1$ or $\det\,L=-1$, nor compact, since the parameterspace is not closed, $0\leq|\vec{v}\,|<1$.\\

Let us start with the connected component of the Lorentz group $\mathcal{L}$, for which \mbox{$L^{\,0}_{\ \ 0}\geq1$,} this class of Lorentz transformations certainly forms a subgroup of $\mathcal{L}$, called the \emph{orthochronous Lorentz group} $\mathcal{L}^{\up}$, and it contains all the Lorentz transformations, which leave the orientation of time invariant. Combining the transformations of $\mathcal{L}^{\up}$ with the time reversal operation $T$, given by the matrix $T=$ diag$(-1,+1,+1,+1)$, one obtains the complementary set $T\,\mathcal{L}^{\up}$, which has an empty intersection with $\mathcal{L}^{\up}$. The union of both sets is again the Lorentz group, $\mathcal{L}^{\up}\,\cup\,T\,\mathcal{L}^{\up}=\mathcal{L}$.\\

Similarly, we can consider the set of Lorentz transformations, which have $\det\,L=+1$, called the \emph{proper Lorentz group} $\mathcal{L}_{+}$ and its intersection with $\mathcal{L}^{\up}$, the \emph{proper, orthochronous Lorentz group} $\mathcal{L}^{\up}_{+}$. The full Lorentz group can then be written as the union of $\mathcal{L}^{\up}_{+}$ with the sets obtained by applying the parity operation $P$, time reversal $T$ and combinations $P\,T$ thereof on it,
\begin{equation}
\mathcal{L}\,=\,\mathcal{L}^{\up}_{+}\,\cup\,T\,\mathcal{L}^{\up}_{+}\,\cup\,P\,\mathcal{L}^{\up}_{+}\,\cup\,P\,T\,\mathcal{L}^{\up}_{+}\ \ ,
\label{eq:decomposition of Lorentz group}
\end{equation}
where the parity operator is given by $P=$ diag$(+1,-1,-1,-1)$. The proper, orthochronous Lorentz group
\begin{equation}
\mathcal{L}^{\up}_{+}\,=\,\left\{\,L\,\in\,\mathcal{L}\,|\,\det\,L=+1\,,\,L^{\,0}_{\ \ 0}\geq1\,\right\}
\label{eq:proper orthochronous lorentz group}
\end{equation}
will be of particular interest here, since it is the subset of $\mathcal{L}$ containing the identity $\mathds{1}_{\!4}$ and all spatial rotations $R\in SO(3)$, which in turn form a subgroup of $\mathcal{L}$ and $\mathcal{L}^{\up}_{+}$.\\

An important fact is, that while the rotations form a subgroup of the (proper, orthochronous) Lorentz group, pure boosts do not. This can be verified by multiplying two boosts, for simplicity consider pure boosts in the $x$-direction \eqref{eq:boost in x direction} and $z$-direction with velocities $|\vec{v}\,|=v_{\mathrm{x}}$ and $|\vec{w}\,|=w_{\mathrm{z}}$ respectively,
\begin{equation}
L_{\mathrm{x}}(|\vec{v}\,|)\,=\,
\begin{pmatrix}   \!\,\gamma_{\mathrm{v}}      &   \!\!\!-\gamma_{\mathrm{v}} v_{\mathrm{x}}   &   0\    &   0\,\\
                        \!-\gamma_{\mathrm{v}} v_{\mathrm{x}}   &   \!\!\!\,\gamma_{\mathrm{v}}      &   0\    &   0\,\\
                        \!\,0           &   \!\!\!\,0           &   1\    &   0\,\\
                        \!\,0           &   \!\!\!\,0           &   0\    &   1\,\\
\end{pmatrix}
\ ,\ L_{\mathrm{z}}(|\vec{w}\,|)\,=\,
\begin{pmatrix}   \!\,\gamma_{\mathrm{w}}      &   0\,   &   0\    &    \!\!\!-\gamma_{\mathrm{w}} w_{\mathrm{z}} \\
                        \!\,0   &    1\,     &   0\    &   0\,\\
                        \!\,0           &   \!\!\!\,0           &   1\    &   0\,\\
                        \!-\gamma_{\mathrm{w}} w_{\mathrm{z}}           &   \!\!\!\,0           &   0\    &   \!\!\!\,\gamma_{\mathrm{w}}\\
\end{pmatrix}.
\label{eq:boosts in x and z direction}
\end{equation}
Their product is
\begin{equation}
L_{\mathrm{z}}(|\vec{w}\,|)\,L_{\mathrm{x}}(|\vec{v}\,|)\,=\,
\begin{pmatrix}
\!\,\gamma_{\mathrm{u}} & \!-\gamma_{\mathrm{u}} v_{\mathrm{x}} & 0\ & \!\!\!-\gamma_{\mathrm{w}} w_{\mathrm{z}}\\
\!\!\!-\gamma_{\mathrm{v}} v_{\mathrm{x}} & \!\!\!\,\gamma_{\mathrm{v}} & 0\ & 0\\
                    \!\,0           &   \!\!\!\,0           &   1\    &   0\,\\
 \!-\gamma_{\mathrm{u}} w_{\mathrm{z}} & \gamma_{\mathrm{u}}v_{\mathrm{x}} w_{\mathrm{z}} & 0\ &   \!\!\!\,\gamma_{\mathrm{w}}
\end{pmatrix}\ ,
\label{eq:product of boost in x and z directions}
\end{equation}
where in this case $\gamma_{\mathrm{u}}\,=\,\gamma_{\mathrm{v}}\gamma_{\mathrm{w}}$. The resulting matrix in \eqref{eq:product of boost in x and z directions} is no longer symmetric and can therefore not be written in the form of \eqref{eq:general lorentz boost}. It now contains a rotation, i.e. it can be written as a combination of a rotation and a boost, such as in \eqref{eq:general lorentz trafo}, where the rotation is about the axis $-\vec{v}\times\vec{w}$, here being the $y$-axis, and the angle of the rotation is given by
\begin{equation}
\cos\delta\,=\,\frac{\gamma(\vec{v}\,)\,+\,\gamma(\vec{w}\,)}{1\,+\,\gamma(\vec{u}\,)}\ \ .
\label{eq:rotation angle for boosts in x and z direction}
\end{equation}
For boosts in arbitrary directions with general velocities $\vec{v}$ and $\vec{w}$, the rotation axis is still given by $\vec{v}\times\vec{w}$, but the angle of rotation is given by
\begin{equation}
\cos\delta\,+\,1\,=\,\frac{(1\,+\,\gamma(\vec{u}\,)\,+\,\gamma(\vec{v}\,)\,+\,\gamma(\vec{w}\,))^{2}}
{(\gamma(\vec{u}\,)\,+\,1)(\gamma(\vec{v}\,)\,+\,1)(\gamma(\vec{w}\,)\,+\,1)}\ \ ,
\label{eq:general thomas precession angle}
\end{equation}
where $\gamma(\vec{u}\,)=\gamma(\vec{v}\,)\gamma(\vec{w}\,)(1+\vec{v}\vec{w})\ $. A derivation of this result, closely connected to the \emph{Thomas precession}, can be found in Ref.~\cite{macfarlane62}.

\newpage		
	\subsection{Relativistic Description of Quantum Systems}\label{sec:Relativistic Description of Quantum Systems}

We started the discussion of special relativity with the principle of relativity, presuming the invariance of the laws of nature under a change of inertial frame and, certainly, we will also assume this for the laws of quantum physics. It is therefore only natural to look for representations of the Lorentz group on Hilbert space. The task of finding all representations, however, is an intricate procedure, which can be found in Ref.~\cite{sexlurbantke} in full detail. Some of the essential results can, amongst other books, also be found in \cite{ryderqft}. We will be content with relativistically describing spin $\tfrac{1}{2}$ particles in this section in order to study the entanglement of such systems in Chapter \ref{chap:entanglement and wigner rotations}.

    	\subsubsection{Representations of the Lorentz group}\label{subsec:reps of lorentz group}		

To find suitable representations of the proper, orthochronous Lorentz group on Hilbert space, we will proceed as discussed in Sec.~\ref{subsec:lie groups and lie algebras}, i.e. we will study the Lie algebra of the Lorentz group by examining Lorentz transformations close to the identity element of the Lorentz group and subsequently construct representations thereof by exponentiating the Lie algebra elements.\\

Consider an infinitesimal transformation $L(\vec{v},\vec{\varphi}\,)$ of the form \eqref{eq:general lorentz boost}, i.e. a general Lorentz transformation with infinitesimal parameters $\vec{\varphi}$ and $\vec{v}$,
\begin{equation}
L(\vec{v},\vec{\varphi}\,)\,\approx\,\mathds{1}_{\!4}\,-\,i\vec{\varphi}\vec{J}\,-\,i\vec{v}\vec{K}\ .
\label{eq:infinitesimal lorentz trafo}
\end{equation}
We have included factors of $i$ to resemble the form of \eqref{eq:exponential rep of 3d rotation}, and the generators are given by the $4\times4$ matrices
\begin{equation}
J^{\,i}\,=\,\begin{pmatrix} \,0 & 0\, \\ \,0 & T^{\,i}\,\end{pmatrix}\ ,\ \
K^{\,i}\,=\,-i\,\begin{pmatrix} \,0 & \vec{e}_{\,i}^{\ T}\, \\ \,\vec{e}_{\,i} & 0\,\end{pmatrix}\ ,
\label{eq:rotation and boost generators}
\end{equation}
where $T^{\,i}$ are the generators of rotations in three dimensions \eqref{eq:so(3) basis compact}, and $\vec{e}_{\,i}$ is the $i$-th unit vector of the three (spatial) coordinate directions. The special cases of pure boosts, e.g. in the $x^{1}$ direction, $L(v_{\mathrm{x}},0)$, and pure rotations $L(0,\vec{\varphi}\,)$, correspond to the infinitesimal versions of \eqref{eq:boost in x direction} and \eqref{eq:rotations in minkowski space} respectively. The six generators $J^{\,i}$ and $K^{\,\mathrm{i}}$ satisfy the commutation relations
\begin{eqnarray}
\left[\,J^{\,l}\,,\,J^{\,m}\,\right] &=& i\,\varepsilon^{\,lmn}\,J^{\,n}\ ,
		\label{eq:lorentz algebra commutation J J}  \\[2mm]
\left[\,J^{\,l}\,,\,K^{\,m}\,\right] &=& i\,\varepsilon^{\,lmn}\,K^{\,n}\ ,
		\label{eq:lorentz algebra commutation J K}  \\[2mm]
\left[\,K^{\,l}\,,\,K^{\,m}\,\right] &=& -\,i\,\varepsilon^{\,lmn}\,J^{\,n}\ .
		\label{eq:lorentz algebra commutation K K}
\end{eqnarray}
The non-commutativity of the boost generators in Eq.~\eqref{eq:lorentz algebra commutation K K} is directly related to the Thomas precession discussed earlier. We see that only boosts along the same direction form a subgroup of the Lorentz group, and also only if the rapidity (see Eq.~\eqref{eq:rapidity}) is used as the group parameter. Although the commutation relations of $J^{\,i}$ and $K^{\,i}$ completely define the Lie algebra of the Lorentz group, this form is not very convenient when looking for other representations, e.g. on Hilbert space. This is much easier, if linear combinations $J^{\,i}_{\pm}$, given by
\begin{equation}
J^{\,i}_{\pm}\,=\,\frac{1}{2}\,\left(J^{\,i}\,\pm\,i\,K^{\,i}\right)\ ,
\label{eq:complexified lorentz algebra}
\end{equation}
of the generators over the complex numbers are being considered. Using \eqref{eq:lorentz algebra commutation J J} - \eqref{eq:lorentz algebra commutation K K} we calculate:
\begin{equation}
\left[\,J^{\,l}_{\pm}\,,\,J^{\,m}_{\pm}\,\right]\,=\,i\,\varepsilon^{\,lmn}\,J^{\,n}_{\pm}\ \ \ \mbox{and}\ \ \ \left[\,J^{\,l}_{+}\,,\,J^{\,m}_{-}\,\right]\,=\,0\ .
\label{eq:complexified lorentz algebra lie bracket}
\end{equation}
Thus the Lie algebra of the Lorentz group decomposes into a direct sum of two 2-dimensional Lie algebras, $L_{+}$ and $L_{-}$, which both have the structure of the rotation algebra \eqref{eq:so(3) structure constants} (see also \eqref{eq:pauli matrices over 2 commutator}). The irreducible representations $D^{(j_{+},j_{-})}$ of $\mathcal{L}^{\up}_{+}$ are therefore described by two irreducible representations of the 3-dimensional rotation group, which in turn are classified by their respective weights $j_{+}$ and $j_{-}$, real, positive, half-integer or integer numbers, corresponding to spin\footnote{We have previously used the letter $s$, see \eqref{eq:spin squared eigenvalue equation}.}. The infinitesimal Lorentz transformation \eqref{eq:infinitesimal lorentz trafo} in terms of the generators $\vec{J}_{+},\,\vec{J}_{-}$ is then
\begin{equation}
L(\vec{v},\vec{\varphi}\,)\,\approx\,\mathds{1}_{\!4}\,-\,i\,(\vec{\varphi}\,-\,i\vec{v}\,)\vec{J}_{+}\,-\,i\,(\vec{\varphi}\,+\,i\vec{v}\,)\vec{J}_{-}\ ,
\label{eq:infinitesimal lorentz trafo in j,jprime rep}
\end{equation}
and it is possible to represent it by
\begin{equation}
D^{(j_{+},j_{-})}(\vec{v},\vec{\varphi}\,)\,=\,D^{(j_{+})}(\vec{\varphi}\,-\,i\vec{v}\,)\otimes D^{(j_{-})}(\vec{\varphi}\,+\,i\vec{v}\,)\ ,
\label{eq:rep of infinitesimal lorentz trafo}
\end{equation}
using the exponential map (see Eq.~\eqref{eq:matrix exponential}), e.g. such as in \eqref{eq:SU(2) exponential rep of 3d rotation}. However, for finite transformations this might not be entirely true. Although $D^{(j_{+})}(\vec{\varphi}\,-\,i\vec{v}\,)$ certainly corresponds to some Lorentz transformations, it might not be  $L(\vec{v},\vec{\varphi}\,)$. This can be understood from the non-commutativity of boosts and rotations \eqref{eq:lorentz algebra commutation J K}. Aside from this, the velocity $\vec{v}$ is not an additive quantity under Lorentz transformations and has to be replaced by the rapidity
\begin{equation}
\vec{u}\,=\,\mathrm{ar}\tanh(|\,\vec{v}\,|)\,\frac{\vec{v}}{|\,\vec{v}\,|}\ .
\label{eq:rapidity vector}
\end{equation}
The representation corresponding to $L(\vec{v},\vec{\varphi}\,)$ for finite velocities and angles is then given by
\begin{equation}
D^{(j_{+},j_{-})}(\vec{v},\vec{\varphi}\,)\,=\,D^{(j_{+})}(\vec{\varphi}\,)\,D^{(j_{+})}(-i\vec{u}\,)\,\otimes\,D^{(j_{-})}(\vec{\varphi}\,)\,D^{(j_{-})}(i\vec{u}\,)\ .
\label{eq:rep of finite lorentz trafo}
\end{equation}

The simplest non-trivial representations are obtained for the pairs $j_{+}=\tfrac{1}{2},\,j_{-}=0$ and $j_{+}=0,\,j_{-}=\tfrac{1}{2}$. These are two inequivalent spinor representations, corresponding to \emph{Weyl spinors}.

    \subsubsection{The Connection of the Lorentz group and SL(2,$\mathbb{C}$)}
    		\label{subsec:lorentz group and SL(2,c)}
    		
Of the representations $D^{(j_{+},j_{-})}$ of $\mathcal{L}_{+}^{\up}$ found in Sec.~\ref{subsec:reps of lorentz group} let us study those with $j_{+}=\tfrac{1}{2},j_{-}=0$ more closely. These are then of the form
\begin{equation}
D^{(\tfrac{1}{2},0)}\,=\,e^{-i\,\vec{\varphi}\frac{\vec{\sigma}}{2}}\,e^{-\,\vec{u}\frac{\vec{\sigma}}{2}}\ ,
\label{eq:j+ half j- zero rep}
\end{equation}
where we discover that this representation, although comprising unimodular matrices, because $-i\,\vec{\varphi}\frac{\vec{\sigma}}{2}$ as well as $-\,\vec{u}\frac{\vec{\sigma}}{2}$ are traceless (see \eqref{eq:determinant of matrix exponential}), is not unitary. Only for pure rotations, $\vec{u}=0$, do we get unitary matrices, which are then elements of $SU(2)$. The transformation group for spin $\frac{1}{2}$ consequently cannot be $SU(2)$, but must contain $SU(2)$ as a subgroup. This is certainly the case for the group $SL(2,\mathbb{C})$,
\begin{equation}
SL(2,\mathbb{C})\,=\,\left\{\,A\,\in\,GL(2,\mathbb{C})\,|\,\det A\,=\,1\,\right\}\ ,
\label{eq:SL(2,C)}
\end{equation}
the group of complex, unimodular, $2\,\times\,2$ matrices. The constraint $\det A=1$ reduces the $8$ real parameters of $A$ to $6$ independent parameters, which shows that $SL(2,\mathbb{C})$ and $\mathcal{L}_{+}^{\up}$ have the same dimension. To establish the physical connection to the (proper, orthochronous) Lorentz group, let us use a similar approach as in Sec.~\ref{subsubsec:SO3 and SU2}, i.e. we introduce a fourth matrix $\sigma^{\,0}=\mathds{1}_{\!2}$ in addition to the Pauli matrices \eqref{eq:pauli matrices}, and write
\begin{equation}
\sigma^{\,\mu}\,=\,\begin{pmatrix} \sigma^{\,0} \\ \vec{\sigma} \end{pmatrix}\ ,\ \
\sigma_{\,\mu}\,=\,\eta_{\,\mu\nu}\,\sigma^{\,\nu}\ .
\label{eq:SL(2,C) coordinate matrix}
\end{equation}

Every Hermitian $2\times 2$ matrix $X$ can then be written as a (real) linear combination of $\{\sigma_{\,\mu}\}$ in an analogous way as in \eqref{eq:3dim vector as matrix},
\begin{equation}
X\,=\,x^{\,\mu}\,\sigma_{\,\mu}\,=\,
\begin{pmatrix}
\ x^{\,0}\,-\,x^{\,3} & -\,x^{\,1}\,+\,i\,x^{\,2}\, \\
-\,x^{\,1}\,\!-\,i\,x^{\,2} & \ x^{\,0}\,+\,x^{\,3}\, \\
\end{pmatrix}\ ,\ \ x\,\in\,\mathbb{R}^{4}\ .
\label{eq:4dim vector as matrix}
\end{equation}

Now consider another matrix $X^{\,\prime}$, related to $X$ via the transformation
\begin{equation}
X\,\rightarrow\,X^{\,\prime}\,=\,A\,X\,A^{\,\dagger}\,=\,x^{\,\prime\,\mu}\,\sigma_{\,\mu}\ ,
\label{eq:SL(2,C) transformation}
\end{equation}
where $A\in SL(2,\mathbb{C})$. The determinants of $X$ and $X^{\prime}$ are then
\begin{equation}
\det X\,=\,(x^{\,0})^{2}\,-\,\vec{x}^{\,2}\,=\,x_{\,\mu}\,x^{\,\mu}\,=\,\det X^{\,\prime}\,=\,x^{\,\prime}_{\,\mu}\,x^{\,\prime\,\mu}\ .
\label{eq:determinant of SL(2,C) trafo}
\end{equation}

This constitutes the property we expected Lorentz transformations to satisfy in \eqref{eq:lorentz invariance of metric}. The matrices $A\in SL(2,\mathbb{C})$ thus induce Lorentz transformations
\begin{equation}
A\,x^{\,\mu}\,\sigma_{\,\mu}\,A^{\,\dagger}\,=\,(L_{\,A})^{\,\mu}_{\ \ \,\nu}\,x^{\,\nu}\,\sigma_{\,\mu}\ .
\label{eq:SL(2,C) lorentz trafo}
\end{equation}

It can be immediately seen here that the matrix $-A$ leads to the same Lorentz transformation $L_{\,A}$ and the map $SL(2,\mathbb{C})\rightarrow\mathcal{L}_{+}^{\up}$ is not bijective, but only surjective, i.e. it maps two elements of $SL(2,\mathbb{C})$ to one element of $\mathcal{L}_{+}^{\up}$.\\

Furthermore $SL(2,\mathbb{C})$ is simply connected. To see this, write $A=U\,H$ as the product of a unitary matrix $U\in SU(2)$ and a positive definite, Hermitian matrix $H$, such as in Eq.~\eqref{eq:j+ half j- zero rep}. Since $H$ is also unimodular, the matrix $H$ can assigned to real 4-vector $h$, which lies on the unit mass shell, i.e. $h^{2}=1$. The unit mass shell carries the topology of $\mathbb{R}^{3}$ and subsequently $SL(2,\mathbb{C})$ can be viewed as the product of two simply connected spaces, $SL(2,\mathbb{C})=\mathbb{R}^{3}\times SU(2)$, and is therefore also simply connected, see Ref.~\cite{sexlurbantke}.\\

We thus conclude that the group $SL(2,\mathbb{C})$ is the universal (double) covering group (Theorem \ref{thm:universal covering}) of $\mathcal{L}_{+}^{\up}$,

\begin{equation}
SL(2,\mathbb{C})/\left\{\mathds{1},-\mathds{1}\right\}\,\cong\,\mathcal{L}_{+}^{\up}\ ,
\label{eq:SL(2,C) covering prop orth lorentz group}
\end{equation}
which both contain the rotation group, $SU(2)$ and $SO(3)$ respectively, as subgroups. The irreducible representation $D^{(0,\tfrac{1}{2})}$ of $\mathcal{L}_{+}^{\up}$, with weights $j_{+}=0,j_{-}=\tfrac{1}{2}$, transforms under a different representation of $SL(2,\mathbb{C})$, called the \emph{conjugate, contragradient representation}, defined by
\begin{equation}
A\,\rightarrow\,A^{-1\,\dagger}\ ,
\label{eq:contragrad rep of SL(2,C)}
\end{equation}
which is inequivalent to the representation $A\rightarrow A$ used previously. We will denote the two-component spinors transforming under the representation $D^{(\tfrac{1}{2},0)}$ by $\psi_{L}$, and those transforming under $D^{(0,\tfrac{1}{2})}$ by $\psi_{R}$. They are called left-handed and right-handed Weyl spinors, respectively.\\

However, neither the representation $D^{(\tfrac{1}{2},0)}$, nor $D^{(0,\tfrac{1}{2})}$ is unitary, in fact, there exist no finite-dimensional, unitary representation of the Lorentz group apart from the trivial (or direct sums thereof). In Ref.~\cite{sexlurbantke}, the connection of $SL(2,\mathbb{C})$ to $\mathcal{L}_{+}^{\up}$ is used to prove this claim.\\

Until now we have only considered representations of the proper, orthochronous Lorentz group $\mathcal{L}_{+}^{\up}$, and we can naturally ask, how we can find representations of other parts of the Lorentz group. The parity symmetry $P$ is certainly of interest in physics, but to accommodate for spatial reflections, we cannot use $SL(2,\mathbb{C})$. The irreducible representations $D^{(\tfrac{1}{2},0)}$ and $D^{(0,\tfrac{1}{2})}$ are not invariant under the parity operations, since $\vec{K}$ transforms like a vector, while $\vec{J}$ transforms like a pseudo-vector, see \eqref{eq:lorentz algebra commutation K K} and \eqref{eq:lorentz algebra commutation J J}. This means that parity reversal changes the handedness of the Weyl spinors and we can construct a $P$-invariant representation as the direct sum of these representations,
\begin{equation}
D^{(\tfrac{1}{2},0)}\oplus D^{(0,\tfrac{1}{2})}\,=\,
\begin{pmatrix} A & \\ & A^{-1\,\dagger} \end{pmatrix} .
\label{eq:dirac spinor representation}
\end{equation}
Objects transforming under this $\mathcal{L}_{+}^{\up}$ reducible (see \eqref{eq:compl reducible matrix rep}) but $\mathcal{L}^{\up}$ irreducible representation are called \emph{Dirac spinors}, which can (in this representation) be written as
\begin{equation}
\psi\,=\,\begin{pmatrix} \psi_{L} \\ \psi_{R} \end{pmatrix}\ .
\label{eq:dirac spinor}
\end{equation}
Other representations of $\mathcal{L}^{\up}$ can be obtained by using the representations of $\mathcal{L}_{+}^{\up}$ with equal weights $j_{+}$ and $j_{-}$, the most important being $D^{(0,0)}$, the scalar representation, and $D^{(\tfrac{1}{2},\tfrac{1}{2})}$, the vector representation.
		
		\subsubsection{Representations of the Poincar$\mathbf{\acute{e}}$ group}\label{subsec:reps of poincare group}
		
So far we have mostly ignored the translations, and with it the largest invariance group of (quantum) physics, the Poincar$\acute{e}$ group $\mathcal{P}$, which is the semidirect product of the Lorentz group and the translation group. As the Lorentz group $\mathcal{L}$, $\mathcal{P}$ decomposes into 4 pieces, of which we will be interested in ($\mathcal{P}_{+}^{\up}$) $\mathcal{P}^{\up}$, the counterpart to the (proper,) orthochronous Lorentz group.\\

The representations of the Lorentz group we have studied, can all be used as representations of the Poincar$\acute{e}$ group as well, by assuming trivial transformation properties under the translation group, i.e. scalars, spinors, and tensors all form representations of $\mathcal{P}$. We have also encountered objects with non-trivial transformation properties under translations, these are the respective fields. Consider a Poincar$\acute{e}$ transformation $(a,\Lambda)$, consisting of a Lorentz transformation $\Lambda$ and a translation $a$, together with a representation $D(a,\Lambda)$ thereof on the space of fields $\Phi(x)$ of a certain kind. The field $\Phi$ is then mapped to a different field $\Phi^{\,\prime}$ by $D(a,\Lambda)$, such that
\begin{equation}
\Phi^{\,\prime}(x)\,=\,D(a,\Lambda)\,\Phi(x)\,=\,D(\Lambda)\,\Phi(\Lambda^{-1}(x-a))\ ,
\label{eq:poincare trafo of field}
\end{equation}
where $D(\Lambda)$ is a (finite dimensional) representation of $\Lambda$. Without further restrictions, this representation is however not irreducible. This is due to the fact, that the solutions of Poincar$\acute{e}$-covariant, linear, homogeneous differential equations form invariant subspaces of the space of all fields of certain types. To gain an irreducible representation, we need to specify the type of field, and the corresponding (invariant) field equation.\\

We could start with simplest case and construct a $\mathcal{P}_{+}^{\up}$ invariant field equation for a scalar field, the \emph{Klein-Gordon equation}, but since we are interested in spin $\tfrac{1}{2}$ fields, we need a stronger equation. Also we could start with finding an equation for two-component spinors. The possible equations of this type turn out to be too restrictive for our analysis here, the \emph{Weyl equation} is only applicable to massless particles, while the \emph{Majorana equation}, although describing massive particles, needs these particles to be identical to their antiparticles. We will therefore investigate the \emph{Dirac equation}, from which all of the above equations can be derived\footnote{For a detailed treatment of Lorentz covariant field equations, e.g. Klein-Gordon-, Weyl-, Majorana- and Dirac-fields, see textbooks on quantum field theory, such as \cite{ryderqft}, or \cite{peskinschroeder}.}.\\

\pa{The Dirac Equation:}\ \\

For the derivation of the Dirac equation we will follow the approach of Ref.~\cite{peskinschroeder}. To this end we try to find a more compact notation for the Lie algebra of the Lorentz group, \eqref{eq:lorentz algebra commutation J J} - \eqref{eq:lorentz algebra commutation K K}. First, reformulate the generators of rotations in $\mathbb{R}^{3}$, Eq.~\eqref{eq:orbital angular momentum i component}, as an antisymmetric tensor,
\begin{equation}
L^{ij}\,=\,-\,i\,(\,x^{\,i}\,\partial^{\,j}\,-\,x^{\,j}\,\partial^{\,i}\,)\ ,
\label{eq:rotation generators as tensor}
\end{equation}
using the the momentum operator $p^{\,k}\,=\,-i\partial^{\,k}$. The generalization of the momentum operator to a relativistic setting is quite natural when looking at Eq.~\eqref{eq:four momentum} and Eq.~\eqref{eq:first quantization}, it is
\begin{equation}
\hat{p}^{\,\mu}\,=\,i\,\partial^{\,\mu}\ ,\ \ \mbox{where}\ \ \partial^{\,\mu}\,=\,\begin{pmatrix} \partial^{\,0} \\ \partial^{\,i}\end{pmatrix}\ .
\label{eq:first quantization relativistic}
\end{equation}
The 6 generators of Lorentz transformations (for scalar fields) can then be written as
\begin{equation}
\mathcal{J}^{\,\mu\nu}\,=\,i\,(\,x^{\,\mu}\,\partial^{\,\nu}\,-\,x^{\,\nu}\,\partial^{\,\mu}\,) \ ,
\label{eq:lorentz trafo generators as tensor}
\end{equation}
and can then be shown (\cite{peskinschroeder}) to satisfy the commutation relations
\begin{equation}
\left[\,\mathcal{J}^{\,\mu\nu}\,,\,\mathcal{J}^{\,\rho\sigma}\,\right]\,=\,i\,(\,		
	\eta^{\,\nu\rho}\,\mathcal{J}^{\,\mu\sigma}\,-\,\eta^{\,\mu\rho}\,\mathcal{J}^{\,\nu\sigma}\,-\,
	\eta^{\,\nu\sigma}\,\mathcal{J}^{\,\mu\rho}\,+\,\eta^{\,\mu\sigma}\,\mathcal{J}^{\,\nu\rho}\,)\ ,
\label{eq:lorentz algebra commutation relations}
\end{equation}
for which we can recover the previous generators \eqref{eq:rotation and boost generators}, by setting
\begin{equation}
K^{\,i}\,=\,\mathcal{J}^{\,i0}\ ,\ \ \mbox{and}\ \ J^{\,i}\,=\,\frac{1}{2}\,\varepsilon^{\,ijk}\,\mathcal{J}^{\,jk}\ .
\label{eq:relation of lorentz trafo generators}
\end{equation}

The notation of the generators as an antisymmetric tensor has a practical advantage, since every set $\{\gamma^{\,\mu}\}$ of $4$\  $n\times n$ matrices $\gamma^{\,\mu}$, satisfying the anti-commutation relation
\begin{equation}
\left\{\,\gamma^{\,\mu}\,,\,\gamma^{\,\nu}\,\right\}_{+}\,=\,2\,\eta^{\,\mu\nu}\ ,
\label{eq:clifford algebra}
\end{equation}
gives rise to an $n$-dimensional representation $\mathcal{S}^{\,\mu\nu}$ of the Lorentz algebra, i.e. satisfying \eqref{eq:lorentz algebra commutation relations}, by
\begin{equation}
S^{\,\mu\nu}\,=\,\frac{1}{2}\,\sigma^{\,\mu\nu}\ ,\ \ \mbox{where}\ \ \sigma^{\,\mu\nu}\,=\,\frac{i}{2}\,\left[\,\gamma^{\,\mu}\,,\,\gamma^{\,\nu}\,\right]\ .
\label{eq:dirac gamma representation}
\end{equation}

One particular realization of these matrices, the so called Weyl representation, is
\begin{equation}
\gamma^{\,0}\,=\,\begin{pmatrix} 0 & \mathds{1}_{\!2} \\ \,\mathds{1}_{\!2} & 0 \end{pmatrix}\ , \ \ \gamma^{\,i}\,=\,\begin{pmatrix} 0 & \sigma^{\,i}\, \\ -\sigma^{\,i} & 0\, \end{pmatrix}\ .
\label{eq:weyl representation}
\end{equation}

The 4-component objects $\psi$, then transforming under Lorentz transformations as
\begin{equation}
\psi\,\rightarrow\,S(\Lambda(\vec{\varphi},\vec{u}\,))\,\psi\,=\,e^{-\tfrac{i}{4}\,\sigma_{\,\mu\nu}\,\omega^{\,\mu\nu}}\,\psi ,
\label{eq:dirac spinor transformation}
\end{equation}
where the parameters of the transformation $\Lambda$ are given by the antisymmetric matrix
\begin{equation}
\omega^{\,\mu\nu}\,=\,\begin{pmatrix}
\raisebox{-0.9mm}{0}	& 												& -\vec{u}^{\,T} 					& 											 \\
 											& \ \raisebox{-0.9mm}{0}	& \ \varphi_{z} 					& -\varphi_{y}					\\
\vec{u} 							& -\varphi_{z}						& \,\raisebox{-0.9mm}{0} 	& \ \ \varphi_{x}				 \\
 											& \ \ \varphi_{y} 				& -\varphi_{x} 						& \ \raisebox{-0.9mm}{0}
\end{pmatrix}\ ,
\label{eq:lorentz trafo parameter matrix}
\end{equation}
are then found to be exactly the Dirac spinors of Eq.~\eqref{eq:dirac spinor}, and the representations \eqref{eq:dirac spinor transformation} and \eqref{eq:dirac spinor representation} coincide. The reason for this approach becomes apparent, when considering the following transformation of the $\gamma$-matrices,
\begin{equation}
S(\Lambda)^{-1}\,\gamma^{\,\mu}\,S(\Lambda)\,=\,\Lambda^{\,\mu}_{\ \ \nu}\,\gamma^{\,\nu}\ ,
\label{eq:gamma matrix trafo}
\end{equation}
which shows, that the index on $\gamma^{\,\mu}$ can be used to construct Lorentz invariant quantities, especially, a Lorentz invariant differential operator, by contracting the index on $\gamma^{\,\mu}$ with the partial derivative operator $\partial_{\,\mu}$. We can then use this operator to write down a Lorentz invariant field equation for the Dirac spinor fields $\psi(x)$.

\begin{equation}
(\,i\,\gamma^{\,\mu}\,\partial_{\,\mu}\,-\,m\,)\,\psi(x)\,=\,0
\label{eq:dirac equation}
\end{equation}

This is the famous \emph{Dirac equation}, it describes, as we awaited, massive spin $\tfrac{1}{2}$ fields. Since we now consider fields, a Lorentz transformation $\Lambda$ will act on $\psi(x)$ as
\begin{equation}
\psi(x)\,\rightarrow\,S(\Lambda)\,\psi(\Lambda^{-1}\,x)\,=\,e^{-\tfrac{i}{2}\,\mathcal{J}_{\,\mu\nu}\,\omega^{\,\mu\nu}}\,\psi(\Lambda^{-1}\,x) \ ,
\label{eq:dirac field transformation}
\end{equation}
where we have now combined the generators of Eq.~\eqref{eq:lorentz trafo generators as tensor}, which we now write as $L^{\,\mu\nu}$, with the generators of Eq.~\eqref{eq:dirac gamma representation}, to form (see Eq.~\eqref{eq:total angular momentum})
\begin{equation}
\mathcal{J}^{\,\mu\nu}\,=\,L^{\,\mu\nu}\,+\,S^{\,\mu\nu}\ .
\label{eq:spin one half field lorentz trafo generators}
\end{equation}
Since the generator $L^{\,\mu\nu}$, corresponding to orbital angular momentum, and $S^{\,\mu\nu}$, corresponding to spin, commute with each other, and separately satisfy the commutation relations of the Lorentz algebra, \eqref{eq:lorentz algebra commutation relations}, also $\mathcal{J}^{\,\mu\nu}$ of \eqref{eq:spin one half field lorentz trafo generators} must satisfy these relations.\\

To see that the Dirac equation is Lorentz invariant, we note that the differential operator $\partial_{\,\mu}$ transforms with the inverse transformation, so transforming \eqref{eq:dirac equation} with $\Lambda$ we get
\begin{equation}
(i\,\gamma^{\,\mu}\,(\Lambda^{-1})^{\,\nu}_{\ \ \mu}\,\partial_{\,\nu}\,-\,m)\,S(\Lambda)\,\psi(\Lambda^{-1}\,x)\ .
\label{eq:lorentz invariance of dirac eq}
\end{equation}
Multiplying \eqref{eq:lorentz invariance of dirac eq} with $S^{-1}(\Lambda)$ from the left and using \eqref{eq:gamma matrix trafo}, as well as the Dirac equation itself, one quickly arrives at
\begin{equation}
(i\,\Lambda^{\,\mu}_{\ \ \rho}\gamma^{\,\rho}\,(\Lambda^{-1})^{\,\nu}_{\ \ \mu}\,\partial_{\,\nu}\,-\,m)\,\psi(\Lambda^{-1}\,x)\,=\,(i\,\gamma^{\,\nu}\,\partial_{\,\nu}\,-\,m)\,\psi(\Lambda^{-1}\,x))\ .
\label{eq:lorentz invariance of dirac eq proven}
\end{equation}



\pa{The Poincar$\mathbf{\acute{e}}$ Algebra:}\ \\

To classify the irreducible representations of the (proper, orthochronous) Poincar$\acute{e}$ group in the invariant subspace of the solutions to the Dirac equation, we again have to find the commutation relations of its generators, which are the generators $\mathcal{J}^{\,\mu\nu}$ of the Lorentz transformations, and $p^{\mu}$ of the translations. Since we already know these relations for the Lorentz group, recall Eq.~\eqref{eq:lorentz algebra commutation relations}, and the commutation relations of the translation group are trivially given by
\begin{equation}
\left[\,p^{\,\mu}\,,\,p^{\,\nu}\,\right]\,=\,0\ ,
\label{eq:translation group commutation relations}
\end{equation}
we are left with determining the commutation relations between the $\mathcal{J}^{\,\mu\nu}$ and the $p^{\,\mu}$, which are
\begin{equation}
\left[\,\mathcal{J}^{\,\mu\nu}\,,\,p^{\,\rho}\,\right]\,=\,i\,(\eta^{\,\rho\,\nu}\,p^{\,\mu}\,-\,\eta^{\,\rho\,\mu}\,p^{\,\nu}\,)\ .
\label{eq:translations and lorentz trafos commutation relations}
\end{equation}

The \emph{Casimir Invariants} of the Poincar$\acute{e}$ group are then the operators, whose eigenvalues are invariant under all transformations of $\mathcal{P}$, and they can therefore be used for the desired classification. The first invariant, i.e. an operator, commuting with all generators, is
\begin{equation}
p_{\,\mu}\,p^{\,\mu}\,=\,-\,\partial^{\,\mu}\,\partial_{\,\mu}\,=\,-\,\raisebox{-0.3mm}{$\Box$}\,=\,M^{2}\ ,
\label{eq:mass casimir operator}
\end{equation}
the operator of squared mass. That $p_{\,\mu}\,p^{\,\mu}$ commutes with all the generators can be quickly seen from the commutation relations \eqref{eq:translations and lorentz trafos commutation relations},
\begin{eqnarray}
\left[\,\mathcal{J}^{\,\mu\nu}\,,\,p^{\,\rho}\,p_{\,\rho}\,\right] &=&
		i\,(\eta^{\,\rho\nu}\,p^{\,\mu}\,-\,\eta^{\,\rho\mu}\,p^{\,\nu}\,)\,p_{\,\rho}\,+\,
		i\,p^{\,\rho}\,(\eta_{\,\rho}^{\ \ \nu}\,p^{\,\mu}\,-\,\eta_{\,\rho}^{\ \ \mu}\,p^{\,\nu}\,)\,= \nonumber\\[2mm]
&=& i\,(p^{\,\mu}\,p^{\,\nu}\,-\,p^{\,\nu}\,p^{\,\mu}\,)\,+\,i\,(p^{\,\nu}\,p^{\,\mu}\,-\,p^{\,\mu}\,p^{\,\nu}\,)\,=\,0\ ,
\label{eq:commutation of mass squared operator with lorentz generators}
\end{eqnarray}
and (trivially) from \eqref{eq:translation group commutation relations}. The interpretation as squared mass operator can be seen as the generalization of the classical 4-momentum \eqref{eq:four momentum}, or from the Dirac equation \eqref{eq:dirac equation}, by multiplying it with $(-\,i\,\gamma^{\,\mu}\,\partial_{\,\mu}\,-\,m)$ from the left,
\begin{eqnarray}
0 &=& (-\,i\,\gamma^{\,\mu}\,\partial_{\,\mu}\,-\,m)\,
			 (\,i\,\gamma^{\,\mu}\,\partial_{\,\mu}\,-\,m\,)\,\psi(x)\,=\nonumber\\[2mm]
&=&	(\,\gamma^{\,\mu}\,\gamma^{\,\nu}\,\partial_{\,\mu}\,\partial_{\,\nu}\,+\,m^{2}\,)\,\psi(x)\,=\nonumber\\[2mm]
&=& (\,\tfrac{1}{2}\left\{\gamma^{\,\mu},\gamma^{\,\nu}\right\}_{+}\,
			\partial_{\,\mu}\,\partial_{\,\nu}\,+\,m^{2}\,)\,\psi(x)\,=\nonumber\\[2mm]
&=&	(\,\eta^{\,\mu\nu}\,\partial_{\,\mu}\,\partial_{\,\nu}\,+\,m^{2}\,)\,\psi(x)\ ,
\label{eq:derivation of KG eq from dirac equation}
\end{eqnarray}
where we used the symmetry of the partial derivatives and the property \eqref{eq:clifford algebra} of the $\gamma$-matrices. This is nothing but the \emph{Klein-Gordon equation},
\begin{equation}
(\,\raisebox{-0.3mm}{$\Box$}\,+\,m^{2}\,)\,\psi(x)\,=\,0\ ,
\label{eq:klein gordon eq}
\end{equation}
which is satisfied by all components of the Dirac spinor individually. The second Casimir operator of $\mathcal{P}$ is the \emph{Pauli-Ljubanski vector}\footnote{Detailed information on the role of the Pauli-Ljubanski vector can be found in Ref.~\cite{ryder99}.} $W_{\,\mu}$
\begin{equation}
W_{\,\mu}\,=\,-\,\tfrac{1}{2}\,\varepsilon_{\,\mu\nu\rho\sigma}\,\mathcal{J}^{\,\nu\rho}\,p^{\,\sigma}\ ,
\label{eq:pauli ljubanski vector}
\end{equation}
whose square commutes with all generators (see Ref.\cite{sexlurbantke}), because $W_{\,\mu}$ is orthogonal to $p^{\,\mu}$,
\begin{equation}
W_{\,\mu}\,p^{\,\mu}\,=\,0 \ ,
\label{eq:pauli ljubanksi momentum orthogonality}
\end{equation}
which follows from the symmetry and antisymmetry of $p^{\,\sigma}p^{\,\mu}$ and $\varepsilon_{\,\mu\nu\rho\sigma}$, respectively. Considering the decomposition of the generators $\mathcal{J}^{\,\nu\rho}$ into orbital angular momentum and spin generators, Eq.~\eqref{eq:spin one half field lorentz trafo generators}, we find that the orbital angular momentum generators, Eq.~\eqref{eq:lorentz trafo generators as tensor}, do not contribute to the Pauli-Ljubanski vector. It is therefore only related to the spin of the representation, and reduces to
\begin{equation}
W_{\,\mu}\,W^{\,\mu}\,=\,-\,\tfrac{1}{2}\,(\tfrac{1}{2}\,+\,1)\,m^{2}
\label{eq:pauli ljubanski eigenvalue}
\end{equation}
for spin $\tfrac{1}{2}$ Dirac fields. This result can be calculated straightforwardly by first using the commutation relation \eqref{eq:translations and lorentz trafos commutation relations} to obtain
\begin{equation}
W_{\,\mu}\,W^{\,\mu}\,=\,-\,\tfrac{1}{4}\,S^{\,\mu\nu}\,S_{\,\mu\nu}\,p^{\,\rho}\,p_{\,\rho}\ ,
\label{eq:square of pauli ljubanski 1}
\end{equation}
and then using the Weyl representation \eqref{eq:weyl representation} of $S^{\,\mu\nu}$, in which
\begin{equation}
S^{\,\mu\nu}\,S_{\,\mu\nu}\,=\,3\,\mathds{1}_{\!4}\ .
\label{eq:spin squared in weyl rep}
\end{equation}

Although there are no other invariant operators of the Poincar$\acute{e}$ group, the characteristic values of the representation, the mass $m$, and the spin $s$, are still not yet enough, to fully describe the irreducible representations of $\mathcal{P}$. In the next section we will find this last piece of information, and construct the unitary, irreducible representations for massive spin $\tfrac{1}{2}$ particles.

		\subsubsection{Wigner's Little Group}\label{subsec:wigners little group}
		
To finally write down unitary representations of $\mathcal{P}_{+}^{\up}$ for state vectors on a Hilbert space, we switch back to the Dirac notation of Sec.~\ref{sec:principles of QM}, and label the basis states $|\,p,\sigma\,\rangle$ by their momentum $p$, and spin\footnote{For spin $\tfrac{1}{2}$, $\sigma$ can take on values $\pm\tfrac{1}{2}$, corresponding to Eq.~\eqref{eq:spin z eigenstates}.} $\sigma$. In doing so we have chosen a basis of plane waves, transforming under translations $U(a)$ as
\begin{equation}
U(a)\,|\,p,\sigma\,\rangle\,=\,e^{-\,i\,p\,a}\,|\,p,\sigma\,\rangle\ ,
\label{eq:plane wave translation trafo}
\end{equation}
i.e. momentum eigenstates (compare Eq.~\eqref{eq:momentum eigenstates}), satisfying
\begin{equation}
\hat{p}^{\,\mu}\,|\,p,\sigma\,\rangle\,=\,p^{\,\mu}\,|\,p,\sigma\,\rangle\ .
\label{eq:rel momentum eigenstates}
\end{equation}

Suppose then, we have found some representation of the Lorentz transformation $\Lambda$ on the space spanned by the $|\,p,\sigma\rangle$, and let us denote it by $U(\Lambda)$, implying that we wish to find it to be unitary. The state $U(\Lambda)|\,p,\sigma\rangle=|\,p,\sigma\rangle^{\Lambda}$ must then be a momentum eigenstate with momentum $\Lambda p$,
\begin{equation}
\hat{p}^{\,\mu}\,|\,p,\sigma\,\rangle^{\Lambda}\,=\,
\hat{p}^{\,\mu}\,U(\Lambda)\,|\,p,\sigma\,\rangle\,=\,\Lambda^{\mu}_{\ \,\nu}\,p^{\,\nu}\,U(\Lambda)\,|\,p,\sigma\,\rangle\ ,
\label{eq:boosted momentum eigenstate}
\end{equation}
because $\hat{p}^{\,\mu}$ transforms like a vector. The transformed state $|\,p,\sigma\,\rangle^{\Lambda}$ must therefore be a linear combination of all states with the same momentum $\Lambda p$, i.e.
\begin{equation}
U(\Lambda)\,|\,p,\sigma\,\rangle\,=\,
\sum\limits_{\sigma^{\prime}}\,Q_{\,\sigma^{\prime}\sigma}(\Lambda,p)\,|\,\Lambda p,\sigma^{\,\prime}\,\rangle\ .
\label{eq:general state lorentz trafo formula}
\end{equation}

An explicit dependence on $p$ has been included here in the transformation matrix $Q(\Lambda,p)$. We could drop this restriction and let $Q$ only depend on $\Lambda$, which basically represents the choice of transformation we encountered in Eq.~\eqref{eq:dirac spinor transformation}. In that transformation, however, it is not immediately clear, how unitarity arises, since the generators \eqref{eq:dirac gamma representation} are not Hermitian.\\

We have made two important assumptions here, which are intuitively clear from a physical point of view. The first is that we assumed the momenta $p$ to satisfy Eq.~\eqref{eq:mass casimir operator}, which is of course what we aimed at in the first place, but it must be emphasized here, that we expect the transformation on Hilbert space to leave invariant $p_{\,\mu}\,p^{\,\mu}$. However, and this is where our second assumption enters, we also expect the sign of $p^{\,0}$ to remain invariant, i.e. we expect the transformation $\Lambda$ to leave invariant the unit mass shell, which is the spacelike hyperboloid in Minkowski space lying in the future light cone. So in addition to the Casimir invariants $s(s+1)$, and $m^{2}$, we also need to specify $sign(p^{\,0})$ to classify the \emph{unitary, irreducible representations} of $\mathcal{P}_{+}^{\up}$.\\

The physically interesting\footnote{The other possible cases, $p=0$ and $p^{2}<0$, have no practical physical interpretation.} four irreducible representations are those with $p^{2}=m^{2}>0$, or $p^{2}=0$, and additionally $sign(p^{\,0})=\pm1$. Here only the case $p^{2}=m^{2}>0\ ,sign(p^{\,0})=+1$ is discussed, but thorough investigations of the other classes can be found in \cite{sexlurbantke} and \cite{alsingmilburn02}.\\

If Eq.~\eqref{eq:general state lorentz trafo formula} describes a representation, it must in particular respect the group multiplication, i.e. it must be a group homomorphism (see Def.~\ref{def:group homomorphism}). This can be achieved (see \cite{sexlurbantke}) by restricting $Q(\Lambda,p)$ to $Q(\mathrm{W},k)$, where $\mathrm{W}$ are Lorentz transformations, which leave invariant a chosen standard momentum $k$. In our case $k^{2}=m^{2}>0$, and $sign(k^{\,0})=+1$. Let us take the standard momentum to be the rest frame momentum of a massive particle,
\begin{equation}
k\,=\,\begin{pmatrix} m \\ 0 \end{pmatrix}\ ,
\label{eq:standard momentum}
\end{equation}
by setting $\vec{v}=0$ in Eq.~\eqref{eq:four momentum}. State vectors to arbitrary momenta $p$ (on the unit mass shell) can then be obtained by the action of the representation of a Lorentz transformation $L(p)$, where $L(p)k=p$, on the states $|\,k,\sigma\,\rangle$, labeled by $\sigma$, which defines a basis,
\begin{equation}
|\,p,\sigma\,\rangle\,=\,U(L(p))\,|\,k,\sigma\,\rangle\ .
\label{eq:boost of rest frame state}
\end{equation}
Afterwards we can act upon \eqref{eq:boost of rest frame state} with the representation of some Lorentz transformation $\Lambda$,
\begin{eqnarray}
U(\Lambda)\,|\,p,\sigma\,\rangle &=& U(\Lambda)\,U(L(p))\,|\,k,\sigma\,\rangle\,=\nonumber\\[2mm]
&=& U(L(\Lambda p)\,L^{-1}(\Lambda p))\,U(\Lambda)\,U(L(p))\,|\,k,\sigma\,\rangle\,=\nonumber\\[2mm]
&=& U(L(\Lambda p))\,U(L^{-1}(\Lambda p)\,\Lambda\,L(p))\,|\,k,\sigma\,\rangle\ ,
\label{eq:unitary rep of lorentz trafo 1}
\end{eqnarray}
where we have inserted the identity in the second line. Obviously, the third line is only true, if we are dealing with a representation. Taking a closer look at the term $U(L^{-1}(\Lambda p)\Lambda L(p))$, we find that the argument, the transformation
\begin{equation}
\mathrm{W}(\Lambda,p)\,:=\,L^{-1}(\Lambda p)\,\Lambda\,L(p)
\label{eq:wigner rotation}
\end{equation}
leaves the standard momentum invariant. It first takes $k$ to $p$, then some Lorentz transformations $\Lambda$, which must be a combination of boosts and rotations, is performed on $p$, before $L^{-1}(\Lambda p)$ takes it back to $k$.\\

Since the operator $U(\mathrm{W})$ does not change $k$, it can only act on the spin degree of freedom, while $U(L(\Lambda p))$ acts on the state according to \eqref{eq:boost of rest frame state}, thus leaving the spin unchanged. Eq.~\eqref{eq:unitary rep of lorentz trafo 1} can then be written as
\begin{equation}
U(\Lambda)\,|\,p,\sigma\,\rangle\,=\,U(\mathrm{W}(\Lambda,p))\,|\,\Lambda p,\sigma\,\rangle\ .
\label{eq:unitary rep of lorentz trafo 2}
\end{equation}

Comparison with \eqref{eq:general state lorentz trafo formula} immediately reveals the connection of $Q(\Lambda,p)$ to $Q(\mathrm{W}(\Lambda,p),k)$ we were looking for,
\begin{equation}
U(\Lambda)\,|\,p,\sigma\,\rangle\,=\,\sum\limits_{\sigma^{\prime}\sigma}\,Q_{\sigma^{\prime}\sigma}(\mathrm{W}(\Lambda,p))\,|\,\Lambda p,\sigma^{\prime}\,\rangle\ .
\label{eq:general state lorentz trafo formula with little group}
\end{equation}

The transformations $\mathrm{W}$, called \emph{Wigner rotations}, and defined by Eq.~\eqref{eq:wigner rotation}, form a subgroup of $\mathcal{L}_{+}^{\up}$, corresponding to the chosen standard momentum, which is called \emph{Wigner's little group}. It is the stabilizing group of the (proper, orthochronous) Lorentz group.\\

The classification problem of the \emph{unitary, irreducible representations} of the (proper, orthochronous) Poincar$\acute{e}$ group can thus be traced back to that of representing the elements of the little group,  which leave invariant a chosen standard momentum, which was first realized by Wigner in his ground breaking paper \cite{wigner39}. As can be seen by our choice of standard momentum, \eqref{eq:standard momentum}, the little group elements for massive particles are 3-dimensional rotations. The corresponding representations of the little group are then simply the representations of $SU(2)$, for which we have already found suitable expressions in Sec.~\ref{subsec:rotations and spin}.\\

The crucial point in which this representation is different from the $SL(2,\mathbb{C})$ transformations in Sec.~\ref{subsec:lorentz group and SL(2,c)}, is that the indices of $Q$ in \eqref{eq:general state lorentz trafo formula with little group} do not refer to spinor indices, but to spin eigenvalues $\sigma=\pm\tfrac{1}{2}$ along some quantization axis. The corresponding spinors can thus be denoted by the usual $\up$, $\down$ notation (see Eq.~\eqref{eq:spin z eigenstates}), and the $SU(2)$ matrices take the same form as in Eq.~\eqref{eq:SU(2) exponential rep of 3d rotation}.\\

\begin{figure}[ht]
	\centering
		\includegraphics[width=0.80\textwidth]{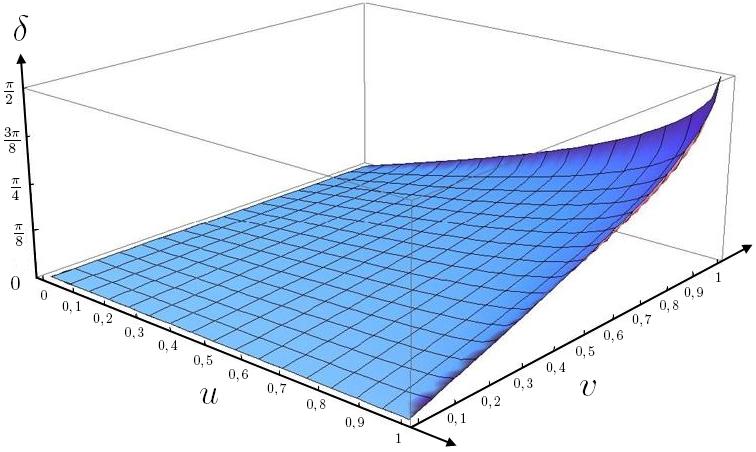}
	\caption{The Wigner rotation angle $\delta$ is displayed as a function of boost velocities $w=\frac{|\vec{w}|}{c}$, and $v=\frac{|\vec{v}|}{c}$, for boosts in perpendicular directions. The rotation angle is very small, if the involved velocities are much smaller than the speed of light and only if both velocities approach the speed of light, the angle becomes $\tfrac{\pi}{2}$.}.
	\label{fig:wignerangle}
\end{figure}

It remains yet to determine the explicit form of the (representations of the) \emph{Wigner rotations} $U(\mathrm{W})$. Since they are constructed from Lorentz transformations, we can use the results about the \emph{Thomas precession} from Sec.~\ref{subsec:lorentz and poincare group}, where we interpret the Lorentz transformations used in \eqref{eq:boosts in x and z direction} as the transformations $L(p)$, and $\Lambda$ respectively. Since the succession, $\Lambda\,L(p)$, of these transformations is a combination of a rotation and a boost, we can get the corresponding pure rotation by applying $L^{-1}(\Lambda p)$. Consider the momentum $p$ to be of the form of \eqref{eq:four momentum}, i.e. $\vec{p}=m\gamma(v)\vec{v}$, while $\Lambda$ is a boost corresponding to an observer with velocity $\vec{w}$. The axis of rotation is then given by $-\vec{v}\times \vec{w}$ and the angle by Eq.~\eqref{eq:general thomas precession angle}. If the boosts are in perpendicular directions, the angle is given by Eq.~\eqref{eq:rotation angle for boosts in x and z direction}, which is illustrated in Fig.~\ref{fig:wignerangle}. The Wigner rotation angle $\delta$ becomes smaller, if the two boosts are not in perpendicular directions.

Of course we could have obtained this result also by choosing Lorentz transformations $L(p)$ and $\Lambda$ in Eq.~\eqref{eq:wigner rotation} and explicitly calculate their product. This was done in Ref.~\cite{alsingmilburn02} for certain interesting cases and their result agrees with that of the Thomas precession.

\newpage\section{Entanglement \& Wigner Rotations}\label{chap:entanglement and wigner rotations}

We are now in a position to analyze the Lorentz transformation properties of entangled states. As in the chapters before, we will focus on massive\footnote{Discussions of the role of Wigner's little group for massless particles, and photons in relativistic quantum information procedures can be found in Ref.~\cite{caban07}, \cite{cabanrembielinski03}, and \cite{lindneretal03}.} spin $\tfrac{1}{2}$ particles, or more precisely on pairs of such particles. We will briefly discuss the general form of the states under consideration, before proceeding to analyze the situation for different special cases in Sec.~\ref{sec:spin and momentum entanglement in inertial frames}. As we will see, the main difficulties arise from the momentum dependence of the Wigner rotations $\mathrm{W}(\Lambda,p)$, see \eqref{eq:wigner rotation}, which has been first discussed by Peres, Scudo, and Terno in Ref.~\cite{peresetal02}, shortly after followed by a detailed review\footnote{In Ref.~\cite{alsingmilburn02} Wigner rotations are also discussed in a second quantized formalism. We refrained from introducing this formalism here, i.e. strictly speaking the Dirac field discussed in Sec.~\ref{subsec:reps of poincare group} is still a classical field. However, the mechanism of Wigner's classification does not depend on the promotion to a quantum field. The notations of Chapter \ref{chap:entanglement and wigner rotations} can therefore be used equivalently when quantizing the Dirac field.} of Wigner rotations by Alsing and Milburn in Ref.~\cite{alsingmilburn02}, and a study of the relation of spin- and momentum entanglement by Gingrich, and Adami \cite{gingrichadami02}.\\

Since then, a number of articles on the subject of \emph{relativistic quantum (information) theory} have been published, addressing questions, from covariantly transforming reduced density operators (Ahn, Lee, and Hwang in Ref.~\cite{ahnetal03b}), over relativistic quantum clock synchronization (Caban, and Rembieli\' nski in Ref.~\cite{cabanrembielinski99}), entanglement distillation (Lamata, and Martin-Delgado in Ref.~\cite{lamataetal06}), generation of entanglement (Pachos, and Solano in Ref.~\cite{pachossolano03}), and quantum cryptography (Czachor, and Wilczewski in Ref.~\cite{czachorwilczewski03}), to spin state transformations (Jordan, Shaji, and Sudarshan in Ref.~\cite{jordanetal06a}), in a relativistic framework. We do however feel that the basic question of entanglement invariance, raised in Ref.~\cite{peresetal02}, was not satisfyingly answered, or simply not addressed.\\

Parallel to, and intermingled with the discussion about relativistic entanglement, clearly, is the question wether or not the violation of Bell inequalities can be maintained in relativistic settings. Work was performed on this subject amongst others by Czachor \cite{czachor97}, Ahn, Lee, and Hwang \cite{ahnetal02}, as well as by Caban, and Rembieli\' nski \cite{cabanrembielinski05},\cite{cabanrembielinski06}. We will further discuss these questions in Sec.~\ref{sec:degree of violation of bell inequalities}. A review of our own main results can be found in Ref.~\cite{friisetal09}, while a review on the role of special relativity in quantum information theory is given in Ref.~\cite{peresterno04}.\\

		\subsection{Relativistic Spin- \& Momentum Entanglement}
			\label{sec:spin and momentum entanglement in inertial frames}

\subsubsection{Single - Particle States}\label{subsec:single particle states}

For our analysis momentum eigenstates, such as in Eq.~\eqref{eq:rel momentum eigenstates}, will be of great interest, but we cannot use the same normalization \eqref{eq:momentum eigenstates orthogonality} as in the non-relativistic case, because of the unitarity of the representation \eqref{eq:unitary rep of lorentz trafo 2}, which requires that
\begin{equation}
\langle\,p^{\,\prime},\sigma^{\,\prime}\,|\,p,\sigma\,\rangle\,=\,\langle\,\Lambda p^{\,\prime},\sigma^{\,\prime}\,|\,\Lambda p,\sigma\,\rangle\ .
\label{eq:unitarity of Lorentz trafos}
\end{equation}
Since the $3$-dimensional $\delta$-function is not Lorentz invariant, an additional normalization factor $2E_{p}$ (see \cite{peskinschroeder}) is chosen here to solve this problem, such that
\begin{equation}
\langle\,p^{\,\prime},\sigma^{\,\prime}\,|\,p,\sigma\,\rangle\,=\,2\,E_{p}\,\delta^{(3)}(\,\vec{p}^{\ \prime}\!-\vec{p}\:)\,\delta_{\,\sigma^{\prime}\sigma}\ ,
\label{eq:lorentz invariant momentum state normalization}
\end{equation}
where $E_{p}\,=\,p^{\,0}>0$ in the chosen representation, and we have
\begin{equation}
\sum\limits_{\sigma\,=\,\pm\tfrac{1}{2}}\,\int\!\frac{d^{3}p}{2\,E_{p}}\,\langle\,p^{\,\prime},\sigma^{\,\prime}\,|\,p,\sigma\,\rangle\,=\,1\ .
\label{eq:invariant integration}
\end{equation}
Since the momenta, which are integrated over, have to satisfy
\begin{equation}
E_{p}\,=\,\sqrt{m^{\,2}\,+\,\vec{p}^{\ 2}},
\label{eq:unit mass shell}
\end{equation}
the domain of integration is the unit mass shell. We will in the following use the abbreviation $d\mu(p)=\frac{d^{3}p}{2\,E_{p}}$. Furthermore we have assumed that the states $|\,p,\sigma\,\rangle$, and $|\,p^{\prime},\sigma^{\prime}\,\rangle$ describe the same sort of particle, e.g. both could describe electrons. If this is not the case, an additional special label $n$ can be added to the states, $|\,p,\sigma,n\,\rangle$, and $|\,p^{\,\prime},\sigma^{\,\prime},n^{\,\prime}\,\rangle$, together with $\delta_{n^{\prime}n}$ in \eqref{eq:lorentz invariant momentum state normalization}, to make this obvious.\\

The spin of the particles is given by the relations
\begin{equation}
\vec{S}^{\,2}\,|\,p,\sigma\,\rangle\,=\,s(s+1)\,|\,p,\sigma\,\rangle\ , \ \mbox{and}\ \ \
S^{z}\,|\,p,\sigma\,\rangle\,=\,\sigma\,|\,p,\sigma\,\rangle\ .
\label{eq:spin in wigner basis}
\end{equation}

If we wish to consider localized particles instead of plane waves, we have to introduce a distribution function $f_{\sigma}(p)$, and write down the single particle ket as
\begin{equation}
|\,\psi\,\rangle_{1-particle}\,=\,\sum\limits_{\sigma}\,\int\!d\mu(p)\,f_{\sigma}(p)\,|\,p,\sigma\,\rangle\ ,
\label{eq:general single particle ket}
\end{equation}
where the distribution function satisfies
\begin{equation}
\sum\limits_{\sigma}\,\int\!d\mu(p_{1})\,|\,f_{\sigma}(p)\,|^{2}\,=\,1\ .
\label{eq:single particle distribution function}
\end{equation}

With a particular choice of basis the spinor can then be displayed as
\begin{equation}
\psi(p)\,=\,\begin{pmatrix} f_{\up}(p) \\ f_{\down}(p) \end{pmatrix}\ ,
\label{eq:single particle state with basis choice}
\end{equation}
and the corresponding density matrix is (Ref.~\cite{peresetal02})
\begin{equation}
\rho\,(p,p^{\,\prime})\,=\,\begin{pmatrix}
\,f_{\up}(p)f^{\,*}_{\up}(p^{\,\prime}) 	& 	f_{\up}(p)f^{\,*}_{\down}(p^{\,\prime})\, \\[3mm]
\,f_{\down}(p)f^{\,*}_{\up}(p^{\,\prime})	&		f_{\down}(p)f^{\,*}_{\down}(p^{\,\prime})\,
\end{pmatrix}\ .
\label{eq:single particle density matrix}
\end{equation}

In quantum information theory often only the spin of a system is considered. Mathematically this means that the momentum degrees of freedom are being traced out from the density matrix, i.e. $p$ is set equal to $p^{\prime}$ and an integration over $p$ is performed in \eqref{eq:single particle density matrix}. This technique is in itself not problematic, it simply corresponds to the fact, that an application, or an experiment, might not incorporate, or require, all degrees of freedom of a physical system. There is however a problem in comparing the remaining reduced spin density matrices of different observers, related by Lorentz transformations.\\

From Eq.~\eqref{eq:unitary rep of lorentz trafo 2} we know how a momentum eigenstate transforms under a Lorentz transformation, i.e. the spin of the state is transformed by a momentum dependent Wigner rotation. If the Lorentz transformation is a pure rotation, the Wigner rotation is as well, and in particular does not depend on the momentum. Since pure rotations do not change the internal structure of a state, these transformations are of no interest. Without loss of generality we will therefore consider only pure boosts from now on. Let $\Lambda$ be such a pure boost, and $U(\Lambda)$ be the corresponding representation for the state \eqref{eq:general single particle ket}, where we will drop the labeling as a single particle state, since this is clear from the ket-notation, then
\begin{equation}
U(\Lambda)\,|\,\psi\,\rangle\,=\,\sum\limits_{\sigma}\,\int\!d\mu(p)\,f_{\sigma}(p)\,U(\mathrm{W}(\Lambda,p))\,|\,\Lambda p,\sigma\,\rangle\ .
\label{eq:general single particle transformation}
\end{equation}

An observer, for whom the state is given by \eqref{eq:general single particle transformation}, will loose information about the spin state, when tracing out over the momentum. This means that the reduced density matrix for the spin degree of freedom becomes a mixed state and therefore two observers, related by a Lorenz boost, will not agree on the entropy of the reduced spin state. In Ref.~\cite{peresetal02} Peres, Scudo, and Terno study such a situation\footnote{See also the comment by Czachor in Ref.~\cite{czachor05}.} for a particle of mass $m$, spin along the $z$ direction, and a Gaussian distribution of width $w$, centered on some fixed value $p$, in momentum space, which is viewed by an observer moving in a perpendicular direction. The authors find, that the von Neumann entropy of the (reduced) spin state increases from $S(\rho_{spin})=0$ in the unboosted frame to
\begin{equation}
S(\rho^{\,\Lambda}_{spin})\,=\,\frac{w^{2}}{8m^{2}}\,\tanh^{2}(\frac{\alpha}{2})\,\left(1\,-\,\ln(\frac{w^{2}}{8m^{2}}\,\tanh^{2}(\frac{\alpha}{2}))\,\right)\
\label{eq:entropy of reduced spin density matrix of peresetal02}
\end{equation}
in the frame related by the boost $\Lambda$, with rapidity $\alpha$. The result only depends on the mass, the strength of the boost, and the width of the distribution, i.e. for a sharp momentum, $w\rightarrow0$, the entropy remains unchanged. Also, even though here $S(\rho^{\,\Lambda}_{spin})\geq S(\rho_{spin})$, this need not generally be the case for all pairs of observers, but we can only conclude, that the reduced density operator has no covariant transformation law.\\

The mass dependence in Eq.~\eqref{eq:entropy of reduced spin density matrix of peresetal02} can be explained by the choice of the momentum (distribution) in the initial state. The angle of the Wigner rotation, \eqref{eq:rotation angle for boosts in x and z direction} depends only on the boost velocities, i.e. on the velocity $\vec{v}$ of the particle in the initial frame, and on the velocity $\vec{w}$ relating this frame to the boosted observer. If a particle of higher mass $m^{\prime}>m$, but with the same momentum $p$ (and momentum distribution) is chosen, the boost taking the rest frame momentum to $p$ requires a smaller velocity $\vec{v}^{\,\prime}$, and consequently also the Wigner rotation angle is smaller.

\subsubsection{Two - Particle States}\label{subsec:two particle states}

The result of Ref.~\cite{peresetal02} clearly suggests that consequences for entangled systems are to be expected when they are viewed by a Lorentz transformed frame. To further investigate entanglement, e.g. in the context of a Bell inequality, two (or more) particles are needed. Therefore we now continue by specifying two-particle states. It is important at this point that we carefully explain the notation used in this work, since it is subtly different than that, used by some other authors, e.g. in Ref.~\cite{alsingmilburn02}. The difference is best illustrated in the two particle case, which can then be used for a generalization to arbitrary particle numbers. Let us therefore write down a two-particle Hilbert space vector as
\begin{equation}
|\,p_{1},\sigma_{1};\,p_{2},\sigma_{2}\,\rangle\,=\,|\,p_{1},\sigma_{1}\,\rangle\,\otimes\,|\,p_{2},\sigma_{2}\,\rangle\ ,
\label{eq:two particle ket}
\end{equation}
where $p_{\,1,2}$ and $\sigma_{\,1,2}$ are the momenta and spins of particle $1$, and particle $2$, respectively. Each of the single particle states on the right hand side is normalized as the states in Eq.~\eqref{eq:lorentz invariant momentum state normalization}, such that
\begin{equation}
\sum\limits_{\sigma_{1},\sigma_{2}}\,\int\!d\mu(p_{1},p_{2})\,\langle\,p_{1},\sigma_{1};\,p_{2},\sigma_{2}\,|\,p_{1},\sigma_{1};\,p_{2},\sigma_{2}\,\rangle\,=\,1\ ,
\label{eq:two particle state normalization}
\end{equation}
where we used the abbreviation $d\mu(p_{1},p_{2})=d\mu(p_{1})d\mu(p_{2})$. However, Eq.~\eqref{eq:two particle ket} is not necessarily a valid state description for arbitrary particles, especially if the particles are indistinguishable. In particular, if we wish to describe two identical fermions, e.g. a pair of electrons, we need to antisymmetrize the state, i.e. a meaningful description of two indistinguishable fermions must then be of the form
\begin{equation}
\frac{1}{\sqrt{2}}\,|\,p_{1},\sigma_{1};\,p_{2},\sigma_{2}\,\rangle\,-\,\frac{1}{\sqrt{2}}\,|\,p_{2},\sigma_{2};\,p_{1},\sigma_{1}\,\rangle\,\ .
\label{eq:two fermion state}
\end{equation}
We will keep this problem in mind when studying the results of our calculations, but will not restrict the analysis to identical particles. Also, we do not explicitly introduce species labels, but will rather assume that for the single particle kets in a state like \eqref{eq:two fermion state}, the ordering of the kets labels the species. As in the single particle case \eqref{eq:general single particle ket}, a general two particle state is a superposition of different vectors \eqref{eq:two particle ket}, i.e. we can write it as
\begin{equation}
|\,\psi\,\rangle_{2-particle}\,=\,
\sum\limits_{\sigma_{1},\sigma_{2}}\,\int\!d\mu(p_{1},p_{2})\,
f_{\sigma_{1}\sigma_{2}}(p_{1},p_{2})\,|\,p_{1},\sigma_{1};\,p_{2},\sigma_{2}\,\rangle\ ,
\label{eq:general two particle state}
\end{equation}
where the distribution function is normalized analogously to \eqref{eq:single particle distribution function},
\begin{equation}
\sum\limits_{\sigma_{1},\sigma_{2}}\,\int\!d\mu(p_{1},p_{2})\,
|\,f_{\sigma_{1}\sigma_{2}}(p_{1},p_{2})\,|^{2}\,=\,1\ .
\label{eq:general two particle state distribution function}
\end{equation}

A similar approach as in Ref.~\cite{peresetal02} (see Sec.~\ref{subsec:single particle states}) is chosen by Gingrich, and Adami in Ref.~\cite{gingrichadami02} for a two-particle state. They consider two particles desribed by a state of the form \eqref{eq:general two particle state} with a distribution function\footnote{We have renamed the quantities corresponding to the notation in this work.}
\begin{equation}
f_{\sigma_{1}\sigma_{2}}(p_{1},p_{2})\,=\,
\frac{1}{\sqrt{2}}\,\delta_{\sigma_{1}\sigma_{2}}\,g(p_{1})\,g(p_{2})\ ,
\label{eq:gingrich adami distribution function}
\end{equation}
where the functions $g(p_{1}),g(p_{2})$ are Gaussian distributions of width $w$, centered around some momentum values for both particles. Such a distribution represents the Bell state $|\,\phi^{+}\,\rangle$ with momenta in a product Gaussian distribution.\\

A Lorentz boost of rapidity $\xi$ is then performed, entangling the spins with the momenta, before subsequently the momentum is traced out of the corresponding density matrix, as explained in Sec.~\ref{subsec:single particle states}. The authors then proceed by quantifying the entanglement remaining in the reduced spin density matrix by the concurrence (see Eq.~\eqref{eq:concurrence}). They find that the concurrence decreases with increasing parameter $\tfrac{w}{m}$, where the mass dependence can be explained as before.\\

This means that for increasing width of the momentum space distribution, the information loss due to the partial trace operation becomes bigger and therefore the spin entanglement decreases. However, for appropriately narrow distributions, $\tfrac{w}{m}\gtrapprox3.377$, the concurrence in the boosted reference frame saturates at nonzero values in the limit $\xi\rightarrow\infty$. This is due to the fact that the Wigner angle $\delta$ is bounded, $0\leq\delta\leq\tfrac{\pi}{2}$.\\

Probably the most important consequence of this example is the entanglement transfer between spin and momentum degrees of freedom due to a Lorentz transformation. Clearly, by applying the inverse transformation to the total state, the initial state can be recovered. This means that the spin entanglement, as measured by the entanglement of the reduced spin density matrix, can increase under a Lorentz transformation. The circumstances under which this can occur, are limited by the following theorem, which was formulated and proven\footnote{The original proof features only two particles, but the number of particles is non-essential to the theorem.} in Ref.~\cite{gingrichadami02}.

\begin{theorem}\label{thm:gingrichadamitheorem}
    \begin{tabbing} \hspace*{3.1cm}\=\hspace*{3cm}\=\kill
            \> The spin entanglement of a pure multi-particle state can\\[2mm]
            \> only increase under Lorentz transformations, if the initial\\[2mm]
            \> entanglement between spin and momentum is nonzero.
    \end{tabbing}
\end{theorem}
\pa{Proof:}\ \\
To proof this, assume the initial state is separable with respect to the partition into spin and momentum, i.e. the $n$-particle state $|\,\Psi\,\rangle^{1\cdots n}$ is given by a product state of momentum degrees of freedom, described by $|\,\psi\,\rangle^{1\cdots n}_{mom}$, and spin degrees of freedom, described by $|\,\phi\,\rangle^{1\cdots n}_{spin}$,
\begin{equation}
|\,\Psi\,\rangle^{1\cdots n}\,=\,|\,\psi\,\rangle^{1\cdots n}_{mom}\,|\,\phi\,\rangle^{1\cdots n}_{spin}\ ,
\label{eq:spin momentum separable state}
\end{equation}
where we omitted the super- and subscripts for ease of notation. Applying a Lorentz transformation on this state and calculating the reduced spin density matrix $\rho^{\,\Lambda}_{spin}$ results in
\begin{equation}
\rho^{\,\Lambda}_{spin}\,=\,\sum\limits_{\mathrm{i}}\,p_{\,\mathrm{i}}\,
U^{\,\mathrm{i}}_{1}\otimes\,\cdots\,\otimes U^{\,\mathrm{i}}_{n}\,
|\,\phi\,\rangle\langle\,\phi\,|\,
U^{\,\mathrm{i}\,\dagger}_{1}\otimes\,\cdots\,\otimes U^{\,\mathrm{i}\,\dagger}_{n}\ ,
\label{eq:red spin density of boosted n particle state}
\end{equation}
where the sum in \eqref{eq:red spin density of boosted n particle state} can be an integral, depending on the choice of $|\,\psi\,\rangle^{1\cdots n}_{mom}$ in \eqref{eq:spin momentum separable state}. Any suitable entanglement measure $E(\rho)$ must be a convex function, which means that the entanglement of the convex sum of density matrices must always be less than, or equal to, the convex sum of the entanglement of the individual density matrices,
\begin{equation}
E(\,\sum\limits_{\mathrm{i}}\,p_{\,\mathrm{i}}\,\rho^{\,\mathrm{i}}\,)\,\leq\,\sum\limits_{\mathrm{i}}\,p_{\,\mathrm{i}}\,E(\rho^{\,\mathrm{i}})\ .
\label{eq:convexity of entanglement measure}
\end{equation}
Using this and the fact that rotations in the state space do not change entanglement measures, we get for the reduced density matrix in Eq.~\eqref{eq:red spin density of boosted n particle state}:
\begin{eqnarray}
E(\rho^{\,\Lambda}_{spin}) &\leq& \sum\limits_{\mathrm{i}}\,p_{\,\mathrm{i}}\,E(\,U^{\,\mathrm{i}}_{1}\otimes\,\cdots\,\otimes
		U^{\,\mathrm{i}}_{n}\,|\,\phi\,\rangle\langle\,\phi\,|\,
		U^{\,\mathrm{i}\,\dagger}_{1}\otimes\,\cdots\,\otimes U^{\,\mathrm{i}\,\dagger}_{n}\,)\,=\nonumber\\[2mm]
&=&	\sum\limits_{\mathrm{i}}\,p_{\,\mathrm{i}}\,E(\,|\,\phi\,\rangle\langle\,\phi\,|\,)\,=\,
		E(\,|\,\phi\,\rangle\langle\,\phi\,|\,)\ ,
\label{eq:gingrichadami proof part 1}
\end{eqnarray}
where the last step follows from the normalization of the reduced density matrix, and since $|\,\phi\,\rangle\langle\,\phi\,|$ is the reduced spin density matrix of the initial state, Theorem \ref{thm:gingrichadamitheorem} is proven.\qed

The question remains wether or not there is some Lorentz invariant entanglement, and if so, how it can be specified. This question was partly addressed by Alsing, and Milburn in Ref.~\cite{alsingmilburn02} and by Czachor in Ref.~\cite{czachor97}, by noting that the overall entanglement of a state with sharp momenta should not change, due to the \emph{local unitary} character of the Wigner rotation, i.e. a Lorentz transformation $\Lambda$ takes the state \eqref{eq:general two particle state} to
\begin{equation}
|\,\psi\,\rangle^{\Lambda}\,=\,\sum\limits_{\sigma_{1},\sigma_{2}}\,\int\!d\mu(p_{1},p_{2})\,
f_{\sigma_{1}\sigma_{2}}(p_{1},p_{2})\,
U(\Lambda,p_{1})\otimes U(\Lambda,p_{2})\,|\,\Lambda p_{1},\sigma_{1};\,\Lambda p_{2},\sigma_{2}\,\rangle \ ,
\label{eq:general lorentz transformed two particle state}
\end{equation}
which can be found to have a similar form as the LOCC twirling operation in Eq.~\eqref{eq:twirling}, which suggests that the entanglement between the two particles cannot be increased by a Lorentz transformation. Apparently because of this Gingrich, and Adami in Ref.~\cite{gingrichadami02} claimed: ``\emph{While spin and momentum entanglement separately are not Lorentz invariant, the joint entanglement of the wave function is.}"\\

The question was however raised again by Jordan, Shaji, and Sudarshan in Ref.~\cite{jordanetal06b}, where the authors analyzed different pure and mixed states under Lorentz transformations, and stated ``\emph{From the entanglements considered, no sum of entanglements is found to be unchanged.}". This conclusion is illustrated with examples, e.g. where the reduced density matrices of momentum and spin, of an initially pure overall state, both become mixed, separable states, which is interpreted as the Lorentz transformation removing both spin and momentum entanglement completely. We will shortly return to this special case and try to solve the apparent dilemma.\\

To simplify the investigation, we want to make a few preliminary assumptions and introduce notational conventions. We have already seen that the width of the distributions in momentum space influences the entanglement of the reduced spin density matrix. Clearly, the finite width of these distributions is a necessity from the physical point of view, since we want to normalize the states, but on the other hand, this makes the formulation of the problem difficult. We will therefore adopt the convention of \cite{jordanetal06b} and assume that all momentum distributions are sufficiently narrow to result in single Wigner rotations. Additionally we assume that all of the chosen distributions are centered around values far enough apart from each other in momentum space, such that the distributions do not overlap and the states of different momenta can formally be said to satisfy the orthogonality relation
\begin{equation}
\langle\,p^{\,\prime}\,|\,p\,\rangle\,=\,\delta_{p^{\,\prime}p}\ .
\label{eq:orthogonality of momentum distributions}
\end{equation}
It must be kept in mind that this is a gross simplification, but the alternative, products of $\delta$-distributions evaluated at the same arguments, is equally ill-defined. We will therefore use this remedy from now on, and treat the momentum states as a discrete\footnote{A review of entanglement in continuous variables can be found in Ref.~\cite{parkerboseplenio00}.} basis, unless mentioned otherwise. We will furthermore assume that the overall state is such, that the spin and momentum degrees of freedom of the two particles are initially separable from each other. i.e.
\begin{equation}
\left|\,\psi\,\right\rangle_{\mathrm{total}}\,=\,\left|\,\psi\,\right\rangle_{\mathrm{mom}}\, \left|\,\psi\,\right\rangle_{\mathrm{spin}}\,.
	\label{eq:initial total state}
\end{equation}
For our analysis two of the above mentioned momentum distributions will suffice, which will be momenta $p_{\pm}$ along the $z$-axis, describing two particles moving in opposite directions (in the rest frame of their source), such that the joint momentum state is given by
\begin{equation}
\left|\,\psi\,\right\rangle_{\mathrm{mom}}\,=\,\cos\alpha\left|\,p_{+},p_{-}\,\right\rangle\,+
\,\sin\alpha\left|\,p_{-},p_{+}\,\right\rangle\ ,
	\label{eq:initial momentum state}
\end{equation}
where the (real) angle $\alpha$ parameterizes all possible momentum states of this kind. The major simplification of this momentum notation now becomes apparent, since only two possible values, corresponding to orthogonal (in the sense of Eq.~\eqref{eq:orthogonality of momentum distributions}) states, are allowed, and we can therefore regard the momentum state in the qubit formalism, where $\left|\,p_{+},p_{-}\,\right\rangle$ can be treated analogously to the qubit state $\left|\,\up\down\,\right\rangle$. Using the momentum state \eqref{eq:initial momentum state}, the initial state \eqref{eq:initial total state} is transformed into  $\left|\,\psi\,\right\rangle^{\,\Lambda}_{\mathrm{total}}$, which is of the form
\begin{eqnarray}
	\left|\,\psi\,\right\rangle^{\,\Lambda}_{\mathrm{total}} &=&
		\,\cos\alpha\left|\,\Lambda p_{+},\Lambda p_{-}\,\right\rangle\,
		\left(U_{+}\otimes\,U_{-}\right)\,\left|\,\psi\,\right\rangle_{\mathrm{spin}} \nonumber \\
	&+& \,\sin\alpha\left|\,\Lambda p_{-},\Lambda p_{+}\,\right\rangle\,
		\left(U_{-}\otimes\,U_{+}\right)\,\left|\,\psi\,\right\rangle_{\mathrm{spin}}\ ,
	\label{eq:boosted general total state}
\end{eqnarray}
where $U_{\pm}=U(\mathrm{W(\Lambda,p_{\pm})})$ are the spin $\tfrac{1}{2}$ representations of the Wigner rotations, corresponding to the momenta $p_{\pm}$, and Lorentz transformation $\Lambda$. It can be immediately seen here that generally, the boosted state $\left|\,\psi\,\right\rangle^{\,\Lambda}_{\mathrm{total}}$ will not factorize into spin and momentum degrees of freedom. Only if the rotated spin states are equal or the parameter $\alpha$ is chosen such that $\sin\alpha$ or $\cos\alpha$ vanish, will the entanglement between spin and momentum remain unchanged. However, since the operation performed on the spin state cannot be written as a single tensor product of unitary operations on the corresponding Hilbert spaces, this conclusion is not unexpected. It presents a similar situation as a double controlled unitary gate\footnote{More information about quantum gates can be found in Ref.~\cite{nielsenchuangQI}}, where the two control qubits as well as the two input qubits can be entangled (for appropriately chosen parameters $\alpha$, and $\beta$).\\

To make this assertion more precise, let us examine it for particular choices of initial spin states $\left|\,\psi\,\right\rangle_{\mathrm{spin}}$, beginning with linear combinations of the spin states $|\!\up\down\,\rangle$, and $|\!\down\up\,\rangle$, which can be viewed as superpositions of the Bell states $\left|\,\psi^{\,\pm}\,\right\rangle$ \eqref{eq:bell psi states} and will therefore be called \emph{Bell $\psi^{\,\pm}$ states}, before continuing to superpositions of the \emph{spin triplet states}. For all states the spin quantization axis is chosen to be the $z$-axis.\\

\pa{Bell $\psi^{\,\pm}$ spin states:}\ \\

For the linear combinations of the states $\left|\,\psi^{\,\pm}\,\right\rangle$ of Eq.~\eqref{eq:bell psi states} a similar parametrization as for the momentum state \eqref{eq:initial momentum state} is utilized
\begin{equation}
\left|\,\psi\,\right\rangle_{\mathrm{spin}}\,=\,\cos\beta\left|\,\uparrow\,\downarrow\,\right\rangle\,+\,
\sin\beta\left|\,\downarrow\,\uparrow\,\right\rangle\ ,
	\label{eq:initial spin state bell type}
\end{equation}
such that the observer in the initial reference frame describes the total state as
\begin{equation}
\left|\,\psi\,\right\rangle_{\mathrm{total}}\,=\,\left(\,\cos\alpha\left|\,p_{+},p_{-}\,\right\rangle\,+\,
\sin\alpha\left|\,p_{-},p_{+}\,\right\rangle\ \right)
\left(\,\cos\beta\left|\,\uparrow\,\downarrow\,\right\rangle\,+\,
\sin\beta\left|\,\downarrow\,\uparrow\,\right\rangle\ \right)\ ,
	\label{eq:initial total state bell type spin}
\end{equation}
with the corresponding density operator
\begin{equation}
\rho\,=\,\left|\,\psi\,\right\rangle\left\langle\,\psi\,\right| \ \ ,
\label{eq:initial state Bell type spin density matrix}
\end{equation}
where $\left|\,\psi\,\right\rangle$ is given by \eqref{eq:initial total state bell type spin}. Since the overall state is a pure state, we can measure the amount of entanglement distributed between the two spin- and the two momentum-qubits by the sum of the linear entropies (see Def.~\ref{def:linear entropy}) of the subsystems in the chosen partition, i.e.
\begin{equation}
E(\rho)\,=\,\sum\limits_{i}\,\left(\,1-\rm{Tr}\,\rho_{\,i}^{\,2}\,\right) \ ,
 \label{eq:linear entropy}
\end{equation}
where $\rho_{\,i}$ is obtained by tracing over all subsystems except the $i$-th. In the following, this measure will be used to quantify the entanglement of the partitions into four individual subsystems, into spin and momentum degrees of freedom, and into the degrees of freedom of the two particles involved.\\

Beginning with the \emph{partition into four individual qubits}, i.e. we calculate the four reduced density matrices of each qubit, by tracing out the respective three other qubits, we get the total amount of entanglement of the state \eqref{eq:initial total state bell type spin}, as quantified by the sum the linear entropies to be
\begin{equation}
E(\rho)\;=\;\tfrac{1}{2}\left(\,2\,-\,\cos(4\alpha)\,-\,\cos(4\beta)\,\right) \ .
\label{eq:linear entropy of unboosted bell type state}
\end{equation}
This result is in accordance with the construction of our state, i.e. if $\alpha=\beta=\tfrac{(2n+1)\pi}{4}$, the entanglement is maximal, which corresponds to Bell states $\left|\,\psi^{\,\pm}\,\right\rangle$ for both spin and momentum, whereas if $\alpha=\beta=\tfrac{n\pi}{2}$, the linear entropy vanishes and the initial state is fully separable.\\

Let us now introduce a second observer moving in the $x$-direction with velocity $\vec{w}$. The associated reference frame is related to the initial reference frame by a boost $\Lambda$, which can be obtained by replacing $v_{\mathrm{x}}$ with $|\vec{w}\,|$ in Eq.~\eqref{eq:boost in x direction}. Since the particles have momenta in the $\pm z$-direction, the resulting Wigner rotations are around the $(-y)$-axis, as discussed at the end of Sec.~\ref{subsec:wigners little group}, about angles $\pm \delta$ (shown in Fig.~\ref{fig:wignerangle}) respectively. Consequently, the Wigner rotation matrices $U_{\pm}$ of Eq.~(\ref{eq:boosted general total state}) are given by
\begin{equation}
U_{\pm}\,=\,\begin{pmatrix} \ \ \,\cos\tfrac{\delta}{2} & \pm\sin\tfrac{\delta}{2}\  \\[3mm]
														\mp\sin\tfrac{\delta}{2} & \ \ \,\cos\tfrac{\delta}{2}\ \end{pmatrix}\ ,
\label{eq:rotation matrices}
\end{equation}
which can be easily seen from Eq.~\eqref{eq:rotated vector as matrix}. Using these rotations in the formula for the boosted overall state, Eq.~\eqref{eq:boosted general total state}, we get
\begin{eqnarray}
&\left|\,\psi\,\right\rangle^{\,\Lambda}_{\mathrm{total}}\,=\,
		\,\cos\alpha\left|\,\Lambda p_{+},\Lambda p_{-}\,\right\rangle\,
		\left[\,c_{1}\,\left|\,\phi^{\,+}\,\right\rangle\,+
		\,c_{2}\,\left|\,\uparrow\downarrow\,\right\rangle\,+
		\,c_{3}\,\left|\,\downarrow\uparrow\,\right\rangle\,\right]+ \nonumber \\[2mm]
\ \ \ 	&+\,\sin\alpha\left|\,\Lambda p_{-},\Lambda p_{+}\,\right\rangle\,
		\left[\,-c_{1}\,\left|\,\phi^{\,+}\,\right\rangle\,+
		\,c_{2}\,\left|\,\uparrow\downarrow\,\right\rangle\,+
		\,c_{3}\,\left|\,\downarrow\uparrow\,\right\rangle\,\right]	\,,
\label{eq:boosted bell type spin total state}
\end{eqnarray}
where $|\,\phi^{\,+}\,\rangle$ is one of the Bell states of Eq.~\eqref{eq:bell phi states}. The constants $c_{1}$, $c_{2}$, and $c_{3}$ in \eqref{eq:boosted bell type spin total state} are given by
\begin{eqnarray}
c_{1} &=& \tfrac{1}{\sqrt{2}}\sin\delta\,(\sin\beta-\cos\beta)\ ,
			\label{eq:boosted bell type state constant 1}\\[2mm]
c_{2} &=& \cos\beta\,\cos^{2}\!\tfrac{\delta}{2}+\sin\beta\,\sin^{2}\!\tfrac{\delta}{2}\ ,
			\label{eq:boosted bell type state constant 2}\\[3mm]
c_{3} &=& \sin\beta\,\cos^{2}\!\tfrac{\delta}{2}+\cos\beta\,\sin^{2}\!\tfrac{\delta}{2}\ .
			\label{eq:boosted bell type state constant 3}
\end{eqnarray}

Calculating the linear entropy of the four reduced density matrices of
\begin{equation}
\rho^{\,\Lambda}\,=\,\left|\,\psi\,\right\rangle^{\Lambda\,\Lambda\!\!}\left\langle\,\psi\,\right|\
\label{eq:boosted bell type spin total density matrix}
\end{equation}
in the same manner, as has been done to gain Eq.~\eqref{eq:linear entropy of unboosted bell type state} for the unboosted state, we find in this case the result
\begin{eqnarray}
E(\rho^{\,\Lambda}) &\;=\;& \tfrac{1}{16}\left(\right.18\,-\,10\cos(4\alpha)\,-\,6\cos(4\beta) 		
		\,-\,2\cos(4\alpha)\cos(4\beta)	\nonumber \\[2mm]
 && -\,8\cos(2\delta)\sin^{2}(2\alpha)\cos^{2}(2\beta)\left.\right)\ .
\label{eq:linear entropy of boosted bell type state}
\end{eqnarray}

It can be easily verified, that this expression reproduces the result of Eq.~\eqref{eq:linear entropy of unboosted bell type state} in the case where the Wigner angle goes to zero, $\delta \rightarrow 0\,$. Since we were sure that the state changes to begin with, we now need to compare the linear entropies of the two reference frames in order to quantify this observation. We do this by calculating the difference of these two measures, which gives the simple formula
\begin{equation}
E(\rho^{\,\Lambda})-E(\rho)\;=\;\sin^{2}\!\delta\,\sin^{2}(2\alpha)\,\cos^{2}(2\beta)\ .
\label{eq:linear entropy difference of boosted and unboosted bell type state}
\end{equation}

As expected, if the Wigner rotation angle vanishes, $\delta=0$, also the difference in linear entropies vanishes.
Analyzing the result of Eq.~\eqref{eq:linear entropy difference of boosted and unboosted bell type state}, we find the peculiar feature that the overall entanglement of this partition of the Hilbert space \emph{does} generally change. Furthermore, this change strongly depends on the choice of initial state, and less surprisingly, on  the strength of the boost, and the velocities of the particles.\\

\begin{figure}
	\centering
		\includegraphics[width=0.80\textwidth]{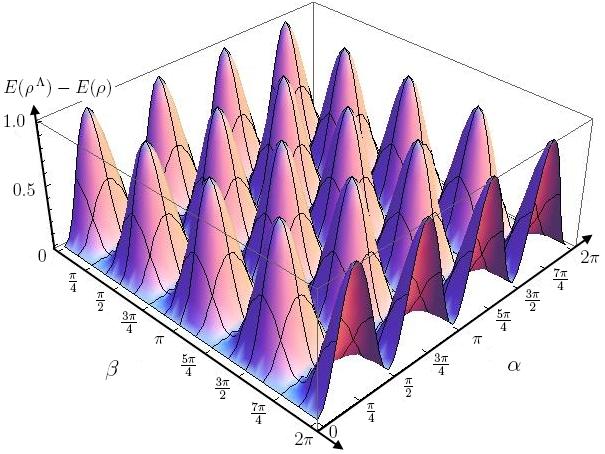}
	\caption{Entanglement-Egg-Tray: Difference of the linear entropies of a $\delta =\pm\tfrac{\pi}{2}$ Wigner rotated Bell-type two particle state in the case of a partition into 4 qubits. Plot of Eq.~\eqref{eq:linear entropy difference of boosted and unboosted bell type state}.}
	\label{fig:entanglementchangebelltype}
\end{figure}

Consider for instance the case, where the initial momentum state is separable, i.e. $\alpha=\tfrac{n\pi}{2}$, where $n\in\mathbb{N}$. The Wigner rotation in this case is just a single product of unitary operations on the spin space and there is no entanglement change due to the boost. On the other hand, the more entangled the state $\left|\,\psi\,\right\rangle_{\mathrm{mom}}$ is initially, the higher the increase in the difference of the linear entropies, producing the striking ``\emph{egg-tray}" pattern in Fig.~\ref{fig:entanglementchangebelltype}.\\

Simultaneously, it must be noticed, that the increase in entanglement is limited by the amount of entanglement, already contained in the initial spin state. If we consider e.g. the maximally entangled Bell states $\left|\,\psi^{\,\pm}\,\right\rangle\,$ for the spins, which we get by choosing $\beta=\tfrac{(2n+1)\pi}{4}$, the entanglement does not change regardless of the Wigner rotation angle or the choice of $\alpha$ in the momentum state. This example was previously given by Chakrabarti in Ref.~\cite{chakrabarti09}, where the momentum state is separable and the spin state is totally antisymmetric. However, by decreasing the spin entanglement of the initial state, an increase in the entropy difference can be obtained (see Fig.~\ref{fig:entanglementchangebelltype}).\\
		
We continue with the \emph{partition into momentum- and spin-degrees of freedom}, i.e. the 4 qubit Hilbert space is separated into two subspaces of two qubits each, one for the momentum qubits, and one for the spin qubits. From here on we can proceed in an analogue manner as before. The reduced density matrices for the two spin- or momentum-degrees of freedom are calculated from the total density matrix \eqref{eq:initial state Bell type spin density matrix}, constructed with the use of \eqref{eq:initial total state bell type spin}, and \eqref{eq:initial state Bell type spin density matrix}, i.e.
\begin{equation}
\rho_{\mathrm{spin}}\,=\,\mathrm{Tr}_{\mathrm{mom}}(\rho)\ ,\ \ \mbox{and}\ \  \rho_{\mathrm{mom}}\,=\,\mathrm{Tr}_{\mathrm{spin}}(\rho)\ .
\label{eq:reduced density matrices for spin and momentum}
\end{equation}

The entanglement of the initial total state with respect to this partition must be identically zero, $E(\rho)=0$, since we constructed \eqref{eq:initial total state bell type spin} to factorize into spin and momentum. When repeating this procedure for the Lorentz transformed reference frame \eqref{eq:boosted bell type spin total state}, we find that the entanglement with respect to this partition does not vanish for all configurations of the initial state, it is
\begin{equation}
E(\rho^{\,\Lambda})\;=\;\frac{1}{2}\,
\sin^{2}\!\delta\,\sin^{2}(2\alpha)\,\left(1\,-\,\sin(2\beta)\right)\,\left[3\,+\,\cos(2\delta)\,+\,
2\sin^{2}\!\delta\sin(2\beta)\right] \ .
\label{eq:spin vs momentum entanglement boosted bell type state}
\end{equation}

\begin{figure}[ht]
	\centering
		\includegraphics[width=0.70\textwidth]{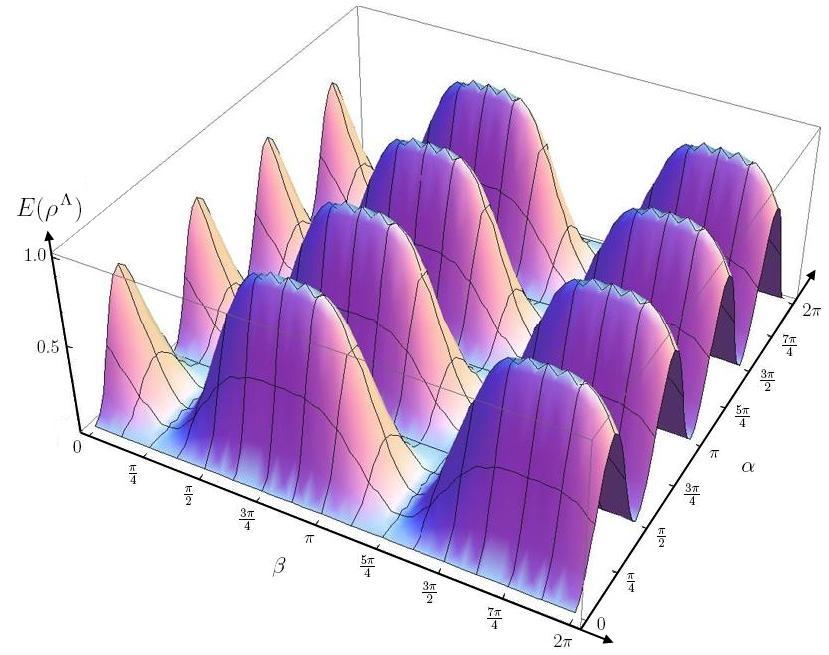}
	\caption{Entanglement change of a $\delta =\pm\tfrac{\pi}{4}$ Wigner rotated spin-Bell-type state in the partition into spin- and momentum-degrees of freedom . Plot of Eq.~\eqref{eq:spin vs momentum entanglement boosted bell type state}.}
	\label{fig:spinmomentumentanglementbelltype}
\end{figure}

Obviously, the result of Eq.~\eqref{eq:spin vs momentum entanglement boosted bell type state} shows a similar dependence on the Wigner rotation angle $\delta$ and the parametrization of the initial momentum state, as the result \eqref{eq:linear entropy difference of boosted and unboosted bell type state} of the partition into four individual qubits. But although the valleys of Fig.~\ref{fig:spinmomentumentanglementbelltype} and Fig.~\ref{fig:entanglementchangebelltype} agree for $\alpha=\frac{n\pi}{2}$, the entanglement change due to the boost is no longer zero for all values $\beta=\frac{(2n+1)\pi}{4}$ in Fig.~\ref{fig:spinmomentumentanglementbelltype}. This points to an imbalance between the Bell states $|\psi^{\,-}\rangle$ and $|\psi^{\,+}\rangle$ in this partition, which was not present in the first partition, since the overall entanglement does not change for either of the two, while the entanglement distributed between momentum and spin is only invariant for $\beta=\frac{(4n+1)\pi}{4}$, i.e. for the symmetric state $|\psi^{\,+}\rangle$.\\

The situation where $\alpha=\beta=\tfrac{\pi}{4}$ corresponds to the pure state example presented by Jordan, Shaji, and Sudarshan in Ref.~\cite{jordanetal06b}. There a Wigner rotation of $\delta=\tfrac{\pi}{4}$ causes the reduced spin and momentum density matrices of the pure state at hand to be mixtures of two products of pure states each. There is no entanglement present in the reduced density matrices in this example, however, the boost entangles the spin degrees of freedom with the momentum degrees of freedom, giving rise to a change in the overall entanglement in this partition, as can be seen in Fig.~\ref{fig:spinmomentumentanglementbelltype}. We therefore have to agree with Jordan, Shaji, and Sudarshan in so far, as the entanglement with respect to the partition into spin and momentum degrees of freedom cannot generally be claimed to be unchanged by Lorentz transformations, but we also soon will find a sum of entanglements, which is unchanged.\\

Another remarkable feature of this partition can be found, by considering the limit of both observer- and particle-velocity approaching the speed of light, corresponding to $\delta\rightarrow\tfrac{\pi}{2}\,$ (see Fig.~\ref{fig:wignerangle}). Then the formulas \eqref{eq:spin vs momentum entanglement boosted bell type state}, and \eqref{eq:linear entropy difference of boosted and unboosted bell type state} coincide, reproducing the entanglement-egg-tray of Fig.~\ref{fig:entanglementchangebelltype}. As of now we cannot tell, wether or not the equality of the entanglement change of the two partitions in this limit is pure coincidence or has some physical meaning.\\

At last we study the physically most appealing partition, which is the separation of the Hilbert space into the subspaces of the two individual particles, which we want to call the \emph{Alice-Bob partition}. Since each particle subspace consists of one spin- and one momentum-qubit, the reduced density matrix for Alice's subsystem is obtained by tracing over the complementary subspace of momentum and spin of Bob's particle, and vice versa,
\begin{equation}
\rho^{\,A}_{\mathrm{mom-spin}}\,=\,\mathrm{Tr}^{\,B}_{\mathrm{mom-spin}}(\rho)\ ,\ \ \ \rho^{\,B}_{\mathrm{mom-spin}}\,=\,\mathrm{Tr}^{\,A}_{\mathrm{mom-spin}}(\rho) \ \ \,.
\label{eq:reduced density matrices for alice and bob}
\end{equation}

The calculation of the corresponding linear entropies is straightforward, giving the result
\begin{equation}
E(\rho)\;=\;\tfrac{1}{8}\left[\,10\,-\,\left(3+\cos(4\alpha)\right)\,\left(3+\cos(4\beta)\right)\,\right] \ .
\label{eq:linear entropy of unboosted bell type spin-mom state}
\end{equation}

As before, a fully separable initial state, obtained by setting $\alpha=\beta=\tfrac{n\pi}{2}$, causes this expression to vanish, while maximal entanglement of spin and momentum, e.g. for the Bell states, where $\alpha=\beta=\tfrac{(2n+1)\pi}{4}$, gives $E=\frac{3}{2}$, corresponding to maximally mixed reduced density matrices for Alice and Bob,
\begin{equation}
\rho^{\,A}_{\mathrm{mix}}\,=\,\rho^{\,B}_{\mathrm{mix}}\,=\,\frac{1}{4}\mathds{1}_{\!4}\ .
\label{eq:maximally mixed reduced density matrices for Alice and Bob}
\end{equation}

Transforming our state to the reference frame of the observer moving in the $x$-direction as before, we find that the \emph{entanglement with respect to the Alice-Bob partition remains unchanged}, i.e. $E(\rho^{\,\Lambda})=E(\rho)$, regardless of the parametrization of the state, the momenta of the particles, or the strength of the boost. This is in complete agreement with the maintained violation of a Bell-inequality (see Sec.~\ref{sec:degree of violation of bell inequalities}), which is sensitive to exactly this partition of the Hilbert space and the local unitarity of the transformation in \eqref{eq:general lorentz transformed two particle state}.\\

A peculiarity of this partition is however that when tracing over spin and momentum it does not matter to which particle the spin and momentum are associated to. We obtain the same result, \eqref{eq:linear entropy of unboosted bell type spin-mom state}, \emph{no change in entanglement}, when we trace over the spin of Alice's particle and momentum of Bob's particle (or the other way around).\\

\pa{Spin triplet states:}\ \\

Although the results we have obtained so far apply to a broad class of states, there are still many possibilities to construct different spin states and it might turn out, that they behave quite differently under the effect of the Wigner rotations. Therefore, in order to test our results, we analyze another class of initial spin states, which we will call \emph{spin triplet states}. These are obtained as a superposition of the three triplet\footnote{The two-particle spin states $|\!\up\up\,\rangle$, $|\,\psi^{\,+}\,\rangle$, and $|\!\down\down\,\rangle$ form a triplet of states to the value $s=1$ of $\vec{S}^{\,2}$, see Eq.~\eqref{eq:spin squared eigenvalue equation}, with $S^{\,z}$ eigenvalues $+1,0$, and $-1$ respectively, while $|\,\psi^{\,-}\,\rangle$ has spin- and magnetic spin quantum numbers equal to zero, forming a singlet, which is already included in the discussion of the Bell $\psi^{\,\pm}$ states.} states, using spherical coordinates in the space of the two particle spin states with the value $s=1$ to parameterize all possible combinations. The spin state $\left|\,\psi\,\right\rangle_{\mathrm{spin}}$ in \eqref{eq:initial total state}, and \eqref{eq:boosted general total state} is thus replaced by
\begin{equation}
\left|\,\psi\,\right\rangle_{\mathrm{spin}}\,=\,\sin\theta\,\cos\phi\left|\,\uparrow\,\uparrow\,\right\rangle\,+\,
\sin\theta\,\sin\phi\,\frac{1}{\sqrt{2}}\left(\left|\,\uparrow\,\downarrow\,\right\rangle\,+\,
\left|\,\downarrow\,\uparrow\,\right\rangle\right)\,+\,
\cos\theta\,\left|\,\downarrow\,\downarrow\,\right\rangle\ ,
	\label{eq:initial spin state triplet type}
\end{equation}
while we use the same momentum state \eqref{eq:initial momentum state} and consider the same reference frames as earlier. To compare the results to the Bell type states, we subsequently analyze the partitions discussed above, and calculate the differences in entanglement for the boosted, and unboosted observers by means of the linear entropy. Starting again with the partition into four individual qubits, we arrive at the analogous expression to \eqref{eq:linear entropy difference of boosted and unboosted bell type state}, which in this case is
\begin{eqnarray}
E(\rho^{\,\Lambda})-E(\rho) &=&  		
	-\frac{1}{4}\sin^{2}\!\delta\sin^{2}(2\alpha)\,\left(\cos\theta+\cos\phi\sin\theta\right)^{2}
	\left[\,-\,5\,+\right.\nonumber \\
&+&\left.\cos(2\theta)+2\sin^{2}\!\theta\,\cos(2\phi)+4\sin(2\theta)\,\cos\phi\right]\ .
\label{eq:linear entropy difference of boosted and unboosted triplet type state}
\end{eqnarray}

\begin{figure}[ht]
	\centering
		\includegraphics[width=0.80\textwidth]{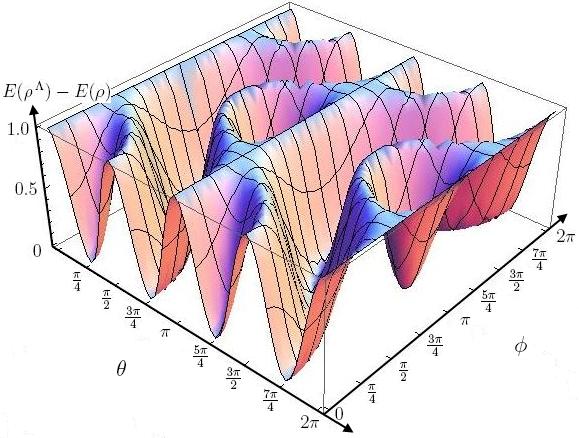}
	\caption[Difference of linear entropy of a $\delta =\pm\tfrac{\pi}{2}$ Wigner rotated spin-triplet-type state with totally symmetric momentum state\newline $(\alpha=\tfrac{\pi}{4})$ for the partition into four individual qubits. Plot of Eq.~\eqref{eq:linear entropy difference of boosted and unboosted triplet type state}.]{Difference of linear entropy of a $\delta =\pm\tfrac{\pi}{2}$ Wigner rotated spin-triplet-type state with totally symmetric momentum state $(\alpha=\tfrac{\pi}{4})$ for the partition into four individual qubits. Plot of Eq.~\eqref{eq:linear entropy difference of boosted and unboosted triplet type state}.}
	\label{fig:entanglementchangetripletttype}
\end{figure}

Due to the different spin parametrization, the result of Eq.~\eqref{eq:linear entropy difference of boosted and unboosted triplet type state} now appears to be much more complicated, although this only arises from the use of the more involved spherical coordinates. We find that the dependence on the Wigner rotation angle $\delta$ and the momentum parameter $\alpha$ is exactly the same as before (see Eq.~\eqref{eq:linear entropy difference of boosted and unboosted bell type state}), and also as earlier, the entanglement change is bigger, the less entangled the initial spin state is.\\

Furthermore, we find a similar saturation of the entanglement, when the initial spin state is chosen maximally entangled, e.g. if $\left|\,\psi\,\right\rangle_{\mathrm{spin}}$ is chosen as a Bell state  $\left|\,\phi^{\,\pm}\right\rangle$ \eqref{eq:bell phi states}, $\left|\,\psi^{\,+}\right\rangle$ \eqref{eq:bell psi states}, or for certain maximally entangled linear combinations of Bell states such as $\frac{1}{N}\left(a\left|\,\Phi^{\,\mp}\right\rangle+b\left|\,\psi^{\,\pm}\right\rangle\right)$ with $N^2=|a|^2+|b|^2$. These states can all\footnote{Note that the antisymmetric Bell state cannot be reached in the current parametrization of the spin state, but we have found the same result for it earlier.} be found on the bottom of the valleys in Fig.~\ref{fig:entanglementchangetripletttype}, indicating that the entanglement of the total states does not change under Lorentz transformations.\\

To further illustrate this, let us consider two examples, first, the maximally entangled initial spin state $\left|\,\Phi^{\,+}\right\rangle$, corresponding to a choice of $\phi=0$ and $\theta=\tfrac{\pi}{4}$, such that the total state is given by
\begin{equation}
\left|\,\psi\,\right\rangle_{\mathrm{total}}\,=\,\left(\,\cos\alpha\left|\,p_{+},p_{-}\,\right\rangle\,+\,\sin\alpha\left|\,p_{-},p_{+}\,\right\rangle\right)\,
\left|\,\Phi^{\,+}\right\rangle\ .
\label{eq:example initial spin state phi plus}
\end{equation}

In our setup, momenta along the $z$-direction, and observer moving in the $x$-direction, this state is transformed to
\begin{eqnarray}
\left|\,\psi\,\right\rangle^{\,\Lambda}_{\mathrm{total}} \ &=& \cos\alpha\left|\,\Lambda p_{+},\Lambda p_{-}\,\right\rangle
		\left(\,\cos\delta\,\left|\,\Phi^{\,+}\right\rangle\,+
		\,\sin\delta\,\left|\,\psi^{\,-}\right\rangle\,\right) \,+\nonumber \\
&+& \,\sin\alpha\left|\,\Lambda p_{-},\Lambda p_{+}\,\right\rangle
		\left(\,\cos\delta\,\left|\,\Phi^{\,+}\right\rangle\,-
		\,\sin\delta\,\left|\,\psi^{\,-}\right\rangle\,\right)\ .
\label{eq:example transformed spin state phi plus}
\end{eqnarray}

In the partition into 4 qubits, the entanglement of the states \eqref{eq:example initial spin state phi plus}, and \eqref{eq:example transformed spin state phi plus} is identical, $E(\rho)=\tfrac{1}{2}(3-\cos(4\alpha))$. Consider, on the other hand, an initially separable spin state, such as $\left|\,\uparrow\uparrow\,\right\rangle$, corresponding to our parametrization $\phi=0$, and $\theta=\tfrac{\pi}{2}$, i.e.
\begin{equation}
\left|\,\psi\,\right\rangle_{\mathrm{total}}\,=\,\left(\,\cos\alpha\left|\,p_{+},p_{-}\,\right\rangle\,+\,\sin\alpha\left|\,p_{-},p_{+}\,\right\rangle\right)\,
\left|\,\uparrow\uparrow\,\right\rangle\ .
\label{eq:example initial spin state upup}
\end{equation}

The difference in entanglement with respect to the 4 qubit partition before and after the boost is
\begin{equation}
E(\rho^{\,\Lambda})-E(\rho)\;=\;\sin^{2}(2\alpha)\sin^{2}\!\delta\ ,
\label{eq:example initial spin state upup entanglement difference}
\end{equation}
such that \eqref{eq:example initial spin state upup} becomes an overall entangled state for a suitable choice of parameters $\alpha$ and $\delta$. The change, represented by \eqref{eq:example initial spin state upup entanglement difference}, becomes maximal for the parameters $\alpha=\tfrac{\pi}{4}$, and $\delta\rightarrow\tfrac{\pi}{2}$, i.e. if the initial momentum state is maximally entangled, and the particles, was well as the observer in the boosted frame, approach the speed of light.\\

As before, we continue with the partition into spin and momentum degrees of freedom, which means we consider the reduced density matrices, obtained by tracing out either the spins, or the momenta. Due to the choice of initial state we have $E(\rho)=0$, and therefore the change in entanglement with respect to this partition is
\begin{eqnarray}
E(\rho^{\,\Lambda}) &\;=\;& 		
	\frac{1}{32}\sin^{2}\!\delta\sin^{2}(2\alpha)\,\left(\cos\theta+\cos\phi\sin\theta\right)^{2}
	\left(\,26+f_{1}-f_{2}\,\right)\nonumber \\
&+& 1-\cos^{4}\!\alpha-\sin^{4}\!\alpha\, -\frac{1}{512}\sin^{2}(2\alpha)\,\left(\,10+f_{1}-f_{2}\,\right)^2 \ ,
\label{eq:spin versus momentum entanglement of boosted triplet type state}
\end{eqnarray}
where we have extracted the functions $f_{1}(\delta,\theta),f_{2}(\delta,\theta,\phi)$, which are given by
\begin{eqnarray}
f_{1}(\delta,\theta) &=& 2\cos(2\delta)\,\left(\,3+\cos(2\theta)\,\right)\,-\,2\cos(2\theta)\\[2mm]
f_{2}(\delta,\theta,\phi) &=& 8\sin^{2}\!\delta\,\left(\,\cos(2\phi)\sin^2\!\theta\,+\,
2\cos\phi\sin(2\theta)\,\right) \ .
\label{eq:functions f1 and f2 for spin versus momentum entanglement of boosted triplet type state}
\end{eqnarray}

\begin{figure}
	\centering
		\includegraphics[width=0.70\textwidth]{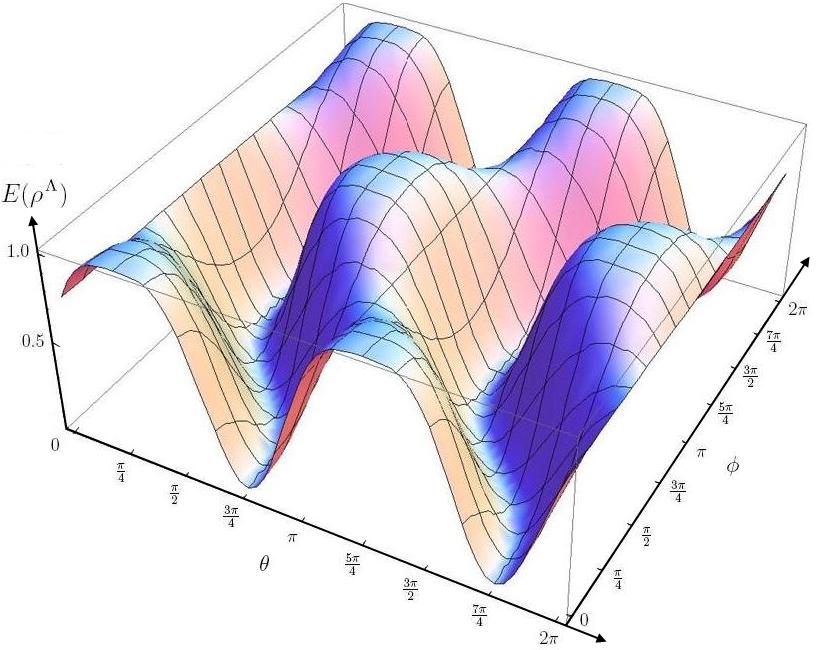}
	\caption[Entanglement difference in the partition into spin- and\newline momentum-degrees of freedom of a $\delta =\pm\tfrac{\pi}{4}$ Wigner rotated spin-triplet-type state with totally symmetric momentum state $(\alpha=\tfrac{\pi}{4})$. Plot of Eq.~\eqref{eq:spin versus momentum entanglement of boosted triplet type state}.]{Entanglement difference in the partition into spin- and momentum-degrees of freedom of a $\delta =\pm\tfrac{\pi}{4}$ Wigner rotated spin-triplet-type state with totally symmetric momentum state $(\alpha=\tfrac{\pi}{4})$. Plot of Eq.~\eqref{eq:spin versus momentum entanglement of boosted triplet type state}.}
	\label{fig:spinmomentumentanglementchangetripletttype}
\end{figure}

This result, even more than the one of the previous partition in \eqref{eq:linear entropy difference of boosted and unboosted triplet type state}, is more involved than its counterpart for the Bell type states in \eqref{eq:spin vs momentum entanglement boosted bell type state}. However, when studying the corresponding three dimensional plot in Fig.~\ref{fig:spinmomentumentanglementchangetripletttype}, we notice that while the valleys, connecting $\theta=\frac{3\pi}{4},\frac{7\pi}{4}, ...\,$, $\phi=0$ with $\theta=\frac{3\pi}{4},\frac{7\pi}{4}, ...\,$, $\phi=2\pi$, remain, there is some change in the ridges in-between (compare Fig.~\ref{fig:entanglementchangetripletttype}). As before there is an imbalance in the entanglement transformation of the spin-momentum partition between the Bell states, in this case $|\phi^{\,+}\rangle$ and $|\psi^{\,+}\rangle\,$, $|\phi^{\,-}\rangle$.\\

Again (as with \eqref{eq:spin vs momentum entanglement boosted bell type state} and \eqref{eq:linear entropy difference of boosted and unboosted bell type state}) we arrive at the same expression as for the 4 qubit partition, Eq.~\eqref{eq:linear entropy difference of boosted and unboosted triplet type state}, when the involved velocities approach the speed of light, $\delta\rightarrow\tfrac{\pi}{2}\,$.\\

At last studying the two particle partition, by tracing over spin and momentum of either Alice, or Bob, we arrive at the rather complicated expression
\begin{eqnarray}
E(\rho) &\;=\;& \frac{1}{256}\Big[\,203\,-\,103\cos(4\alpha)\,+\,
\left(\,3+\cos(4\alpha)\,\right)\ \nonumber \\
&&\,\times\,\big(\,-12\cos(2\theta)\,-\,13\cos(4\theta)\,+\,
16\left(\,3+5\cos(2\theta)\,\right)\cos(2\phi)\sin^2\!\theta \nonumber \\
&&\,+\,8\cos(4\phi)\sin^4\!\theta\,-\,256\cos\theta\cos\phi\sin^3\!\theta\sin^2\!\phi\big)\Big] \ ,
\label{eq:linear entropy of Alice-Bob partition of unboosted triplet type state}
\end{eqnarray}
which, despite its unappealing form, satisfies the criteria of vanishing for fully separable states, i.e. for the parameters $\alpha=\tfrac{n\pi}{2}$, and $\theta=n\pi$, whereas it becomes maximal, $E=\frac{3}{2}$, if both the initial spin state and the initial momentum state are Bell states, i.e. for $\alpha=\tfrac{(2n+1)\pi}{4}$ and  either $\theta=\phi=\tfrac{(2n+1)\pi}{2}$, or $\theta=\tfrac{(2n+1)\pi}{4},\, \phi=n\pi$. Viewing the situation from the perspective of the boosted observer, we get the same expression \eqref{eq:linear entropy of Alice-Bob partition of unboosted triplet type state}, such that we can conclude that there is \emph{no change in the entanglement of the Alice-Bob partition}, as before.

		\subsection{Relativistic Bell Inequalities}
			\label{sec:degree of violation of bell inequalities}

In Sec.~\ref{subsec:two particle states} we found that the entanglement, distributed between the degrees of freedom of a two-particle system, behaves differently under Lorentz transformations, depending on a number of specifications, first of all the partitions of Hilbert space considered, e.g. the total entanglement between the particles remains unchanged, but if only certain degrees of freedom, momentum or spin, are considered, we cannot claim that there is any invariance of the entanglement of the reduced systems. Clearly, we can start out with a state of two particles, where the spin state is entangled, but when considering only the spin degrees of freedom of this state in a different inertial frame, this might not be the case. Moreover, even if we also considered the momenta in the last example, we would find a change in entanglement, if spins and momenta were considered separately. So how can we make assertions about applications of entanglement, if we are not sure, which definition to apply?\\

Clearly, this leads to the question, wether a change of the reference frame will influence the possible violation of a Bell inequality (see Sec.~\ref{subsec:bell theorem}). Surely this question can be addressed by a simple calculation of the expectation values featuring in such an inequality, e.g. the CHSH inequality \eqref{eq:chsh inequality}. Approaches of this kind have been made, e.g. by Czachor in Ref.~\cite{czachor97}, Ahn, Lee, Moon, and Hwang in Ref.~\cite{ahnetal02},\cite{ahnetal03a}, and \cite{moonetal04}, and Chakrabarti in Ref.~\cite{chakrabarti09}, and the authors all find that the violation of Bell inequalities is reduced due to the Lorentz boosts applied.\\

These results do however only consider measurements, where the original measurement directions are maintained. It is not surprising then that a changed spin orientation, originating from the Wigner rotation, will change the expectation values of spin measurements along a fixed direction. Even in a non-relativistic setting it is always possible to choose measurement directions such that a Bell inequality is satisfied, even for a maximally entangled state. In Ref.~\cite{leechangyoung04} Lee, and Chang-Young then show that it is possible to obtain the maximal violation of a Bell inequality in every reference frame, by carefully adjusting the measurement directions. We will go on to show how this can be achieved for arbitrary states in a physically intuitive way.\\

In order to do a meaningful calculation, we need to first find the appropriate observables. As suggested in Ref.~\cite{czachor97}, we will use an observable derived\footnote{A detailed construction can be found in Ref.~\cite{czachor97}.} from the \emph{Pauli-Ljubanski vector} $W^{\,\mu}$ \eqref{eq:pauli ljubanski vector}. Since its square \eqref{eq:pauli ljubanski eigenvalue} is a Casimir operator of the Poincar$\acute{e}$ group, corresponding to spin, it is straightforward to use its spatial part $\overrightarrow{W}$ as a spin observable, where we divide by the component $p^{\,0}$ of the four momentum, to eliminate the additional factor $m$, and we write
\begin{equation}
\vec{\sigma}_{p}\,=\,\frac{\overrightarrow{W}}{p^{\,0}}\,=\,
\sqrt{1-\vec{v}^{\:2}}\,\vec{\sigma}_{\perp}\,+\,\vec{\sigma}_{\parallel}\ ,
\label{eq:relativistic spin observable}
\end{equation}
where $\vec{\sigma}_{\perp}$, and $\vec{\sigma}_{\parallel}$ are the spin components perpendicular, and parallel to the momentum $\,\vec{p}=m\gamma(\vec{v}\,)\vec{v}\ $ of the particle, respectively, i.e.
\begin{equation}
\vec{\sigma}_{\parallel}\,=\,\frac{(\vec{\sigma}\cdot\vec{p}\,)\,\vec{p}}{|\,\vec{p}\,|^{2}}\ ,\ \ \mbox{and}\ \ \
\vec{\sigma}_{\perp}\,=\,\vec{\sigma}\,-\,\vec{\sigma}_{\parallel}\ .
\label{eq:spin perp and parallel to momentum}
\end{equation}

To further construct a binary observable for spin $\tfrac{1}{2}$ particles analogous to those, \eqref{eq:QM bell exp value calculation}, used in the CHCH inequality, we have to evaluate $\vec{\sigma}_{p}$ along a chosen measurement direction $\vec{a}$, resulting in the hermitian observable $\hat{a}(p)$, given by
\begin{equation}
\hat{a}(p)\,=\,\frac{\vec{a}\cdot\vec{\sigma}_{p}}{|\lambda(\vec{a}\cdot\vec{\sigma}_{p})|} \ \ \,,
\label{eq:normalized relativistic spin observable}
\end{equation}
where $\lambda(\vec{a}\cdot\vec{\sigma}_{p})$ is the eigenvalue of the operator $\vec{a}\cdot\vec{\sigma}_{p}$. It can be re-expressed in the form
\begin{equation}
\hat{a}(p)\,=\,\vec{a}_{p}\cdot\vec{\sigma} \quad \; \mbox{with} \quad \; \vec{a}_{p}\,=\,\frac{\sqrt{1-\vec{v}^{\:2}}\,\vec{a}_{\perp}\,+\,\vec{a}_{\parallel}}
{\sqrt{1\,+\,\vec{v}^{\:2}(\vec{a}_{\parallel}^{\,2}\,-\,1)}} \ .
\label{eq:pauli-lubanski-direction}
\end{equation}			

Let us briefly see what the physical meaning of this observable is, by analyzing an idealized spin measurement. Consider a non-relativistic spin $\tfrac{1}{2}$ particle, moving through a Stern-Gerlach apparatus. Clearly, we would say that the measurement device has a well defined orientation with respect to our lab, and by sending the particle through the apparatus, we measure the spin of the particle along that direction. In this non-relativistic case, when changing our point of view to that of the (rest frame of the) particle, we can picture the situation as the other way around, i.e. the detector is moving towards the particle. However, in both situations the lab frame, and the rest frame of the particle will agree on the orientation of the measurement device.\\

Translating this situation to special relativity, this is no longer the case, in the particle rest frame, the orientation of the apparatus will be Lorentz contracted. So how can these two frames generally agree on the outcome of the measurement? The observable in Eq.~\eqref{eq:pauli-lubanski-direction} takes care of exactly this problem. If a measurement direction $\vec{a}$ is chosen in a frame, where the measured particle has momentum $p$, then the direction $\vec{a}_{p}$ is the direction $\vec{a}$, as viewed from the particle rest frame, where it is clear how to calculate the expectation value of a spin measurement, simply dot $\vec{a}_{p}$ into $\vec{\sigma}$, the usual vector of Pauli matrices \eqref{eq:pauli matrices}. With this interpretation of Eq.~\eqref{eq:pauli-lubanski-direction} at hand, we can write it as
\begin{equation}
\hat{a}(p)\,=\,\vec{a}_{p}\cdot\vec{\sigma}\,=\,\frac{(L^{-1}(p)a)^{i}\sigma_{i}}{|(L^{-1}(p)a)^{j}|} \ \ \,,
\label{eq:pauli-lubanski-direction from rest frame}
\end{equation}
where $|(L^{-1}(p)a)^{j}|$ is the norm of the spatial part of the Lorentz transformed orientation vector $L^{-1}(p)a\,$, and $L(p)$, as before, is a Lorentz boost taking the rest frame momentum to $p$. We can now establish the connection to the Wigner rotation. To do this, we basically repeat the arguments, used in the interpretation of Eq.~\eqref{eq:pauli-lubanski-direction} for three different inertial frames, which we will call $S$ (the rest frame of the particle), $S^{\,\prime}$, and $S^{\,\prime\prime}$.\\

Let us start with the particle description in its \emph{rest frame} $S$. There the particle momentum is given by the standard momentum $k$ \eqref{eq:standard momentum}, and the particle state is simply denoted as $|\,k,\sigma\,\rangle\,$. As described above, a spin measurement along the direction $\vec{a}$ (as seen in the rest frame) is represented by the observable
\begin{equation}
\hat{a}\,=\,\frac{\vec{a}\cdot\vec{\sigma}}{|\vec{a}|} \ .
\label{eq:rest frame observable}
\end{equation}

Now let us switch to the perspective of reference frame $S^{\,\prime}$, where the particle is observed to have momentum $p=L(p)k$, and the state vector $|\,p,\sigma\,\rangle$ is obtained such as in Eq.~\eqref{eq:boost of rest frame state}. Since the particle is now no longer at rest, we use the spin observable of Eq.~\eqref{eq:pauli-lubanski-direction from rest frame}. For any chosen measurement direction $a^{\,\prime}$ in the frame $S^{\,\prime}$, the corresponding observable is given by $\hat{a}^{\,\prime}$, in particular, if we want to measure along the same direction as the observer in $S$, which in $S^{\,\prime}$ is given by $a^{\,\prime}=L(p)a$. Therefore we get
\begin{equation}
\hat{a}^{\,\prime}\,=\,\frac{(L^{-1}(\mathrm{p})a^{\,\prime})^{i}\sigma_{i}}
{|(L^{-1}(\mathrm{p})a^{\,\prime})^{j}|}\,=\,
\frac{a^{i}\sigma_{i}}{|a^{j}|}\,=\,\hat{a} \ ,
\label{eq:frame S prime observable}
\end{equation}
which is obvious, since the boost from the rest frame $S$ to the moving frame $S^{\,\prime}$ included no Wigner rotation and consequently, no change of the spin state. Thus the expectation values of $\hat{a}$ in $S$, and $\hat{a}^{\,\prime}$ in $S^{\,\prime}$ agree, both observers obtain the same measurement result, if the observer in the rest frame chooses the direction $a$ and the observer in the moving frame chooses to measure along (the spatial part of) $a^{\,\prime}=L(p)a$.\\

Of course we will expect a similar result, if we yet again change our frame of reference to $S^{\,\prime\prime}$, which we assume to be moving in a perpendicular direction to the particle momenta with respect to $S^{\,\prime}$. Let the corresponding Lorentz transformation be called $\Lambda$, then we obtain the particle momentum $\,\Lambda p\,$ in $S^{\,\prime\prime}$ as
\begin{equation}
\Lambda\,p\,=\,\Lambda\,L(p)\,k \ .
\label{eq:frame S primeprime momentum}
\end{equation}

The state of the particle in this frame is then given by \eqref{eq:unitary rep of lorentz trafo 2},
\begin{equation}
U(\Lambda)\,|\,p,\sigma\,\rangle\,=\,U(\Lambda)\,U(L(p))\,|\,k,\sigma\,\rangle\,=\,U(\mathrm{W}(\Lambda,p))\,|\,\Lambda p,\sigma\,\rangle\ ,
\label{eq:lorentz trafo to frame S primeprime}
\end{equation}
which includes the unitary representation $U(\mathrm{W}(\Lambda,p))$ of the Wigner rotation \eqref{eq:wigner rotation}. Now the observer in $S^{\,\prime\prime}$ wants to measure the spin of the particle along the same direction as the observer in $S^{\,\prime}$, which he obtains by Lorentz transforming the direction $a^{\,\prime}$ to his frame, i.e.
\begin{equation}
a^{\,\prime\prime}\,=\,\Lambda\,a^{\,\prime}\,=\,\Lambda\,L(p)\,a \ .
\label{eq:frame S primeprime measurement direction}
\end{equation}
The correct observable for the measurement in his frame,
\begin{equation}
\hat{a}^{\,\prime\prime}\,=\,\frac{(L^{-1}(\Lambda p)a^{\,\prime\prime})^{i}\sigma_{i}}
{|(L^{-1}(\Lambda p)a^{\,\prime\prime})^{j}|}\,=\,
\frac{(\mathrm{W}(\Lambda ,p)a)^{i}\sigma_{i}}{|(\mathrm{W}(\Lambda ,p)a)^{j}|} \ ,
\label{eq:frame S primeprime observable}
\end{equation}
can then easily be rewritten, to include the expression of the Wigner rotation, using \eqref{eq:frame S primeprime measurement direction}, and \eqref{eq:wigner rotation}. Since $\mathrm{W}(\Lambda ,p)$ is a spatial rotation, it will certainly leave the norm of $\vec{a}$, the spatial part of $a$, invariant,
\begin{equation}
|(\mathrm{W}(\Lambda ,p)a)^{j}|\,=\,|R(\mathrm{W}(\Lambda ,p))\,\vec{a}\,|\,=\,|\,\vec{a}\,|\ ,
\label{eq:norm of wigner rotation of measurement direction}
\end{equation}
and using \eqref{eq:su(2) matrix transformation} we further get
\begin{equation}
\hat{a}^{\,\prime\prime}\,=\,\left(R(\mathrm{W}(\Lambda ,p))\frac{\vec{a}}{|\vec{a}|}\right)\cdot\vec{\sigma}
\;\,=\,\;U(\mathrm{W}(\Lambda ,p))\,
\left[\frac{\vec{a}\cdot\vec{\sigma}}{|\vec{a}|}\right]\,
U^{\dagger}(\mathrm{W}(\Lambda ,p)) \ .
\label{eq:frame S primeprime observable related to S}
\end{equation}
The expectation value of this observable in frame $S^{\,\prime\prime}$ then immediately reveals its equivalence to the expectation value of $\hat{a}$ in frame $S$, when considering Eq.~\eqref{eq:lorentz trafo to frame S primeprime},
\begin{equation}
\langle\,\Lambda p,\sigma\,|\,U^{\dagger}(\mathrm{W}(\Lambda,p))\ \hat{a}^{\,\prime\prime}\ U(\mathrm{W}(\Lambda,p))\,|\,\Lambda p,\sigma\,\rangle\,=\,
\langle\,\sigma\,|\,\frac{\vec{a}\cdot\vec{\sigma}}{|\vec{a}|}\,|\,\sigma\,\rangle\ ,
\label{eq:exp value in S primeprime}
\end{equation}
where we already used the orthogonality \eqref{eq:orthogonality of momentum distributions} of the momentum states. Certainly, using the observable of Eq.~\eqref{eq:normalized relativistic spin observable}, every observer will agree on the expectation values of spin measurements along directions, specified in the rest frame of the particle. Let us quickly sketch how this works for more than one particle, e.g. with a momentum state given by \eqref{eq:initial momentum state}. Consider an observer in a frame, where the two-particle state $\,\left|\,\psi\,\right\rangle\,$ is given by
\begin{equation}
\left|\,\psi\,\right\rangle\,=\,\left|\,p_{+},p_{-}\,\right\rangle\,\left|\,\phi\,\right\rangle_{\mathrm{spin}} ,
	\label{eq:two particle relativistic spin measurement}
\end{equation}
and two measurement devices, called Alice and Bob, are resting far apart from each other. Furthermore, assume that the momenta $p_{+},p_{-}$ are directed at Alice, and Bob respectively. If then the measurement directions for a combined spin measurement are chosen to be $a$ and $b$ on the respective sides, with corresponding observables $\hat{a}(p)$, and $\hat{b}(p)$, we calculate the expectation value as
\begin{eqnarray}
\left\langle\,\psi\,\right|\,\hat{a}(p)\otimes\hat{b}(p)\,\left|\,\psi\,\right\rangle &=&
	\left\langle\,\phi\,\right|\,\left\langle\,p_{+},p_{-}\,\right|\,
	\hat{a}(p)\otimes\hat{b}(p)\,\left|\,p_{+},p_{-}\,\right\rangle\,\left|\,\phi\,\right\rangle\,= \ \nonumber\\[2mm]
&=& \left\langle\,p_{+},p_{-}\,\right|\left.p_{+},p_{-}\,\right\rangle\,\left\langle\,\phi\,\right|\,
	\hat{a}(p_{+})\otimes\hat{b}(p_{-})\,\left|\,\phi\,\right\rangle\,= \nonumber\\[2mm]
&=& \left\langle\,\phi\,\right|\,
	\hat{a}(p_{+})\otimes\hat{b}(p_{-})\,\left|\,\phi\,\right\rangle\ .
	\label{eq:two particle relativistic spin measurement exp value}
\end{eqnarray}

This construction ensures, that the expectation values of all observers will agree, if the measurement directions are appropriately Lorentz transformed. In particular the maximally possible violation of any Bell inequality can be achieved in any reference frame, if there is one reference frame, in which this is possible. The use of the Pauli-Ljubanski spin observable and the calculation of the expectation values as shown above, therefore strongly supports the point of view, obtained from the analysis of the Alice-Bob partition of the two-particle Hilbert space (see Sec.~\ref{subsec:two particle states}). The Bell inequality seems to be sensitive to the entanglement between the particles, and not only on that between their spins.\\

\newpage\section*{Conclusion}\label{conclusion}\addtocontents{toc}{\vspace{-1ex}}\addcontentsline{toc}{section}{Conclusion}		 

We have studied \emph{entanglement}, which has been well analyzed and classified in (non-relativistic) quantum information theory, \emph{in the realm of a relativistic description} of quantum physics. These considerations are strongly influenced by the \emph{Wigner rotations}, which, in turn, arise from group theoretical approaches to the subject. By constructing unitary, irreducible representations of the Poincar$\acute{e}$ group, we find that the transformation properties of the quantum states complicate the definition of entanglement, since not all partitions of the two-particle Hilbert space show the same behavior. Again by resorting to arguments from group theory, i.e. by invoking a spin observable constructed from one of the Casimir invariants of the Poincar$\acute{e}$ group, we find that the partition into particle subspaces matches the predictions for the transformation of a Bell inequality under a change of the inertial reference frame.\\

Can this be interpreted as a \emph{favored role of the particle description}? Or can we find other observables, sensitive to the change of entanglement of a different partition? To investigate these questions in an experiment, is certainly no easy task. Even ignoring the large velocities involved, the question remains, how to produce the states used in our calculations in a laboratory, and how to test the entanglement between spin and momentum degrees of freedom.\\

While there are still open questions about the change in entanglement, the \emph{maximally possible violation of a Bell inequality is maintained} throughout all inertial frames. This means, in particular, that special relativity does not open up chances for local-realistic descriptions, where those were not given in non-relativistic quantum mechanics, but obviously, the relativistic description does not favor any non-local interactions. Although we have not explicitly introduced the second quantized operator formalism, this is clearly done (see e.g. \cite{peskinschroeder}) to obtain local models of nature, in which causality is upheld.\\

Furthermore, when introducing quantum fields, the particle concept becomes less fundamental, and although a violation of Bell inequalities is predicted (for certain parameters), the results for the expectation values \eqref{eq:QM bell exp value calculation} are obtained as approximations only \cite{tommasini02}. We are therefore convinced that it is our \emph{understanding of reality}, which conflicts with the principles of quantum physics.\\

The topic of \emph{relativistic quantum information theory} surely still presents many puzzles, and although much interesting work has been done on that field, e.g. when investigating entanglement in non-inertial frames, which was done by Adesso, Alsing, Ericsson, Fuentes-Schuller, Mann, McMahon, Milburn, and Tessier in Ref.~\cite{alsingmilburn03}, \cite{alsingmcmahonmilburn03}, \cite{fuentesschullermann05}, \cite{alsingfuentesschullermanntessier06}, and \cite{adessofuentesschullerericsson07}, or in the context of general relativity, as was investigated by Adesso, Ball, Borzeszkowski, Fuentes-Schuller, Mensky, and Schuller in Ref.~\cite{borzeszkowskimensky00}, \cite{ballfuentesschullerschuller06}, and \cite{adessofuentesschuller09}, we believe that more intriguing insights will follow about the character of entanglement, when it is studied in a relativistic framework.\\

\newpage\listoffigures

\addtocontents{toc}{\vspace{-1ex}}

\newpage
\section*{Abstract}

One of the most fundamental, but nonetheless hard to grasp phenomena of quantum physics is \emph{entanglement}. It describes an inseparable connection between quantum systems, and the properties thereof. In a quantum mechanical description of the physical world even systems far apart from each other can share a common state, which does not allow for an equivalent description as multiple, distinct, independent objects. This entanglement of the subsystems, although arising from mathematical principles, is no mere abstract concept, but can be tested in experiment, and be utilized in modern quantum information theory procedures, such as quantum teleportation. In particular, entangled states play a crucial role in testing our understanding of reality, by violating \emph{Bell inequalities}.\\

While the role of entanglement is well studied in the realm of non-relativistic quantum mechanics, where detection, classification, and quantification of entanglement are investigated in great detail, its significance in a relativistic quantum theory is a relatively new field of interest. In this work the consequences of a \emph{relativistic description of quantum entanglement} are discussed. To this end, we analyze the representations of the symmetry groups of special relativity, i.e. of the Lorentz group, and the \emph{Poincar}$\mathit{\acute{e}}$ group, on the Hilbert space of quantum states. We desribe how \emph{unitary, irreducible representations} of the Poincar$\acute{e}$ group for massive spin $\tfrac{1}{2}$ particles are constructed from representations of Wigner's little group. We then proceed to investigate the role of the resulting \emph{Wigner rotations} in the transformation of quantum states under a change of inertial reference frame.\\

By considering different partitions of the Hilbert space of two particles, we find that the \emph{entanglement} of the quantum states \emph{appears different in different inertial frames}, depending on the form of the states, the chosen inertial frames, and the particular choice of partition. It is explained, how, despite of this, the maximally possible \emph{violation of Bell inequalities} is \emph{frame independent}, when using appropriate spin observables, which are related to the Pauli-Ljubanski vector, a Casimir operator of the Poincar$\acute{e}$ group.

\newpage
\section*{Zusammenfassung}

\emph{Verschränkung} ist eines der fundamentalsten Phänomene der Quantenphysik, und ist dennoch schwer zu erfassen. Es beschreibt eine untrennbare Verbindung von Quantensystemen und deren Eigenschaften. In einer quantenmechanischen Beschreibung der physikalischen Welt können sogar weit voneinander entfernte Systeme durch gemeinsame Zustände beschrieben werden, welche sich nicht äquivalent als Zerlegung in mehrere verschiedene, von einander unabhängige, Teilsysteme darstellen lassen. Diese Verschränkung der Subsysteme ist nicht bloß ein abstraktes Konzept, obwohl sie aus mathematischen Prinzipien hervorgeht, sondern kann in Experimenten überprüft, und sogar in Anwendungen der Quanteninformationstheorie, wie etwa Quantenteleportation, eingesetzt werden. Verschränkung spielt vor allem eine entscheidende Rolle bei der Verletzung von \emph{Bell-Ungleichungen}, welche unser Verständnis der physikalischen Realität testen.\\

Während sowohl Verschränkung, als auch ihre Detektion, Klassifizierung und auch Quantifizierung, im Rahmen der nicht-relativistischen Quantenmechanik detailliert untersucht wurden, ist die Bedeutung von Verschränkung in dem Kontext relativistischer Quantentheorie erst seit kurzem von Interesse. In dieser Arbeit diskutieren wir die Konsequenzen der relativistischen Beschreibung für verschränkte Quantensysteme. Zu diesem Zweck analysieren wir die Darstellungen der Symmetriegruppen der Speziellen Relativitätstheorie, d.h. der Lorentzgruppe und der \emph{Poincar}$\mathit{\acute{e}}$\emph{gruppe}, auf dem Hilbertraum der Quantenzustände. Weiters wird beschrieben, wie die \emph{unitären, irreduziblen Darstellungen} der Poincar$\acute{e}$gruppe für massive Spin $\tfrac{1}{2}$ Teilchen, durch Aufsuchen der Darstellungen von Wigners kleiner Gruppe, konstruiert werden können. Die Rolle der dabei resultierenden \emph{Wigner-Rotationen} in der Transformation der Quantenzustände bei einem Wechsel des Inertialsystems wird besprochen.\\

Durch Betrachtung verschiedener Partitionen des Hilbertraums zweier Teilchen, kommen wir zu dem Schluss, dass die \emph{Verschränkung} von Quantenzuständen \emph{in unterschiedlichen Inertialsystemen verschieden} erscheint. Dies hängt von der Form der Zustände, den gewählten inertialen Beobachtern und insbesondere auch von der betrachteten Partition ab. Schließlich wird erklärt, warum die maximal mögliche \emph{Verletzung von Bell-Ungleichungen} dennoch \emph{invariant} unter einem solchen Beobachterwechsel ist, wenn die passenden Spin-Observablen herangezogen werden. Letztere stehen in engem Zusammenhang mit dem Pauli-Ljubanski-Vektor, einem Casimir-Operator der Poincar$\acute{e}$gruppe.

\newpage
\vspace*{-2cm}
\section*{Curriculum vitae}
Ich, Nicolai Friis, wurde am 17.Dezember 1984 als Sohn von Elisabeth und Hans Petter Friis in Wien geboren.\\

\hspace*{-0.9cm}
\begin{tabular}{rl}
September \textbf{1995} 	&\ Besuch des Realgymnasiums BRG 17 Geblergasse in\\[1mm]
- Juni \textbf{2003}        &\ Wien, mit Schwerpunkt in Mathematik und\\[1mm]
                            &\ Naturwissenschaften.\\[2mm]
Juni \textbf{2003}          &\ Abschluss der Reifeprüfung mit ausgezeichnetem Erfolg.\\[2mm]
Oktober \textbf{2003}       &\ Zivildienst im Nachbarschaftszentrum 17, des \\[1mm]
- September \textbf{2004}   &\ Wiener Hilfswerks.\\[2mm]
Oktober \textbf{2004} 	    &\ Beginn des Diplomstudiums Physik an der \\[1mm]
                            &\ Universität Wien.\\[2mm]
Juli \textbf{2005}          &\ Abschluss der ersten Diplomprüfung.\\[3mm]
August \textbf{2005}        &\ Teilnahme am Europäischen Forum Alpbach als\\[1mm]
\& August \textbf{2006}     &\ Volontariatsstipendiat.\\[2mm]
Februar \textbf{2007}	    &\ Zuerkennung des Leistungsstipendiums der Universität Wien.\\[2mm]
Jänner \textbf{2008}        &\ Abschluss der zweiten Diplomprüfung mit ausgezeichnetem\\[1mm]
                            &\ Erfolg.\\[2mm]
Februar \textbf{2008}       &\ Zuerkennung des Leistungsstipendiums der Universität Wien.\\[3mm]
Juni \textbf{2008}          &\ Übernahme des Diplomarbeitsthemas.\\[3mm]
September \textbf{2008}     &\ Teilnahme an der ``DPG - School of Physics - Foundations of\\[1mm]
                            &\ Quantum Physics" der Deutschen Physikalischen Gesellschaft.\\[2mm]
Jänner \textbf{2009}        &\ Zuerkennung des Leistungsstipendiums der Universität Wien.\\[2mm]
Oktober \textbf{2008}       &\ Tutor für theoretische Physik an der Universität Wien,\\[1mm]
- Juli \textbf{2009}        &\ Betreuung der Lehrveranstaltungen ``Übungen zu theoretischer\\[1mm]
\& Oktober \textbf{2009}    &\ Physik 1 - Klassische Mechanik" \& ``Übungen zu theoretischer \\[1mm]
- Februar \textbf{2010}     &\ Physik 2 - Quantenmechanik"\\[3mm]
Juni \textbf{2009}          &\ Teilnahme am ``2nd Vienna Symposium on the Foundations of\\[1mm]
                            &\ Modern Physics" an der Fakultät für Physik der Universität\\[1mm]
                            &\ Wien.\\[2mm]
September \textbf{2009}     &\ Teilnahme am Intensivkurs ``The black hole information\\[1mm]
                            &\ paradox" am SISSA in Triest.
\end{tabular}

\end{document}